\documentclass[journal]{IEEEtran}
\usepackage{amsmath,amsfonts}
\usepackage{mathrsfs}
\usepackage{algorithmic}
\usepackage{algorithm}
\usepackage{array}
\usepackage[caption=false,font=normalsize,labelfont=sf,textfont=sf]{subfig}
\usepackage{textcomp}
\usepackage{stfloats}
\usepackage{url}
\usepackage{verbatim}
\usepackage{graphicx}
\usepackage{cite}
\usepackage{balance}
\newtheorem{proposition}{Proposition}
\newtheorem{theorem}{Theorem}
\newtheorem{lemma}{Lemma}
\newtheorem{definition}{Definition}
\newtheorem{corollary}{Corollary}

\newcommand{\argmin}{\mathop{\mathrm{argmin}}\limits}

\begin{document}

\title{Direct and Converse Theorems in Estimating Signals with Sublinear Sparsity}

\author{Keigo~Takeuchi,~\IEEEmembership{Member,~IEEE}
\thanks{
This work was supported in part by the Grant-in-Aid for Scientific Research~(B) (Japan Society for the Promotion of Science (JSPS) KAKENHI) under Grant 26K00948. An earlier version of this paper was submitted in part to 2026 Int. Symp. Inf. Theory and its Appl.
}
\thanks{K.~Takeuchi is with the Department of Electrical and Electronic Information Engineering, Toyohashi University of Technology, Toyohashi 441-8580, Japan (e-mail: takeuchi@ee.tut.ac.jp).}
}

\markboth{IEEE transactions on information theory}%
{Takeuchi: Direct and Converse Theorems in Estimating Signals with Sublinear Sparsity}

\IEEEpubid{0000--0000/00\$00.00~\copyright~xxxx IEEE}

\maketitle

\begin{abstract}
This paper addresses the estimation of signals with sublinear sparsity sent over the additive white Gaussian noise channel. This fundamental problem arises in designing denoisers used in message-passing algorithms for sublinear sparsity. From a theoretical perspective, the main results are direct and converse theorems in the sublinear sparsity limit, where the signal sparsity grows sublinearly in the signal dimension as the signal dimension tends to infinity. As a direct theorem, the maximum likelihood estimator is proved to achieve vanishing square error in the sublinear sparsity limit if the noise variance is smaller than a threshold. This threshold is known to be achievable by an existing separable Bayesian estimator. As a converse theorem, all estimators cannot achieve square errors smaller than the signal power under a mild condition if the noise variance is larger than another threshold. In particular, the two thresholds coincide with each other when non-zero signals have constant amplitude. These results imply the asymptotic optimality of the existing separable Bayesian estimator used in approximate message-passing for sublinear sparsity. From a numerical perspective, a non-separable estimator is proposed via a heuristic approximation of the true posterior mean estimator. Numerical simulations show that the ML estimator and the proposed non-separable estimator outperform the separable Bayesian estimator for high signal-to-noise ratio (SNR). In the low SNR regime, on the other hand, 
the two estimators are inferior to the separable Bayesian estimator while 
the proposed non-separable estimator slightly outperforms the ML estimator.   
\end{abstract}


\section{Introduction}
\subsection{Motivation}
\IEEEPARstart{E}{stimate} an unknown $N$-dimensional and $k$-sparse signal 
vector $\boldsymbol{x}$ sent over the additive white Gaussian noise (AWGN) 
channel, 
\begin{equation} \label{AWGN}
\boldsymbol{y} = \boldsymbol{x} + \boldsymbol{\omega}, 
\quad \boldsymbol{\omega}\sim\mathcal{N}(\boldsymbol{0}, 
\sigma_{N/k}^{2}\boldsymbol{I}_{N}),
\end{equation}
with noise variance $\sigma_{N/k}^{2}=\sigma^{2}/\log(N/k)$ for some 
$\sigma^{2}>0$. 
The signal support $\mathcal{S}_{\boldsymbol{x}}=\{n\in\mathcal{N}=\{1,\ldots,N\}: 
x_{n}\neq0\}$ is assumed to be sampled from all possible 
sets $\mathfrak{S}_{k}^{N}=\{\mathcal{S}\subset\mathcal{N}: |\mathcal{S}|=k\}$ of size~$k$ 
uniformly and randomly. 
In particular, this paper focuses on signals with sublinear sparsity: 
The signal sparsity~$k$---number of non-zero elements in 
$\boldsymbol{x}$---grows sublinearly in $N$ as $N\to\infty$. 
The goal is estimation of the sparse signal vector $\boldsymbol{x}$ based on 
the observation $\boldsymbol{y}$. 
This simple problem arises in designing denoisers used in message-passing 
algorithms for sublinear sparsity~\cite{Takeuchi251,Takeuchi261}. 

The noise power scaling $\sigma_{N/k}^{2}=\sigma^{2}/\log(N/k)$ in (\ref{AWGN}) 
is a key point in estimation of the sparse signal vector. Conventional 
research for sublinear sparsity~\cite{Donoho92,Johnstone04,Pas14,Rockova18} 
postulated $N^{-1}\mathbb{E}[\|\boldsymbol{\omega}\|_{2}^{2}]=\Omega(1)$. 
This normalization implies that error-free estimation is impossible as long as 
the non-zero signal power is ${\cal O}(1)$. In evaluating the performance 
of an estimator $\hat{\boldsymbol{x}}\in\mathbb{R}^{N}$, the normalized 
square error $N^{-1}\|\boldsymbol{x} - \hat{\boldsymbol{x}}\|_{2}^{2}$ was 
used. For the scaling in this paper, on the 
other hand, $k^{-1}\|\boldsymbol{x} - \hat{\boldsymbol{x}}\|_{2}^{2}$ 
normalized by sparsity~$k$---simply called square error---should be 
considered~\cite{Takeuchi251, Takeuchi261}. 

We know that the posterior mean estimator 
$\hat{\boldsymbol{x}}_{\mathrm{opt}}=\mathbb{E}[\boldsymbol{x} | \boldsymbol{y}]$ 
minimizes the mean square error $k^{-1}\mathbb{E}[\|\boldsymbol{x} 
- \hat{\boldsymbol{x}}_{\mathrm{opt}}\|_{2}^{2}]$. However, 
the true posterior mean estimator 
requires high complexity. As a suboptimal low-complexity estimator, 
the element-wise posterior mean estimator 
$\mathbb{E}[x_{n} | y_{n}]$---called separable Bayesian estimator---was 
proposed in \cite{Takeuchi251}. When the essential infimum $u_{\mathrm{min}}$ 
of non-zero absolute signals $\{|x_{n}|: n\in\mathcal{S}_{\boldsymbol{x}}\}$ 
is strictly positive, this estimator was proved to achieve 
asymptotically zero square error if and only if $\sigma^{2}$ is smaller 
than $u_{\mathrm{min}}^{2}/2$~\cite{Takeuchi251}. 

When the separable Bayesian estimator is used as a denoiser, 
both approximate message-passing (AMP)~\cite{Takeuchi251} and orthogonal 
AMP (OAMP)~\cite{Takeuchi261} for sublinear sparsity require strictly 
larger sample complexity than information-theoretically optimal  
complexity~\cite{Wainwright09,Fletcher09,Aeron10,Scarlett17,Aksoylar17,Gamarnik17,Reeves20}. Early research~\cite{Wainwright09,Fletcher09,Aeron10} 
elucidated the scaling of the optimal complexity for the linear model 
in compressed sensing while 
\cite{Scarlett17,Aksoylar17} tackled generalized linear models. The prefactor 
in the optimal scaling was evaluated in \cite{Gamarnik17,Reeves20} for 
constant non-zero signals. These existing results depend strongly 
on the assumption of independent and identically 
distributed (i.i.d.) sensing matrices, so that the linear model in compressed 
sensing cannot be reduced to the AWGN channel~(\ref{AWGN}). 
Thus, the existing results are not directly applicable 
to signal estimation from the observation in the AWGN channel~(\ref{AWGN}).  

\IEEEpubidadjcol

The suboptimality for sublinear sparsity is in contrast to the optimality of 
AMP and OAMP for 
conventional linear sparsity $k/N\to\rho$ for some $\rho\in(0, 1]$: 
For linear sparsity, AMP~\cite{Donoho09,Rangan11} was proved to be 
Bayes-optimal~\cite{Reeves191,Barbier201,Barbier19} via state 
evolution~\cite{Bayati11,Bayati15,Javanmard13,Takeuchi242} when sparse 
signals are compressed with zero-mean i.i.d.\ sub-Gaussian matrices. 
Similarly, OAMP~\cite{Ma17} or 
equivalently vector AMP (VAMP)~\cite{Rangan192} was proved to be 
Bayes-optimal~\cite{Barbier18,Li24} for right-orthogonally invariant 
matrices via state evolution~\cite{Rangan192,Takeuchi20}. See 
\cite{Wang24,Dudeja24} for a generalization of right-orthogonally 
invariant matrices. In contrast to these optimality for linear sparsity, 
both AMP and OAMP are suboptimal for sublinear 
sparsity~\cite{Takeuchi251,Takeuchi261}.

To investigate room for improvement in AMP and OAMP for sublinear sparsity, 
this paper presents information-theoretic analysis for estimation of the sparse 
signal vector in the AWGN channel~(\ref{AWGN}). In the analysis, non-zero 
signals are assumed to be sampled from a discrete set uniformly and randomly. 
The aims of this simplifying assumption are twofold: 
\begin{itemize}
\item In the direct part, i.e.\ achievability of vanishing square error, 
Gallager's bound~\cite{Gallager68} is available 
for evaluating the performance of the maximum likelihood (ML) estimator. 
Gallager's bound is a general technique that was utilized to prove 
achievability in sparse linear regression~\cite{Aksoylar17}. 
In contrast to a long proof for the separable Bayesian 
estimator~\cite{Takeuchi251} via a heuristic strategy, 
the proof of the direct part is simplified for the ML estimator. 

\item Sparse signals with discrete non-zero elements can be regarded 
as index-modulated signals in 
communication~\cite{Mesleh08,Jeganathan08,Basar13,Renzo14}. Thus, 
OAMP~\cite{Takeuchi261} is available for detection of index-modulated 
signals while AMP~\cite{Takeuchi251} requires a strong assumption on the 
channel matrix. The latter aim is outside the scope of this paper, so that 
this communication issue is not discussed anymore.  
\end{itemize}

The converse part, i.e.\ impossibility of vanishing square error is proved 
via a hypothesis testing~\cite{Polyanskiy10,Polyanskiy13}. 
In the hypothesis testing, we test whether the null hypothesis 
$\boldsymbol{y}\sim q(\boldsymbol{y})$ holds for some probability 
density function (pdf) $q(\boldsymbol{y})$ different from 
$p(\boldsymbol{y})$ or the alternative 
hypothesis $\boldsymbol{y}\sim p(\boldsymbol{y})$. When 
the true marginal pdf $p(\boldsymbol{y})$ of $\boldsymbol{y}$ in (\ref{AWGN}) 
is indistinguishable from $q(\boldsymbol{y})$ independent of 
the channel model~(\ref{AWGN}), it is intuitively impossible to estimate 
the signal vector $\boldsymbol{x}$. Inspired by \cite{Reeves20}, this paper 
utilizes a general relationship~\cite{Guo05} between mutual information and 
square error to derive a lower bound on the square error, which depends on the 
Kullback-Leibler (KL) divergence $D(p \| q)$ of $q(\boldsymbol{y})$ from the 
truth $p(\boldsymbol{y})$. In this paper, a Gaussian pdf 
$q(\boldsymbol{y})$ is selected. By proving that $D(p \| q)$ converges to zero, 
this paper proves a converse result. 

\subsection{Contributions}
The contributions of this paper are threefold: achievability, 
converse, and numerical simulations. Part of these contributions were  
presented in \cite{Takeuchi262}.  

Before presenting theoretical results of this paper, additional assumptions 
are summarized. Throughout this paper, non-zero elements of the signal vector 
are sampled from the set of $M$ discrete non-zero points 
$\mathcal{U}=\{u_{m}\in\mathbb{R}\setminus\{0\}: 
m\in\mathcal{M}=\{1,\ldots,M\}\}$ uniformly and 
randomly. Let $u_{\mathrm{min}}=\min_{m\in\mathcal{M}}|u_{m}|$ and 
$u_{\mathrm{max}}=\max_{m\in\mathcal{M}}|u_{m}|$.  
The set of all possible $k$-sparse signal vectors is written as 
$\mathcal{X}_{k}^{N}(\mathcal{U})=\{\boldsymbol{x}\in\mathbb{R}^{N}: 
|\mathcal{S}_{\boldsymbol{x}}|=k, x_{n}\in\mathcal{U}~\hbox{for all 
$n\notin\mathcal{S}_{\boldsymbol{x}}$}\}$.   

\begin{theorem}[Achievability] \label{theorem_Gallager} 
Let $\hat{\boldsymbol{x}}_{\mathrm{ML}}\in\mathcal{X}_{k}^{N}(\mathcal{U})$ 
denote the ML estimator of 
$\boldsymbol{x}$ based on the observation $\boldsymbol{y}$ in (\ref{AWGN}). 
If $\sigma^{2}<u_{\mathrm{min}}^{2}/2$ holds, then the square error 
$k^{-1}\mathbb{E}[\|\boldsymbol{x} - \hat{\boldsymbol{x}}_{\mathrm{ML}}\|_{2}^{2}]$ 
converges to zero in the sublinear sparsity limit: $N$ tends to 
infinity with $\log k/\log N\to\gamma$ for some $\gamma\in[0, 1)$. 
\end{theorem}

Theorem~\ref{theorem_Gallager} implies that the ML estimator satisfies the 
same achievability as the separable Bayesian estimator~\cite{Takeuchi251} in the 
low noise regime $\sigma^{2}<u_{\mathrm{min}}^{2}/2$. 
See Appendix~\ref{appen_separable_Bayesian_estimator} for a review of 
the separable Bayesian estimator. Since the ML estimator 
can be suboptimal in terms of the square error, Theorem~\ref{theorem_Gallager} 
makes no claims for high noise regime. This paper presents the following 
converse result for high noise regime:  

\begin{theorem}[Converse] \label{theorem_converse}
Let $\hat{\boldsymbol{x}}(\boldsymbol{y})\in\mathbb{R}^{N}$ denote any 
estimator of $\boldsymbol{x}$ given $\boldsymbol{y}$ in (\ref{AWGN}). 
If $\sigma^{2}>u_{\mathrm{max}}^{2}/2$ is satisfied, then 
\begin{equation} \label{MSE_lower_bound}
\liminf_{N\to\infty}\frac{1}{k}\left(
 \mathbb{E}[\|\boldsymbol{x} - \hat{\boldsymbol{x}}(\boldsymbol{y})\|_{2}^{2}] 
 - \mathbb{E}[\|\boldsymbol{x}\|_{2}^{2}]
\right) \geq 0
\end{equation}
holds in the sublinear sparsity limit for all 
\begin{equation} \label{gamma_regime}
\gamma < \min\left\{
 \frac{C_{0}}{C_{0} + \sigma^{2}/(u_{\mathrm{max}}u_{\mathrm{min}})},\; 
 \frac{C_{1}}{C_{1} + 3}
\right\},
\end{equation}
with 
\begin{equation} 
C_{0} = \frac{\sigma^{2}}{u_{\mathrm{max}}^{2}}\left(
 1 - \frac{u_{\mathrm{max}}^{2} }{2\sigma^{2}}
\right), 
\end{equation}
\begin{equation}
C_{1} = \frac{\sigma^{2}}{2u_{\mathrm{max}}^{2}}\left(
 1 - \frac{u_{\mathrm{max}}^{2} }{2\sigma^{2}}
\right)^{2}
+ \frac{(u_{\mathrm{max}} - u_{\mathrm{min}}/2)^{2}}{2\sigma^{2}}. 
\end{equation}
\end{theorem}

Theorem~\ref{theorem_converse} claims a stronger result than the strict 
positivity of the square error: It is impossible to reduce the square error 
from the prior value $k^{-1}\mathbb{E}[\|\boldsymbol{x}\|_{2}^{2}]$. The 
condition $\sigma^{2}>u_{\mathrm{max}}^{2}/2$ is required to prove this strong 
converse result. Weak converse results need to be considered to investigate 
the intermediate regime $\sigma^{2}\in(u_{\mathrm{min}}^{2}/2, 
u_{\mathrm{max}}^{2}/2]$. 

\begin{figure}[t]
\centering
\includegraphics[width=\hsize]{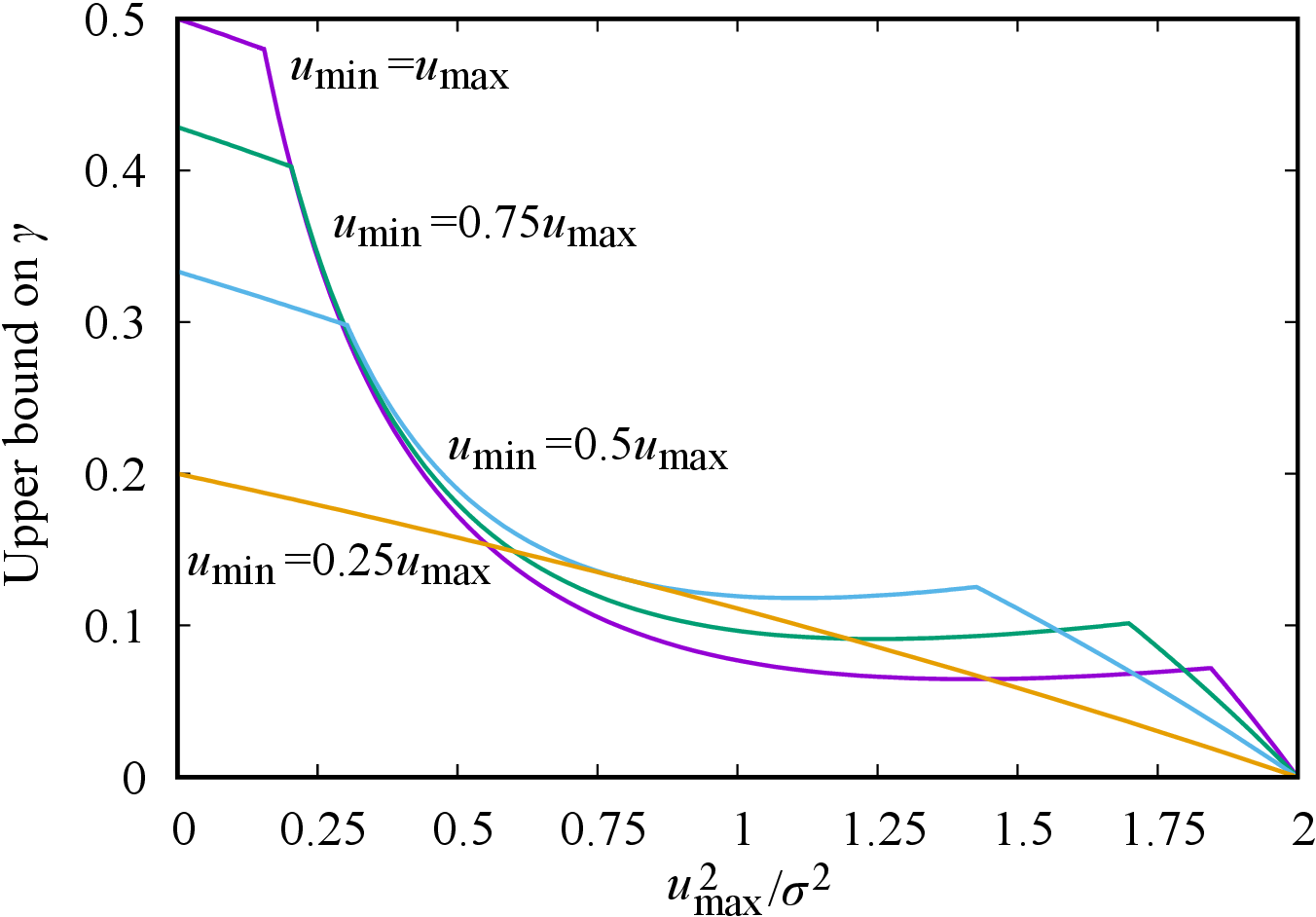}
\caption{Upper bound~(\ref{gamma_regime}) on $\gamma$ versus 
$u_{\mathrm{max}}^{2}/\sigma^{2}$. The lower bound~(\ref{MSE_lower_bound}) 
in Theorem~\ref{theorem_converse} holds when $(u_{\mathrm{max}}^{2}/\sigma^{2}, 
\gamma)$ is located below each curve. 
}
\label{fig0}
\end{figure}

We can obtain a tight result for constant non-zero signals, i.e.\ 
$u_{\mathrm{min}}=u_{\mathrm{max}}$. The upper bound in (\ref{gamma_regime}) is 
strictly positive as long as $\sigma^{2}>u_{\mathrm{min}}^{2}/2$ is satisfied. 
See Fig.~\ref{fig0} for the quantitative evaluation of 
$\gamma$ satisfying (\ref{gamma_regime}). 
Combining Theorems~\ref{theorem_Gallager} and \ref{theorem_converse}, this 
paper concludes the existence of small $\gamma$ such that the square error 
converges to zero in the sublinear sparsity limit if and only if 
$\sigma^{2}<u_{\mathrm{min}}^{2}/2$ holds. In other words, the separable 
Bayesian estimator~\cite{Takeuchi251} is asymptotically optimal 
for small $\gamma$. 

The last contribution is numerical results. A non-separable estimator is 
proposed via a heuristic approximation of the posterior mean 
estimator in the noiseless limit. This estimator is called non-separable 
approximate Bayesian estimator or more simply non-separable estimator. 
The proposed non-separable estimator is numerically shown 
to outperform the ML estimator for all signal-to-noise ratios (SNRs) in terms 
of the square error. Furthermore, it is superior to the separable Bayesian 
estimator~\cite{Takeuchi251} in the high SNR regime. Unfortunately, 
the non-separable estimator cannot improve the performance of 
AMP~\cite{Takeuchi251} for sublinear sparsity compared to the separable 
Bayesian estimator. The last result suggests that AMP for sublinear sparsity 
has a performance bottleneck in another part rather than the type of 
denoisers. 

\subsection{Organization}
The remainder of this paper is organized as follows: After summarizing 
the notation used in this paper, Theorem~\ref{theorem_Gallager} is proved 
in Section~\ref{sec2}. The ML estimator is defined in this section. 
Section~\ref{sec3} presents the proof of 
Theorem~\ref{theorem_converse}. The two theorems are proved with several 
technical lemmas proved in Appendices. Section~\ref{sec4} presents numerical 
results for AMP~\cite{Takeuchi251}. 
Section~\ref{sec5} concludes this paper. 

\subsection{Notation}
Throughout this paper, $\boldsymbol{A}^{\mathrm{T}}$ denotes the transpose of 
a matrix $\boldsymbol{A}$. The notation $\boldsymbol{1}$ represents a 
vector of which the elements are all one while $1(\cdot)$ denotes 
the indicator function. The $p$-norm is written as $\|\cdot\|_{p}$.  
For a vector $\boldsymbol{v}$, its $n$th element is written as $v_{n}$. 
The notation $\boldsymbol{v}_{\setminus n}$ denotes the vector obtained by 
eliminating the element $v_{n}$ from $\boldsymbol{v}$. 
For a set of indices $\mathcal{I}$, the notation $\boldsymbol{v}_{\mathcal{I}}$ 
represents the vector obtained by extracting the elements 
$\{v_{n}: n\in\mathcal{I}\}$ from $\boldsymbol{v}$. 
The $n$th maximum element of a vector $\boldsymbol{v}$ is written as 
$v_{(n)}$ or $[\boldsymbol{v}]_{(n)}$. 
The notational convention $\sum_{n=i}^{j}\cdots = 0$ is used for $i>j$. 

The notation $p(X)$ denotes the pdf of 
an absolutely continuous random variable $X$ while $\mathbb{P}(\cdot)$ 
represents a probability measure. The Gaussian distribution with 
mean $\boldsymbol{\mu}$ and covariance $\boldsymbol{\Sigma}$ is denoted by 
$\mathcal{N}(\boldsymbol{\mu}, \boldsymbol{\Sigma})$. 
The function $Q(x)$ represents the complementary cumulative distribution 
function of the standard Gaussian distribution. 
See Table~\ref{table1} at the end of this paper 
for the list of the other special notation used throughout this paper. 

\section{Direct Part} \label{sec2}
\subsection{Maximum Likelihood Estimation}
Before proving Theorem~\ref{theorem_Gallager}, the ML estimator is presented. 
The ML estimator $\hat{\boldsymbol{x}}_{\mathrm{ML}}
\in\mathcal{X}_{k}^{N}(\mathcal{U})$ of $\boldsymbol{x}$ based on 
$\boldsymbol{y}$ in (\ref{AWGN}) is defined as 
\begin{equation} \label{ML}
\hat{\boldsymbol{x}}_{\mathrm{ML}} 
= \argmin_{\boldsymbol{x}\in\mathcal{X}_{k}^{N}(\mathcal{U})}
p(\boldsymbol{y} | \boldsymbol{x}). 
\end{equation}
Since $\boldsymbol{x}$ is uniformly distributed on 
$\mathcal{X}_{k}^{N}(\mathcal{U})$, the ML estimator~(\ref{ML}) is equivalent 
to the maximum a posteriori (MAP) estimator 
\begin{equation}
\hat{\boldsymbol{x}}_{\mathrm{MAP}}
=\argmin_{\boldsymbol{x}\in\mathcal{X}_{k}^{N}(\mathcal{U})}
p(\boldsymbol{x} | \boldsymbol{y}),
\end{equation} 
which minimizes the error probability 
$\mathbb{P}(\hat{\boldsymbol{x}}_{\mathrm{MAP}}\neq\boldsymbol{x})$. 
In terms of AMP~\cite{Takeuchi251} and OAMP~\cite{Takeuchi261} for sublinear 
sparsity, the posterior mean estimator $\hat{\boldsymbol{x}}_{\mathrm{opt}}
=\mathbb{E}[\boldsymbol{x} | \boldsymbol{y}]$ should be used to minimize 
the mean square error $k^{-1}\mathbb{E}[\|\boldsymbol{x} 
- \hat{\boldsymbol{x}}_{\mathrm{opt}}\|_{2}^{2}]$. Nonetheless, this paper 
focuses on the ML estimator~(\ref{ML}) because the optimality of 
the ML estimator is partially proved in the converse part. 


The ML estimator~(\ref{ML}) can be represented via 
a partition of indices $\mathcal{N}=\{1,\ldots,N\}$: For some 
$k_{*}\in\{0,\ldots,k\}$ decompose $\mathcal{N}$ into three disjoint sets 
$\mathcal{N}_{1}\subset\mathcal{N}$, 
$\mathcal{N}_{2}\subset\mathcal{N}\setminus\mathcal{N}_{1}$, and 
$\mathcal{N}_{3}=\mathcal{N}\setminus(\mathcal{N}_{1}\cup\mathcal{N}_{2})$ 
with cardinality $k_{*}$, $k - k_{*}$, and $N-k$, respectively. The set 
$\mathcal{N}_{1}$ consists of indices that correspond to the top $k_{*}$ 
maximum received signal values while $\mathcal{N}_{2}$ is composed of indices 
for the top $(k - k_{*})$ minimum received signal values. 
The set $\mathcal{N}_{3}$ consists of the remaining indices. 

Let $[\boldsymbol{y}]_{(n)}$ denote the $n$th maximum element of 
$\boldsymbol{y}$, i.e.\ $[\boldsymbol{y}]_{(1)}\geq \cdots 
\geq [\boldsymbol{y}]_{(N)}$. Furthermore, we define 
the permutation $\sigma: \mathcal{N}\to\mathcal{N}$ satisfying 
$y_{\sigma(n)}=[\boldsymbol{y}]_{(n)}$ for all $n\in\mathcal{N}$. 
Using these notations, we have the representation 
$\mathcal{N}_{1} =\{\sigma(n): n\in\{1,\ldots, k_{*}\}\}$, 
$\mathcal{N}_{2} =\{\sigma(n): n\in\{N - (k - k_{*}) + 1,\ldots, N\}\}$, and 
$\mathcal{N}_{3} =\{\sigma(n): n\in\{k_{*} + 1,\ldots, N - (k - k_{*})\}\}$. 
\begin{proposition} \label{proposition_ML}
Determine $\mathcal{N}_{1}$, $\mathcal{N}_{2}$, and $\mathcal{N}_{3}$ with 
\begin{equation} \label{k_minimization}
k_{*}= \argmin_{k_{0}\in\{0,\ldots,k\}}\xi_{k}^{N}(k_{0}; \boldsymbol{y}), 
\end{equation}
where $\xi_{k}^{N}(k_{0}; \boldsymbol{y})$ is given by 
\begin{align} 
\xi_{k}^{N}(k_{0}; \boldsymbol{y}) 
&=  \sum_{i=1}^{k_{0}}\min_{u\in\mathcal{U}}([\boldsymbol{y}]_{(i)} - u)^{2} 
+ \sum_{i=k_{0}+1}^{N - (k - k_{0})}[\boldsymbol{y}]_{(i)}^{2}
\nonumber \\
&+ \sum_{i=N - (k - k_{0})+1}^{N}
\min_{u\in\mathcal{U}}([\boldsymbol{y}]_{(i)} - u)^{2}. 
\label{square_error_ML}
\end{align}
Then, the ML estimator~(\ref{ML}) satisfies $\hat{x}_{\mathrm{ML},n}=0$ 
for all $n\in\mathcal{N}_{3}$. For the remaining elements we have 
\begin{equation} \label{ML_non_zero}
\hat{x}_{\mathrm{ML},n} = \argmin_{u\in\mathcal{U}}(y_{n} - u)^{2}
\end{equation}
for all $n\in\mathcal{N}_{1}\cup\mathcal{N}_{2}$. 
\end{proposition}
\begin{IEEEproof}
See Appendix~\ref{proof_proposition_ML}.
\end{IEEEproof}

The partition $\mathcal{N}=\mathcal{N}_{1}\cup\mathcal{N}_{2}\cup
\mathcal{N}_{3}$ can be computed in ${\cal O}(N)$ time. Furthermore, 
the minimizer~(\ref{k_minimization}) does not depend on 
$\sum_{i=k+1}^{N-k}[\boldsymbol{y}]_{(i)}^{2}$, so that the minimization 
problem~(\ref{k_minimization})  
can be solved in ${\cal O}(k)$ time. Thus, the complexity 
of the ML estimator is linear in $N$.


\subsection{Proof of Theorem~\ref{theorem_Gallager}} 
\label{proof_theorem_Gallager}
To prove Theorem~\ref{theorem_Gallager}, we focus on 
the number of incorrectly detected non-zero 
positions in support recovery and the number of incorrectly detected 
non-zero signals after support recovery. 
For the former number, let $w=|\mathcal{S}_{\boldsymbol{x}}
\setminus\mathcal{S}_{\hat{\boldsymbol{x}}_{\mathrm{ML}}}|\in\{0,\ldots,k\}$. The 
number of correctly detected positions is 
$|\mathcal{S}_{\boldsymbol{x}}\cap\mathcal{S}_{\hat{\boldsymbol{x}}_{\mathrm{ML}}}|=k-w$. 
Furthermore, we have $|\mathcal{S}_{\hat{\boldsymbol{x}}_{\mathrm{ML}}}
\setminus\mathcal{S}_{\boldsymbol{x}}| = w$. When the ML estimator detects 
$(k-w)$ positions correctly, the error in support recovery increases 
the square error by at most $2u_{\mathrm{max}}^{2}w$, with $u_{\mathrm{max}}
=\max_{m\in\mathcal{M}}|u_{m}|$. 

For the latter number, we write the number of incorrectly detected non-zero 
elements in correctly detected positions 
$n\in\mathcal{S}_{\boldsymbol{x}}
\cap\mathcal{S}_{\hat{\boldsymbol{x}}_{\mathrm{ML}}}$ as 
$N_{\boldsymbol{x},\hat{\boldsymbol{x}}_{\mathrm{ML}}} = 
|\mathcal{S}_{\boldsymbol{x}}\cap\mathcal{S}_{\hat{\boldsymbol{x}}_{\mathrm{ML}}}|
- |\{n\in\mathcal{S}_{\boldsymbol{x}}
\cap\mathcal{S}_{\hat{\boldsymbol{x}}_{\mathrm{ML}}}: x_{n}=\hat{x}_{\mathrm{ML}, n}\}|$.  
When $N_{\boldsymbol{x},\hat{\boldsymbol{x}}_{\mathrm{ML}}} = w'$ holds for 
$w'\in\{0,\ldots,k-w\}$, the error 
in non-zero signal detection increases the square error by at most 
$d_{\mathrm{max}}^{2}w'$, 
with $d_{\mathrm{max}}=\max_{m,m'\in\mathcal{M}: m\neq m'}|u_{m} - u_{m'}|$. 
These two error events imply that the square error is bounded from above by 
$2u_{\mathrm{max}}^{2}w + d_{\mathrm{max}}^{2}w'\leq
\max\{2u_{\mathrm{max}}^{2}, d_{\mathrm{max}}^{2}\}k$. 

We evaluate the square error 
$k^{-1}\mathbb{E}[\|\boldsymbol{x} - \hat{\boldsymbol{x}}_{\mathrm{ML}}\|_{2}^{2}]$, 
in which the expectation is over $\boldsymbol{x}$ and $\boldsymbol{y}$ 
in (\ref{AWGN}). The signal vector $\boldsymbol{x}$ is uniformly distributed 
on $\mathcal{X}_{k}^{N}(\mathcal{U})$. 
Let $\mathcal{E}_{d}=\{\|\boldsymbol{x} 
- \hat{\boldsymbol{x}}_{\mathrm{ML}}\|_{2}^{2} \geq 
\max\{2u_{\mathrm{max}}^{2}, d_{\mathrm{max}}^{2}\}d\}$ with  
$d = \lceil k/\sqrt{\log(N/k)} \rceil$.
Since $\|\boldsymbol{x} - \hat{\boldsymbol{x}}_{\mathrm{ML}}\|_{2}^{2}$ 
is bounded from above by $\max\{2u_{\mathrm{max}}^{2}, d_{\mathrm{max}}^{2}\}k$, 
we use the representation 
$\mathbb{E}[\|\boldsymbol{x} - \hat{\boldsymbol{x}}_{\mathrm{ML}}\|_{2}^{2}]
= \mathbb{E}[\|\boldsymbol{x} - \hat{\boldsymbol{x}}_{\mathrm{ML}}\|_{2}^{2} 
| \mathcal{E}_{d}]\mathbb{P}(\mathcal{E}_{d}) 
+ \mathbb{E}[\|\boldsymbol{x} - \hat{\boldsymbol{x}}_{\mathrm{ML}}\|_{2}^{2} 
| \mathcal{E}_{d}^{\mathsf{c}}]\mathbb{P}(\mathcal{E}_{d}^{\mathsf{c}})$ to 
obtain the following upper bound: 
\begin{equation} \label{square_error_upper_bound}
\frac{1}{k}\mathbb{E}\left[
 \|\boldsymbol{x} - \hat{\boldsymbol{x}}_{\mathrm{ML}}\|_{2}^{2}
\right]
< \max\{2u_{\mathrm{max}}^{2}, d_{\mathrm{max}}^{2}\}\left\{
 \mathbb{P}(\mathcal{E}_{d})
 + \frac{d}{k}
\right\}. 
\end{equation}
Thus, $\mathbb{P}(\mathcal{E}_{d})\to0$ implies 
$k^{-1}\mathbb{E}[\|\boldsymbol{x} - \hat{\boldsymbol{x}}_{\mathrm{ML}}\|_{2}^{2}]
\to0$ in the sublinear sparsity limit.  

The main challenge is evaluation of $\mathbb{P}(\mathcal{E}_{d})$. 
To evaluate this probability, we classify the signal space into types.  
Let $\boldsymbol{u}_{\boldsymbol{m}}=[u_{m_{1}},\ldots, u_{m_{k}}]^{\mathrm{T}}
\in\mathcal{U}^{k}$ denote the vector of non-zero signal elements, 
with an index vector $\boldsymbol{m}\in\mathcal{M}^{k}$. 
For a subset $\mathcal{S}\subset\mathfrak{S}_{k}^{N}$ of size~$k$,  
$\boldsymbol{m}\in\mathcal{M}^{|\mathcal{S}|}$, and 
$w\in\{0,\ldots,|\mathcal{S}|\}$,  
define the type $\mathcal{T}_{w,\boldsymbol{m}}^{N}(\mathcal{S})$ 
of $|\mathcal{S}|$-sparse vectors in 
$\mathcal{X}_{|\mathcal{S}|}^{N}(\mathcal{U})$ as 
\begin{equation} 
\mathcal{T}_{w,\boldsymbol{m}}^{N}(\mathcal{S})
= \left\{
 \boldsymbol{x}'\in\mathcal{X}_{|\mathcal{S}|}^{N}(\mathcal{U}): 
 |\mathcal{S}_{\boldsymbol{x}'}\setminus\mathcal{S}| = w,\; 
 \boldsymbol{x}_{\mathcal{S}_{\boldsymbol{x}'}}'
 = \boldsymbol{u}_{\boldsymbol{m}}  
\right\}. \label{type}
\end{equation}
The type $\mathcal{T}_{w,\boldsymbol{m}}^{N}(\mathcal{S}_{\boldsymbol{x}})$ in 
(\ref{type}) contains all possible $k$-sparse vectors that have 
non-zero elements $\boldsymbol{u}_{\boldsymbol{m}}$ and $w$ incorrectly 
detected positions for the signal vector $\boldsymbol{x}$. 
By definition, we have the following partition 
of the signal space $\mathcal{X}_{k}^{N}(\mathcal{U})$: 
\begin{equation} \label{X_set_representation}
\mathcal{X}_{k}^{N}(\mathcal{U})
=\bigcup_{w=0}^{k}\bigcup_{\boldsymbol{m}\in\mathcal{M}^{k}}
\mathcal{T}_{w,\boldsymbol{m}}^{N}(\mathcal{S}).
\end{equation}

We reduce the evaluation of $\mathbb{P}(\mathcal{E}_{d} | \boldsymbol{x})$ 
to that of $\mathbb{P}(\hat{\boldsymbol{x}}_{\mathrm{ML}}
\in\mathcal{T}_{w,\boldsymbol{m}}^{N}(\mathcal{S}_{\boldsymbol{x}}), 
N_{\boldsymbol{x},\hat{\boldsymbol{x}}_{\mathrm{ML}}} = w' | \boldsymbol{x})$ for 
$w'\in\{0,\ldots, k-w\}$. 
The integer $w'\in\{0,\ldots,k-w\}$ in 
$N_{\boldsymbol{x},\hat{\boldsymbol{x}}_{\mathrm{ML}}} = w'$ represents 
the number of incorrectly detected non-zero elements in 
correctly detected positions 
$n\in\mathcal{S}_{\boldsymbol{x}}\cap \mathcal{S}_{\hat{\boldsymbol{x}}_{\mathrm{ML}}}$. 
Thus, this joint event implies 
$\|\boldsymbol{x} - \hat{\boldsymbol{x}}_{\mathrm{ML}}\|_{2}^{2}
\leq 2u_{\mathrm{max}}^{2}w + d_{\mathrm{max}}^{2}w'
\leq\max\{2u_{\mathrm{max}}^{2},d_{\mathrm{max}}^{2}\}(w + w')$. 
By definition, $\mathbb{P}(\mathcal{E}_{d} | \boldsymbol{x})$ is 
affected only by the events satisfying $w + w'\geq d$ for all 
$w\in\{0,\ldots,k\}$ and $w'\in\{0,\ldots,k-w\}$: 
\begin{align}
&\mathbb{P}(\mathcal{E}_{d} | \boldsymbol{x}) \nonumber \\
=& \sum_{w+w'\geq d}\sum_{\boldsymbol{m}\in\mathcal{M}^{k}}
\mathbb{P}(\hat{\boldsymbol{x}}_{\mathrm{ML}}
\in\mathcal{T}_{w,\boldsymbol{m}}^{N}(\mathcal{S}_{\boldsymbol{x}}), 
N_{\boldsymbol{x},\hat{\boldsymbol{x}}_{\mathrm{ML}}} = w' | \boldsymbol{x})
\nonumber \\
&\cdot\mathbb{P}\left(
 \left.
  \mathcal{E}_{d}
 \right| \hat{\boldsymbol{x}}_{\mathrm{ML}}
 \in\mathcal{T}_{w,\boldsymbol{m}}^{N}(\mathcal{S}_{\boldsymbol{x}}), 
 N_{\boldsymbol{x},\hat{\boldsymbol{x}}_{\mathrm{ML}}} = w', \boldsymbol{x}
\right) \nonumber \\
\leq& \sum_{w + w' \geq d}\sum_{\boldsymbol{m}\in\mathcal{M}^{k}}
\mathbb{P}(\hat{\boldsymbol{x}}_{\mathrm{ML}}
\in\mathcal{T}_{w,\boldsymbol{m}}^{N}(\mathcal{S}_{\boldsymbol{x}}),
N_{\boldsymbol{x},\hat{\boldsymbol{x}}_{\mathrm{ML}}} = w' | \boldsymbol{x}), 
\label{probability_event}
\end{align}
where the last inequality follows from the trivial upper bound 
$\mathbb{P}(\mathcal{E}_{d} | \hat{\boldsymbol{x}}_{\mathrm{ML}}
\in\mathcal{T}_{w,\boldsymbol{m}}^{N}(\mathcal{S}_{\boldsymbol{x}}), 
N_{\boldsymbol{x},\hat{\boldsymbol{x}}_{\mathrm{ML}}} = w', \boldsymbol{x})\leq1$. 

We use Gallager's bound~\cite{Gallager68} to obtain the following result: 
\begin{lemma} \label{lemma_Gallager_bound} 
Let $d_{\mathrm{min}}=\min_{m,m'\in\mathcal{M}: m\neq m'}|u_{m} - u_{m'}|>0$ and 
$u_{\mathrm{min}}=\min_{m\in\mathcal{M}}|u_{m}|$. Then, in the case of $w=0$ 
we have 
\begin{align}
&\sum_{\boldsymbol{m}\in\mathcal{M}^{k}}\mathbb{P}(
\hat{\boldsymbol{x}}_{\mathrm{ML}}
\in\mathcal{T}_{0,\boldsymbol{m}}^{N}(\mathcal{S}_{\boldsymbol{x}}),
N_{\boldsymbol{x},\hat{\boldsymbol{x}}_{\mathrm{ML}}} = w'  
| \boldsymbol{x}) \nonumber \\
&\leq \binom{k}{w'}(M - 1)^{w'}
(N/k)^{-\frac{d_{\mathrm{min}}^{2}w'}{8\sigma^{2}}} 
\label{Gallager_bound_w0}
\end{align}
for all $w'\in\{0,\ldots,k\}$ and 
$\boldsymbol{x}\in\mathcal{X}_{k}^{N}(\mathcal{U})$. 
On the other hand, in the case of $w>0$ we have 
\begin{align}
&\sum_{\boldsymbol{m}\in\mathcal{M}^{k}}\mathbb{P}(
\hat{\boldsymbol{x}}_{\mathrm{ML}}
\in\mathcal{T}_{w,\boldsymbol{m}}^{N}(\mathcal{S}_{\boldsymbol{x}}),
N_{\boldsymbol{x},\hat{\boldsymbol{x}}_{\mathrm{ML}}} = w'  
| \boldsymbol{x}) \nonumber \\
&\leq(ek/w)^{w}\binom{k - w}{w'}M^{w}(M - 1)^{w'}
(N/k)^{-E_{w,w'}(\sigma^{2})} \label{Gallager_bound_w}
\end{align}
for all $w'\in\{0,\ldots,k-w\}$ and  
$\boldsymbol{x}\in\mathcal{X}_{k}^{N}(\mathcal{U})$, 
in which $E_{w,w'}(\sigma^{2})$ is given by 
\begin{equation} \label{reliability_function}
E_{w,w'}(\sigma^{2}) = \left\{
 \begin{array}{cl}
 E_{w,w'}^{(1)} & \hbox{for $\sigma^{2}< u_{\mathrm{min}}^{2}/8$,} \\
 E_{w,w'}^{(2)} & \hbox{for $\sigma^{2}\in[u_{\mathrm{min}}^{2}/8, 
u_{\mathrm{min}}^{2}/2),$} \\
 E_{w,w'}^{(3)} & \hbox{for $\sigma^{2}\in\left[u_{\mathrm{min}}^{2}/2, 
\frac{u_{\mathrm{min}}^{2}w + d_{\mathrm{min}}^{2}w'}{2w}\right),$} \\ 
 0 & \hbox{for $\sigma^{2}\geq 
 \frac{u_{\mathrm{min}}^{2}w + d_{\mathrm{min}}^{2}w'}{2w}$,} 
 \end{array}
\right. 
\end{equation}
with
\begin{equation}
E_{w,w'}^{(1)}(\sigma^{2}) = 
\left(
 \frac{u_{\mathrm{min}}^{2}}{4\sigma^{2}} - 1
\right)w
+ \frac{d_{\mathrm{min}}^{2}}{8\sigma^{2}}w',
\end{equation}
\begin{equation} \label{reliability_function2}
E_{w,w'}^{(2)}(\sigma^{2}) 
= \frac{\{2(u_{\mathrm{min}}^{2} - \sqrt{2\sigma^{2}u_{\mathrm{min}}^{2}})w
+ d_{\mathrm{min}}^{2}w'\}^{2}}
{8\sigma^{2}(u_{\mathrm{min}}^{2}w
+ d_{\mathrm{min}}^{2}w')}, 
\end{equation}
\begin{equation}
E_{w,w'}^{(3)}(\sigma^{2}) 
= \frac{\{(u_{\mathrm{min}}^{2} - 2\sigma^{2})w + d_{\mathrm{min}}^{2}w'\}^{2}}
{8\sigma^{2}(u_{\mathrm{min}}^{2}w + d_{\mathrm{min}}^{2}w')}. 
\end{equation}
\end{lemma}
\begin{IEEEproof}
See Appendix~\ref{proof_lemma_Gallager_bound}. 
\end{IEEEproof}

Consider $w>0$ and $w'=0$. 
The function $E_{w,0}(\sigma^{2})/w$ in (\ref{reliability_function}) 
corresponds to Gallager's reliability function while $w$---more precisely 
$w\log(N/k)$---is analogous to 
the code length. Furthermore, the threshold $u_{\mathrm{min}}^{2}/2$ 
corresponds to the channel capacity while $u_{\mathrm{min}}^{2}/8$ corresponds 
to the so-called cutoff rate. When support recovery and signal estimation 
are considered jointly, in general, $E_{w,w'}(\sigma^{2})$ is not 
linear in $w$ or $w'$. 

We utilize the following proposition to evaluate the upper 
bound~(\ref{probability_event}). 
\begin{proposition} \label{proposition_geometric}
Define $d=\lceil k/\sqrt{\log(N/k)} \rceil$ and 
let $r_{k,N}$ denote a variable that depends on $k$ and $N$. 
If $\limsup|r_{k,N}|=0$ is satisfied in the sublinear sparsity limit, then 
$\sum_{w=d}^{k}r_{k,N}^{w}$ converges to zero in the sublinear sparsity limit. 
\end{proposition}
\begin{IEEEproof}
We evaluate the geometric series as 
\begin{equation}
\sum_{w=d}^{k}r_{k,N}^{w} = r_{k,N}^{d}\frac{1 - r_{k,N}^{k - d + 1}}{1 - r_{k,N}} \to 0,
\end{equation}
where the last convergence follows from the definition 
$d=\lceil k/\sqrt{\log(N/k)} \rceil$ and the assumption 
$\limsup|r_{k,N}|=0$. 
\end{IEEEproof}

We are ready to prove Theorem~\ref{theorem_Gallager}. 
From the upper bound~(\ref{square_error_upper_bound}), 
it is sufficient to prove the convergence 
$\mathbb{P}(\mathcal{E}_{d} | \boldsymbol{x})\to0$ 
in the sublinear sparsity limit. 

For $M=1$ we apply Lemma~\ref{lemma_Gallager_bound} to 
the upper bound~(\ref{probability_event}). Since (\ref{Gallager_bound_w}) 
for $w'\neq0$ is zero for $M=1$, we use the linearity 
$E_{w,0}(\sigma^{2})=wE_{1,0}(\sigma^{2})$ to have 
\begin{equation}
\mathbb{P}(\mathcal{E}_{d} | \boldsymbol{x})
\leq \sum_{w=d}^{k}(ek/d)^{w}  
(N/k)^{-wE_{1,0}(\sigma^{2})} \to 0,  
\label{probability_convergence}
\end{equation}
where the last convergence follows from 
Proposition~\ref{proposition_geometric}  
with $r_{k,N} = (ek/d)(N/k)^{-E_{1,0}(\sigma^{2})}$, which satisfies 
\begin{equation}
\log r_{k,N} < - E_{1,0}(\sigma^{2})\log(N/k) 
+ \frac{1}{2}\log\log(N/k) + 1
\to - \infty, 
\end{equation}
because of $d=\lceil k/\sqrt{\log(N/k)} \rceil$ and 
the strict positivity of $E_{1,0}(\sigma^{2})$ 
under the assumption $\sigma^{2}<u_{\mathrm{min}}^{2}/2$. 
See Appendix~\ref{evaluation_probability_event} for the proof of 
$\mathbb{P}(\mathcal{E}_{d} | \boldsymbol{x})\to0$ in the case of $M>1$.

\section{Converse Part} \label{sec3}
\subsection{Proof Overview}
The proof of Theorem~\ref{theorem_converse} consists of two steps: 
A first step is a hypothesis testing with respect to the distribution of 
$\boldsymbol{y}$ in (\ref{AWGN}). As a null hypothesis, the distribution is 
selected to that of $\boldsymbol{y}$ in (\ref{AWGN}) with the signal 
vector $\boldsymbol{x}$ replaced by capacity-achieving Gaussian signaling.  
The alternative hypothesis is the true distribution. The two distributions 
are proved to be asymptotically indistinguishable from each other 
in terms of the KL divergence. 

In the second step, we connect the KL divergence to the square error. 
Inspired by \cite{Reeves20}, we utilize a general relationship~\cite{Guo05} 
between mutual information and square error to derive a lower bound on 
the square error. The lower bound depends on the KL divergence and is useful 
only when the distribution in the null hypothesis is the capacity-achieving 
output distribution and indistinguishable from the true distribution. 
Since these conditions are justified in the first step, we use the lower 
bound to prove Theorem~\ref{theorem_converse}.  

\subsection{Hypothesis Testing}
A hypothesis testing is considered. 
For some pdf $q(\boldsymbol{y})$ on $\mathbb{R}^{N}$ different from the 
true marginal pdf $p(\boldsymbol{y})$ of $\boldsymbol{y}$, 
consider the null hypothesis $\boldsymbol{y}\sim q(\boldsymbol{y})$ 
for the received vector $\boldsymbol{y}$ in (\ref{AWGN}). The alternative 
hypothesis is the true marginal pdf $\boldsymbol{y}\sim 
p(\boldsymbol{y})$. 
Intuitively, it is impossible to estimate the signal vector $\boldsymbol{x}$ 
from the received vector $\boldsymbol{y}$ if the true pdf $p(\boldsymbol{y})$ 
is asymptotically indistinguishable from $q(\boldsymbol{y})$. 
The difference between $p(\boldsymbol{y})$ and $q(\boldsymbol{y})$ is 
measured with the KL divergence.  
\begin{definition}
The KL divergence of $q(\boldsymbol{y})$ from the true pdf 
$p(\boldsymbol{y})$ is defined as 
\begin{equation} \label{KL_divergence}
D(p(\boldsymbol{y}) \| q(\boldsymbol{y})) 
= \int p(\boldsymbol{y})\log\frac{p(\boldsymbol{y})}{q(\boldsymbol{y})}
d\boldsymbol{y}. 
\end{equation}
The KL divergence is simply written as $D(p\|q)$. 
\end{definition}

As the null hypothesis, we select the zero-mean Gaussian pdf 
$q(\boldsymbol{y})$ with covariance 
$N^{-1}\mathbb{E}_{p}[\|\boldsymbol{y}\|_{2}^{2}]\boldsymbol{I}_{N}$, 
in which $\mathbb{E}_{p}$ denotes the expectation with respect to the true 
pdf $p(\boldsymbol{y})$. We know that this pdf $q(\boldsymbol{y})$ 
is the capacity-achieving output pdf for the AWGN 
channel~(\ref{AWGN})~\cite{Cover06}. 
This property is important to prove Theorem~\ref{theorem_converse} in the 
second step. Since the expectation $\mathbb{E}_{q}[\|\boldsymbol{y}\|_{2}^{2}]$ 
with respect to $q(\boldsymbol{y})$ is equal to the true expectation 
$\mathbb{E}_{p}[\|\boldsymbol{y}\|_{2}^{2}]$,  
we are not allowed to distinguish $p(\boldsymbol{y})$ 
from $q(\boldsymbol{y})$ via the power difference. 

\begin{lemma} \label{lemma_KL_divergence}
Suppose $\sigma^{2}>u_{\mathrm{max}}^{2}/2$. For the zero-mean Gaussian pdf 
$q(\boldsymbol{y})$ with 
covariance $N^{-1}\mathbb{E}_{p}[\|\boldsymbol{y}\|_{2}^{2}]\boldsymbol{I}_{N}$, 
the KL divergence $D(p \| q)$ 
converges to zero in the sublinear sparsity limit for 
\begin{equation} \label{gamma_constraint}
\gamma < \min\left\{
 \frac{C_{0}}{C_{0} + \sigma^{2}/(u_{\mathrm{max}}u_{\mathrm{min}})}, 
 \frac{C_{1,1} + C_{1,2}}{C_{1,1} + C_{1,2} + 3}
\right\},
\end{equation} 
with 
\begin{equation} \label{constants}
C_{0} = \frac{\sigma^{2}}{u_{\mathrm{max}}^{2}}\left(
 1 - \frac{u_{\mathrm{max}}^{2} }{2\sigma^{2}}
\right), \quad
C_{1,1} = \frac{\sigma^{2}}{2u_{\mathrm{max}}^{2}}\left(
 1 - \frac{u_{\mathrm{max}}^{2} }{2\sigma^{2}}
\right)^{2},
\end{equation}
\begin{equation} \label{constant2}
C_{1,2} = \frac{(u_{\mathrm{max}} - u_{\mathrm{min}}/2)^{2}}{2\sigma^{2}}. 
\end{equation}
\end{lemma}
\begin{IEEEproof}
See Appendix~\ref{proof_lemma_KL_divergence}. 
\end{IEEEproof}

The pdf $q(\boldsymbol{y})$ in Lemma~\ref{lemma_KL_divergence} is the 
marginal pdf of $\boldsymbol{y}$ in the AWGN channel~(\ref{AWGN}) for 
capacity-achieving Gaussian signaling $\boldsymbol{x}\sim
\mathcal{N}(\boldsymbol{0}, N^{-1}\mathbb{E}[\|\boldsymbol{x}\|_{2}^{2}])$. 
Lemma~\ref{lemma_KL_divergence} implies that, in the high noise regime 
$\sigma^{2} > u_{\mathrm{max}}^{2}/2$, it is impossible to judge 
from the observation $\boldsymbol{y}$ whether sparse signaling has been 
used or Gaussian signaling. In other words, the prior information on 
signal sparsity is useless in the high noise regime. 

The condition $\sigma^{2} > u_{\mathrm{max}}^{2}/2$ cannot be weakened. 
Extreme value theory~\cite[p.~302]{David03} implies that the expected 
maximum of $N$ independent zero-mean Gaussian noise samples with variance 
$\sigma_{N/k}^{2}$ is
$\sqrt{2\sigma_{N/k}^{2}\log N}\{1 + o(1)\}=\sqrt{2\sigma^{2}(1-\gamma)^{-1}}$
$\cdot\{1 + o(1)\}$. For $\sigma^{2} < u_{\mathrm{max}}^{2}/2$, thus, there is 
small $\gamma>0$ such that the maximum non-zero signal is strictly larger 
than all Gaussian noise samples. In this case, it is possible to detect the 
positions of the maximum non-zero signals from the observation 
$\boldsymbol{y}$.  

The condition~(\ref{gamma_constraint}) for $\gamma$ may be weakened. The 
former condition in $\min\{\cdots\}$ is required to control typical Gaussian 
noise samples while the latter condition is associated with non-typical noise 
samples. These two conditions are essential in the proof strategy of 
Lemma~\ref{lemma_KL_divergence}. The condition~(\ref{gamma_constraint}) for 
$\gamma$ may be weakened by using another sophisticated proof strategy. 

\subsection{Square Error}
We prove a lemma to connect the KL divergence $D(p \| q)$ with 
the square error. For notational convenience, 
we write the conditional pdf $p(\boldsymbol{y} | \boldsymbol{x})$ representing 
the AWGN channel~(\ref{AWGN}) as $p_{\boldsymbol{y} | \boldsymbol{x}}$. This 
notation is useful in utilizing the permutation invariance of pdfs. 
We know that the capacity of the AWGN channel~(\ref{AWGN})---the supremum 
of the mutual information $I(\boldsymbol{x}; \boldsymbol{y})$ over 
the distribution of $\boldsymbol{x}$ under the average power constraint 
$N^{-1}\mathbb{E}[\|\boldsymbol{x}\|_{2}^{2}]$---is given by 
\begin{equation} \label{capacity}
C_{\mathrm{AWGN}}
= \frac{1}{2}\log\left(
 1 + \frac{\mathbb{E}[\|\boldsymbol{x}\|_{2}^{2}]}{N\sigma_{N/k}^{2}} 
\right),   
\end{equation}
which is achieved by $\boldsymbol{x}\sim\mathcal{N}(\boldsymbol{0}, 
N^{-1}\mathbb{E}[\|\boldsymbol{x}\|_{2}^{2}]\boldsymbol{I}_{N})$. In this case, 
the capacity-achieving output pdf is equal to the zero-mean 
Gaussian pdf with covariance 
$N^{-1}\mathbb{E}_{p}[\|\boldsymbol{y}\|_{2}^{2}]\boldsymbol{I}_{N}$, i.e.\ 
$q(\boldsymbol{y})$ in Lemma~\ref{lemma_KL_divergence}.  

For this Gaussian pdf $q(\boldsymbol{y})$, define  
\begin{align} \label{J_function}
J(q(\boldsymbol{y})) 
&=C_{\mathrm{AWGN}}
-\mathbb{E}\left[
 \log\frac{p_{\boldsymbol{y} | \boldsymbol{x}}(\boldsymbol{y} | \boldsymbol{x})}
 {q(\boldsymbol{y})}
\right]
+ D(p(\boldsymbol{y}) \| q(\boldsymbol{y})) \nonumber \\
&= D(p(\boldsymbol{y}) \| q(\boldsymbol{y})),  
\end{align}
where the second equality holds because $q(\boldsymbol{y})$ is equal to 
the capacity-achieving output pdf for the AWGN channel~(\ref{AWGN}). 
The following lemma is proved for the capacity-achieving output pdf 
$q(\boldsymbol{y})$ and the AWGN channel~(\ref{AWGN}), while it is possible 
to generalize $q(\boldsymbol{y})$ and the conditional pdf 
$p_{\boldsymbol{y}|\boldsymbol{x}}$ to any pdf and any permutation-invariant 
conditional pdf, i.e.\ $p_{\boldsymbol{y}|\boldsymbol{x}}
(\boldsymbol{P}\boldsymbol{y} | \boldsymbol{P}\boldsymbol{x})
=p_{\boldsymbol{y} | \boldsymbol{x}}(\boldsymbol{y} | \boldsymbol{x})$ for any 
permutation matrix $\boldsymbol{P}$.  

\begin{lemma} \label{lemma_square_error}
Let $\hat{\boldsymbol{x}}(\boldsymbol{y})\in\mathbb{R}^{N}$ denote any 
estimator of $\boldsymbol{x}$ given $\boldsymbol{y}$ in the AWGN 
channel~(\ref{AWGN}). 
Then, for the zero-mean Gaussian pdf $q(\boldsymbol{y})$ with 
covariance $N^{-1}\mathbb{E}_{p}[\|\boldsymbol{y}\|_{2}^{2}]\boldsymbol{I}_{N}$ 
we have 
\begin{equation} \label{mse_lower_bound}
\frac{1}{k}\mathbb{E}[\|\boldsymbol{x} 
- \hat{\boldsymbol{x}}(\boldsymbol{y})\|_{2}^{2}] 
\geq \frac{1}{\alpha}\left\{
 C_{N}(\alpha)e^{-2D(p(\boldsymbol{y}) \| q(\boldsymbol{y}))} - 1
\right\}
\end{equation}
for all $\alpha\in(0, (2u_{\mathrm{max}}^{2})^{-1})$, with
\begin{equation}
C_{N}(\alpha) = \frac{(1 - u_{\mathrm{max}}^{4}\alpha^{2})e^{
\frac{\alpha}{k}\mathbb{E}[\|\boldsymbol{x}\|_{2}^{2}]} }
{\{1 + f(\alpha)[ (1 - k/N)^{-k} - 1 ]\}^{2}},
\end{equation} 
\begin{equation} \label{function_f}
f(\alpha) = \left(
 \frac{1 - u_{\mathrm{max}}^{4}\alpha^{2}}
 {1 - 2u_{\mathrm{max}}^{2}\alpha}
\right)^{1/2} - 1. 
\end{equation}
\end{lemma}
\begin{IEEEproof}
See Appendix~\ref{proof_lemma_square_error}. 
\end{IEEEproof}

Lemma~\ref{lemma_square_error} only provides a trivial bound 
when $D(p(\boldsymbol{y}) \| q(\boldsymbol{y}))$ 
tends to infinity in the sublinear sparsity limit. 
In this case, the lower bound~(\ref{mse_lower_bound}) reduces to 
$k^{-1}\mathbb{E}[\|\boldsymbol{x} 
- \hat{\boldsymbol{x}}(\boldsymbol{y})\|_{2}^{2}]\geq -\alpha^{-1}$ 
in the sublinear sparsity limit. When 
$D(p(\boldsymbol{y}) \| q(\boldsymbol{y}))\to0$ holds in the sublinear 
sparsity limit, on the other hand, we have 
\begin{equation} 
\frac{1}{k}\mathbb{E}[\|\boldsymbol{x} 
- \hat{\boldsymbol{x}}(\boldsymbol{y})\|_{2}^{2}] 
\geq \frac{1}{\alpha}\left\{
 (1 - u_{\mathrm{max}}^{4}\alpha^{2})e^{
\frac{\alpha}{k}\mathbb{E}[\|\boldsymbol{x}\|_{2}^{2}]}  - 1
\right\} + o(1)
\end{equation}
in the sublinear sparsity limit under the additional assumption $\gamma<1/2$. 
Taking the limit $\alpha\to0$ results in the non-trivial lower bound 
$k^{-1}\mathbb{E}[\|\boldsymbol{x} 
- \hat{\boldsymbol{x}}(\boldsymbol{y})\|_{2}^{2}] 
\geq k^{-1}\mathbb{E}[\|\boldsymbol{x}\|_{2}^{2}] + o(1)$. 

\begin{IEEEproof}[Proof of Theorem~\ref{theorem_converse}]
Let $q(\boldsymbol{y})$ denote the zero-mean Gaussian 
pdf with covariance $N^{-1}\mathbb{E}[\|\boldsymbol{y}\|_{2}^{2}]
\boldsymbol{I}_{N}$. Lemma~\ref{lemma_KL_divergence} implies 
$D(p(\boldsymbol{y}) \| q(\boldsymbol{y}))\to0$ in the sublinear sparsity 
limit under the assumptions in Theorem~\ref{theorem_converse}. 
Under the additional assumption $\gamma<1/2$, 
Lemma~\ref{lemma_square_error} implies Theorem~\ref{theorem_converse}. 
Thus, it is sufficient to confirm that the assumption~(\ref{gamma_regime}) 
in Theorem~\ref{theorem_converse} implies $\gamma<1/2$. 

Under the change of variables $u=u_{\mathrm{max}}/\sigma$ 
and $v=u_{\mathrm{min}}/\sigma$, we write the first argument of the 
minimum operator in the assumption~(\ref{gamma_regime}) as  
\begin{equation}
\gamma_{1}(u, v)
= \frac{C_{0}}{C_{0} + (uv)^{-1}}, \quad 
C_{0}= \frac{1}{u^{2}}
\left(
 1 - \frac{u^{2}}{2}
\right). 
\end{equation}
We prove $\gamma_{1}(u, v)< 1/2$ for all $u\in(0, \sqrt{2})$ and 
$v\in(0, u]$. From this inequality, we find that 
the assumption~(\ref{gamma_regime}) implies $\gamma<1/2$. 
Since $\gamma_{1}(u, v)$ is monotonically increasing with respect to 
$v\in(0, u]$ for fixed $u\in(0, \sqrt{2})$, we obtain 
$\gamma_{1}(u, v)\leq \gamma_{1}(u, u)\equiv \gamma_{1}(u)$, given by 
\begin{equation}
\gamma_{1}(u) = \frac{C_{0}u^{2}}{C_{0}u^{2} + 1}
= \frac{2 - u^{2}}{4 - u^{2}} < \frac{1}{2}
\end{equation} 
for all $u^{2}\in(0, 2)$. Thus, $\gamma_{1}(u, v)< 1/2$ holds for all 
$u\in(0, \sqrt{2})$ and $v\in(0, u]$. 
\end{IEEEproof}

\section{Numerical Results} \label{sec4}
\subsection{Non-Separable Approximate Bayesian Estimator}
As shown shortly, the ML estimator~(\ref{ML}) in 
Proposition~\ref{proposition_ML} outperforms the separable Bayesian 
estimator~\cite{Takeuchi251} in the low noise regime. However, 
it is inferior to the separable Bayesian estimator in the high noise 
regime. To improve the performance of the ML estimator in the high noise 
regime, we propose the following non-separable approximate Bayesian estimator 
of $x_{n}$---simply called non-separable estimator: 
\begin{equation} \label{non_separable_denoiser}
\hat{x}_{n} = \frac{\sum_{m\in\mathcal{M}}
u_{m}e^{u_{m}y_{n}/\sigma_{N/k}^{2} - \ell_{m}(\boldsymbol{y}_{\setminus n})/(2\sigma_{N/k}^{2})}}
{1 + \sum_{m\in\mathcal{M}}
e^{u_{m}y_{n}/\sigma_{N/k}^{2} - \ell_{m}(\boldsymbol{y}_{\setminus n})/(2\sigma_{N/k}^{2})}}, 
\end{equation}
where $\ell_{m}(\boldsymbol{y}_{\setminus n})$ is given by 
\begin{align}
\ell_{m}(\boldsymbol{y}_{\setminus n})
&= u_{m}^{2}
+ \min_{k_{0}\in\{0,\ldots,k-1\}}\xi_{k-1}^{N-1}(k_{0}; \boldsymbol{y}_{\setminus n}) 
\nonumber \\
&- \min_{k_{0}\in\{0,\ldots,k\}}\xi_{k}^{N-1}(k_{0}; \boldsymbol{y}_{\setminus n}),  
\end{align}
with $\xi_{k}^{N}(k_{0}; \boldsymbol{y})$ given in (\ref{square_error_ML}). 
See Appendix~\ref{derivation_denoiser} for a heuristic derivation of 
(\ref{non_separable_denoiser}). 

\begin{figure}[t]
\centering
\includegraphics[width=\hsize]{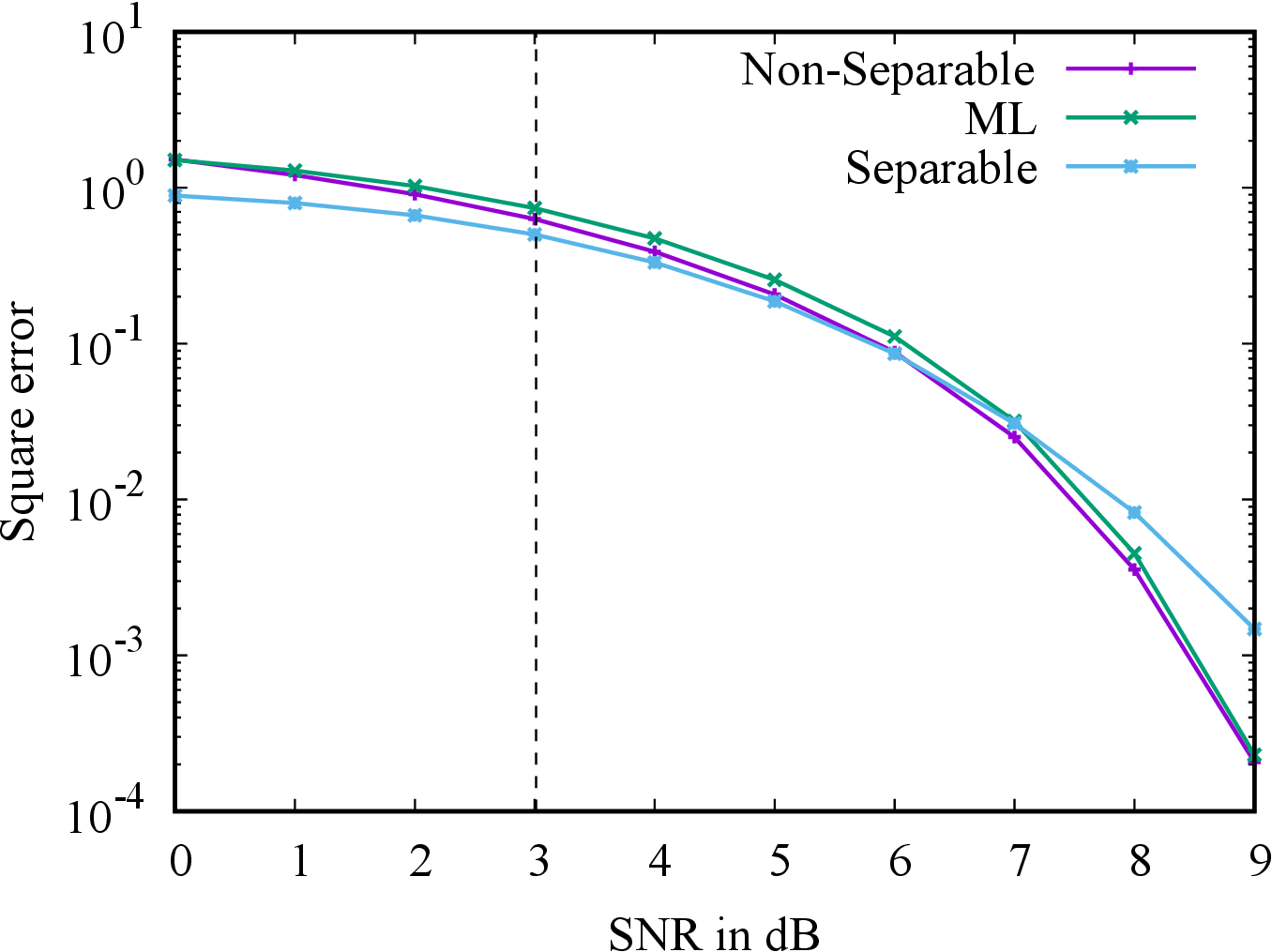}
\caption{Square error versus SNR $1/\sigma^{2}$ in dB for $N=2^{16}$, $k=16$, 
constant non-zero signals $\mathcal{U}=\{1\}$, and 
the AWGN channel~(\ref{AWGN}). 
$10^{5}$ independent trials were simulated for the non-separable 
estimator~(\ref{non_separable_denoiser}), separable Bayesian 
estimator~\cite{Takeuchi251}, and ML estimator~(\ref{ML}). The vertical 
dashed line shows the threshold $1/\sigma^{2}=2$ in 
Theorems~\ref{theorem_Gallager} and \ref{theorem_converse}.   
}
\label{fig1}
\end{figure}

The constant non-zero signals $\mathcal{U}=\{1\}$ were considered in 
numerical simulations to compare the three estimators for the AWGN 
channel~(\ref{AWGN}).  
Figure~\ref{fig1} shows the square error normalized by $k$ versus 
SNR $1/\sigma^{2}$. The ML estimator outperforms 
the separable Bayesian estimator~\cite{Takeuchi251} in the high SNR regime 
while it is inferior to the separable Bayesian estimator in the low SNR 
regime. The non-separable estimator~(\ref{non_separable_denoiser}) 
outperforms the ML estimator~(\ref{ML}) for all SNRs. However, it is 
inferior to the separable Bayesian estimator in the low SNR regime. 
The high and low SNR regimes are separated at approximately 6~dB.  
The suboptimality of the non-separable 
estimator~(\ref{non_separable_denoiser}) is 
due to the max-log approximation in the high SNR regime. 

As shown in \cite[Fig.~1]{Takeuchi251}, the sublinear sparsity limit is 
too far for the threshold $1/\sigma^{2}=2$ to predict a sharp result for 
finite $N$. The threshold predicts a rough position around which the square 
error decreases rapidly.  

\subsection{Compressed Sensing}
The separable Bayesian estimator and the non-separable 
estimator~(\ref{non_separable_denoiser}) are applied to 
AMP~\cite{Takeuchi251} for compressed sensing with sublinear sparsity. 
The ML estimator is not considered, because of its non-differentiability.  
In compressed sensing, the $M$-dimensional measurement vector 
$\boldsymbol{y}\in\mathbb{R}^{M}$ is given by 
\begin{equation} \label{compressed_sensing}
\boldsymbol{y} = k^{-1/2}\boldsymbol{A}\boldsymbol{x} + \boldsymbol{w}, 
\quad \boldsymbol{w}\in\mathcal{N}(\boldsymbol{0}, \sigma^{2}
\boldsymbol{I}_{N}), 
\end{equation}  
where the sensing matrix $\boldsymbol{A}\in\mathbb{R}^{M\times N}$ has 
independent standard Gaussian elements. In the sublinear sparsity limit, 
$M/\{k\log(N/k)\}\to\delta>0$ should hold while $\log k/\log N\to 
\gamma\in[0, 1)$ is satisfied as $k$ and $N$ tend to infinity.  

AMP using the separable Bayesian estimator was proved to achieve vanishing 
square error in the sublinear sparsity 
limit~\cite[Proposition~1]{Takeuchi251} if $\delta$ is larger than the 
threshold $\delta_{\mathrm{AMP}}=2(1 + \sigma^{2})$. This threshold 
$\delta_{\mathrm{AMP}}$ is strictly larger than the optimal 
threshold~\cite{Reeves20} $\delta_{\mathrm{opt}}=2/\log(1 + \sigma^{-2})$.

To improve the convergence property of AMP for finite-sized systems, 
AMP with damping~\cite{Takeuchi251} was simulated. The common damping factor 
for all iterations was optimized via exhaustive search. 

\begin{figure}[t]
\centering
\includegraphics[width=\hsize]{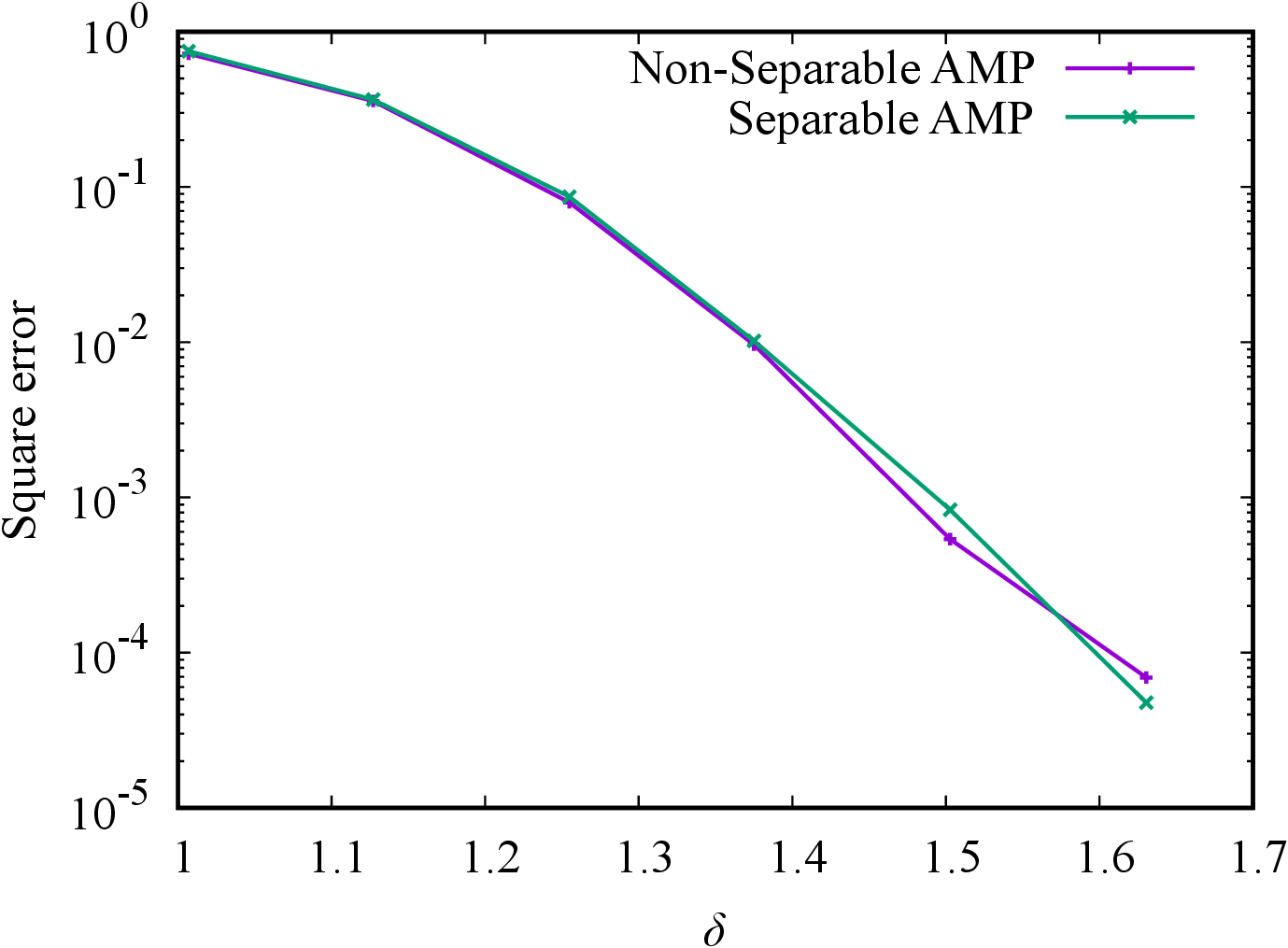}
\caption{Square errors of AMP using the separable Bayesian estimator and 
the non-separable estimator versus $\delta=M/\{k\log(N/k)\}$ for 
$N=2^{16}$, $k=16$, constant non-zero signals $\mathcal{U}=\{1\}$, 
$1/\sigma^{2}=40$~dB, $30$ iterations, and 
the measurement model~(\ref{compressed_sensing}). 
According to the numerical simulations shown in Fig.~\ref{fig1}, 
in non-separable AMP, the non-separable 
estimator~(\ref{non_separable_denoiser}) was used 
when the input SNR to the denoiser is larger than 6~dB. Otherwise, 
the separable Bayesian estimator was used.    
$10^{5}$ independent trials were simulated. The optimal 
threshold~\cite{Reeves20} is approximately $\delta_{\mathrm{opt}}\approx0.22$ 
while the threshold for AMP~\cite{Takeuchi251} is 
$\delta_{\mathrm{AMP}}\approx 2.0$.  
}
\label{fig2}
\end{figure}

Figure~\ref{fig2} shows the square errors for AMP using the separable 
Bayesian estimator and the non-separable 
estimator~(\ref{non_separable_denoiser}). The square errors reduce rapidly 
as $\delta$ increases toward the threshold $\delta_{\mathrm{AMP}}\approx 2$ 
for AMP~\cite{Takeuchi251}. In contrast to the superiority 
of the non-separable estimator in the high SNR regime, the two 
AMP algorithms are comparable to each other for all $\delta$. 

\begin{figure}[t]
\centering
\includegraphics[width=\hsize]{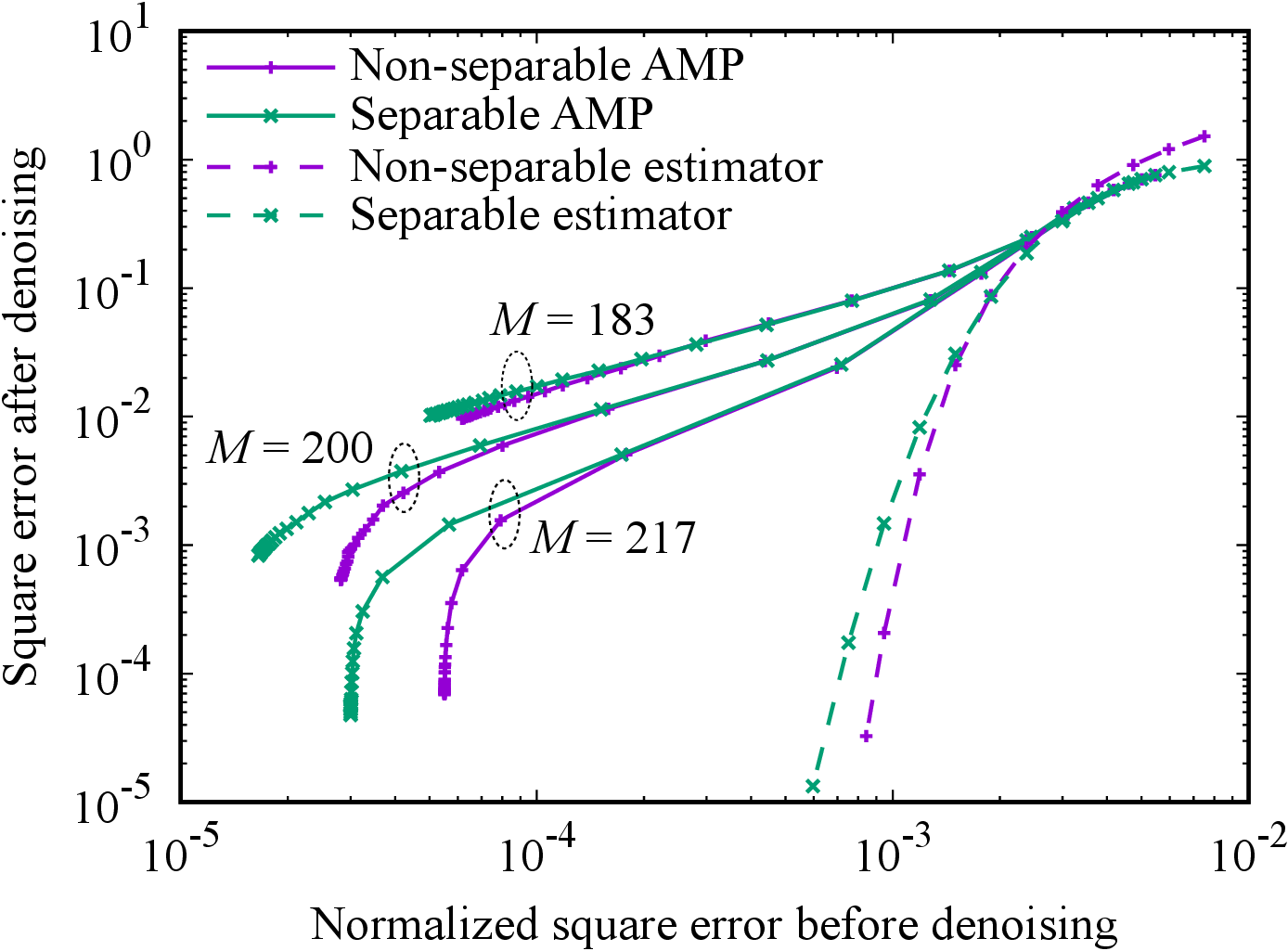}
\caption{Square errors after denoising versus the square error normalized 
by $N$ before denoising for $N=2^{16}$, $k=16$, constant non-zero signals 
$\mathcal{U}=\{1\}$, and $1/\sigma^{2}=40$~dB. The measurement 
model~(\ref{compressed_sensing}) was postulated for the 
separable/non-separable AMP while the AWGN channel~(\ref{AWGN}) was assumed 
for the separable Bayesian/non-separable estimators. 
In non-separable AMP, the non-separable 
estimator~(\ref{non_separable_denoiser}) was used 
when the input SNR to the denoiser is larger than 6~dB. Otherwise, 
the separable Bayesian estimator was used.    
$10^{5}$ independent trials were simulated. 
}
\label{fig3}
\end{figure}

To understand this unexpected result, we focus on the input-output 
relationship for denoising shown in Fig.~\ref{fig3}. The iteration 
trajectories for AMP run from the upper right toward the lower left. 
If perfect Gaussianity held for estimation errors before denoising 
in AMP, the two trajectories for AMP would overlap the corresponding 
input-output relationship for the separable Bayesian estimator and the 
non-separable estimator. 
However, AMP outputs much worse square errors than the 
corresponding estimators for small input square errors while 
it overlaps the separable Bayesian estimator for large input square errors. 
Furthermore, AMP using the non-separable estimator is trapped in  
worse input square errors than the separable counterpart on the final stage of 
convergence. As a result, the two AMP algorithms achieve almost the same 
performance. 

These observations suggest that the performance bottleneck is in 
non-ideal properties for estimation errors before denoising, 
rather than the type of denoisers. Because of finite size effect, the 
estimation errors cannot be regarded as zero-mean i.i.d. Gaussian random 
variables. This non-Gaussianity is a dominant factor for the degradation 
of the convergence property in finite-sized systems. 

\section{Conclusions} \label{sec5}
This paper has proved a sharp result for the estimation of signals 
with sublinear sparsity sent over the AWGN channel: For non-zero signals with 
unit amplitude, the ML estimator achieves vanishing 
square error in the sublinear sparsity limit if $\sigma^{2}$ in the noise 
variance is smaller than $1/2$. Conversely, all estimators result in square 
errors larger than or equal to $1$ in the sublinear sparsity limit with 
small $\gamma$ if $\sigma^{2}$ is larger than $1/2$. These theoretical 
results prove the optimality of the separable Bayesian 
estimator~\cite{Takeuchi251} in the sublinear sparsity limit 
for small $\gamma$. 

Numerical results for AMP suggest that AMP for sublinear sparsity has a 
performance bottleneck in another part---such as finite size effect---rather 
than the type of denoisers. To improve the convergence property of AMP, 
we need to deal with non-Gaussianity for estimation errors before denoising.  

\appendices 

\section{Estimators}
\subsection{Separable Bayesian Estimator}
\label{appen_separable_Bayesian_estimator}
We review the separable Bayesian estimator~\cite{Takeuchi251} for the case 
of the signal support $\mathcal{S}_{\boldsymbol{x}}$ 
sampled from all possible subsets $\mathfrak{S}_{k}^{N}$ of size~$k$ 
uniformly and randomly. When this signal prior is used, as well as 
the true prior of non-zero signals, the posterior mean estimator 
$\hat{\boldsymbol{x}}_{\mathrm{opt}}=\mathbb{E}[\boldsymbol{x} | \boldsymbol{y}]$ 
given $\boldsymbol{y}$ in (\ref{AWGN}) requires high complexity. 
To reduce the complexity, \cite{Takeuchi251} approximated the signal prior 
with the so-called spike and slab prior~\cite{Johnstone04}---
$\boldsymbol{x}$ is assumed to have i.i.d.\ elements that take non-zero 
values and zero with probabilities $k/N$ and $1-k/N$, respectively. 
This approximate signal prior makes the posterior mean estimator separable: 
$\mathbb{E}[x_{n} | \boldsymbol{y}] = \mathbb{E}[x_{n} | y_{n}]$. 
Since it is postulated only in deriving the estimator, the true signal 
prior should be used for performance evaluation. 

Let $U\in\mathbb{R}\setminus\{0\}$ denote a random variable representing 
i.i.d.\ non-zero signals. As derived in \cite[Section~IV-A]{Takeuchi251}, 
the element-wise posterior mean estimator $\mathbb{E}[x_{n} | y_{n}]$ is 
given by 
\begin{equation} \label{separable_Bayesian_estimator}
\mathbb{E}[x_{n} | y_{n}] 
= \frac{k\mathbb{E}_{U}[e^{-(y_{n} - U)^{2}/(2\sigma_{N/k}^{2})}]
\mathbb{E}[U | U + \omega_{n} = y_{n}]}
{k\mathbb{E}_{U}[e^{-(y_{n} - U)^{2}/(2\sigma_{N/k}^{2})}] 
+ (N - k)e^{-y_{n}^{2}/(2\sigma_{N/k}^{2})}}, 
\end{equation}
with $\omega_{n}\sim\mathcal{N}(0, \sigma_{N/k}^{2})$ independent of $U$. 
For the constant non-zero signals $\mathbb{P}(U=1)=1$, the element-wise 
posterior mean estimator reduces to 
\begin{equation}
\mathbb{E}[x_{n} | y_{n}] 
= \frac{ke^{-(y_{n} - 1)^{2}/(2\sigma_{N/k}^{2})}}
{ke^{-(y_{n} - 1)^{2}/(2\sigma_{N/k}^{2})} + (N - k)e^{-y_{n}^{2}/(2\sigma_{N/k}^{2})}}. 
\end{equation}

To compare the main results in this paper with 
an existing result~\cite[Lemma 1]{Takeuchi251} for the separable Bayesian 
estimator, we focus on discrete non-zero signals 
$\mathcal{U}=\{u_{m}\in\mathbb{R}\setminus\{0\}: m\in\mathcal{M}\}$. 
We do not necessarily assume the uniform distribution on $\mathcal{U}$. 
The separable Bayesian estimator~(\ref{separable_Bayesian_estimator}) was 
proved to achieve the following square error in the sublinear sparsity 
limit:  
\begin{theorem}[\cite{Takeuchi251}] 
\label{theorem_separable_Bayesian_estimator}
For all $\sigma^{2}>0$, we have 
\begin{equation}
\frac{1}{k}\sum_{n\in\mathcal{N}}\mathbb{E}\left[
 \left(
  x_{n} -  \mathbb{E}[x_{n} | y_{n}] 
 \right)^{2}
\right] \to \mathbb{E}\left[
 U^{2}1(U^{2} < 2\sigma^{2})
\right] 
\end{equation}
in the sublinear sparsity limit. 
\end{theorem}
\begin{IEEEproof}
Theorem~\ref{theorem_separable_Bayesian_estimator} is equivalent to 
\cite[Lemma~1]{Takeuchi251}. More precisely, \cite[Lemma~1]{Takeuchi251} 
claimed the convergence in probability to treat random small perturbation. 
When the random small perturbation is eliminated, however, the convergence 
in mean is what was actually proved in \cite{Takeuchi251}. 

To use \cite[Lemma~1]{Takeuchi251}, we need to confirm Assumption~7 in 
\cite[p.~4608]{Takeuchi251}. For the discrete random variable 
$U\in\mathcal{U}$, Assumption~7 is trivial with the exception of the uniform 
Lipschitz-continuity for $\mathbb{E}[U | U + \omega_{n}=y_{n}]$ with respect 
to $\sigma_{N/k}^{2}>0$. In fact, this uniform Lipschitz-continuity is broken 
for discrete $U\in\mathcal{U}$ as $\sigma_{N/k}^{2}\to0$. Nonetheless, 
we can still use \cite[Lemma~1]{Takeuchi251}: The uniform 
Lipschitz-continuity assumption was used in \cite{Takeuchi251} only to 
justify that there are some constants $L>0$ and $C>0$ such that 
$|\mathbb{E}[U | U + \omega_{n}=y_{n}]| \leq L|y_{n}| + C$ is satisfied for 
all $y_{n}\in\mathbb{R}$. Since $\mathbb{E}[U | U + \omega_{n}=y_{n}]$ 
is uniformly bounded for discrete $U\in\mathcal{U}$, the inequality 
to be justified is trivial for discrete $U\in\mathcal{U}$.   
\end{IEEEproof}

Theorem~\ref{theorem_separable_Bayesian_estimator} implies that the 
separable Bayesian estimator achieves vanishing square error 
for all $\sigma^{2}<u_{\mathrm{min}}^{2}/2$, as proved in 
Theorem~\ref{theorem_Gallager} for the ML estimator.  
For all $\sigma^{2}>u_{\mathrm{max}}^{2}/2$, on the other hand, the square error 
is equal to $\mathbb{E}[U^{2}]$. The latter result is consistent to 
Theorem~\ref{theorem_converse}. In the intermediate region 
$2\sigma^{2}\in(u_{\mathrm{min}}^{2}, u_{\mathrm{max}}^{2})$, 
Theorem~\ref{theorem_separable_Bayesian_estimator} indicates that the 
square error is in the open interval $(0, \mathbb{E}[U^{2}])$. 

\subsection{Proof of Proposition~\ref{proposition_ML}}
\label{proof_proposition_ML}
We represent the $k$-sparse vector 
$\boldsymbol{x}\in\mathcal{X}_{k}^{N}(\mathcal{U})$ with 
the support set $\mathcal{S}_{\boldsymbol{x}}\in\mathcal{X}_{k}^{N}(\{1\})$ and 
the non-zero values $\boldsymbol{u}=\boldsymbol{x}_{\mathcal{S}_{\boldsymbol{x}}}
\in\mathcal{U}^{k}$. 
Using the definition of $\boldsymbol{y}$ in (\ref{AWGN}), we write 
the ML estimator~(\ref{ML}) as 
\begin{equation} \label{ML_tmp}
(\mathcal{S}_{\hat{\boldsymbol{x}}_{\mathrm{ML}}}, \hat{\boldsymbol{u}})
= \argmin_{(\mathcal{S}, \boldsymbol{u})\in\mathcal{X}_{k}^{N}(\{1\})\times \mathcal{U}^{k}}
\left\{
 \|\boldsymbol{y}_{\mathcal{S}} - \boldsymbol{u}\|_{2}^{2}
 + \sum_{n\notin\mathcal{S}}y_{n}^{2}
\right\},
\end{equation}
where $\hat{\boldsymbol{u}}$ corresponds to non-zero elements 
in the ML estimator. 

Let $\boldsymbol{u}_{+}$ and $\boldsymbol{u}_{-}$ denote the vectors 
obtained by extracting the positive and negative elements from 
$\boldsymbol{u}$, respectively. 
For the dimension $k_{0}\in\{0,\ldots,k\}$ of $\boldsymbol{u}_{+}$, 
we decompose $\mathcal{U}^{k}$ into $k+1$ disjoint subsets 
$\mathcal{U}^{k}(k_{0})=\{\boldsymbol{u}\in\mathcal{U}^{k}: 
\boldsymbol{u}_{+}\in\mathcal{U}_{+}^{k_{0}}\}$ with 
$\mathcal{U}_{+}=\{u_{m}\in\mathcal{U}: u_{m}>0\}$ and 
$\mathcal{U}_{-}=\{u_{m}\in\mathcal{U}: u_{m}<0\}$. 
We represent the objective function in the 
minimization problem~(\ref{ML_tmp}) as 
\begin{align}
&\|\boldsymbol{y}_{\mathcal{S}} - \boldsymbol{u}\|_{2}^{2}
+ \sum_{n\notin\mathcal{S}}y_{n}^{2} \nonumber \\
&= \|\boldsymbol{u}\|_{2}^{2} 
- 2\boldsymbol{u}_{+}^{\mathrm{T}}\boldsymbol{y}_{\mathcal{S}_{+}}
- 2\boldsymbol{u}_{-}^{\mathrm{T}}\boldsymbol{y}_{\mathcal{S}_{-}}
+ \|\boldsymbol{y}\|_{2}^{2}, 
\end{align}
where $\mathcal{S}_{\pm}$ consists of the indices of 
$\boldsymbol{y}_{\mathcal{S}}$ that correspond to $\boldsymbol{u}_{\pm}$. 
Minimizing this expression over the support sets $\mathcal{S}_{+}$ and 
$\mathcal{S}_{-}$, we find that the minimizers $\mathcal{S}_{+}^{*}$ and 
$\mathcal{S}_{-}^{*}$ are the indices that correspond to the top $k_{0}$ 
maximum and top $(k-k_{0})$ minimum elements of $\boldsymbol{y}$, 
respectively. Thus, $\hat{x}_{\mathrm{ML},n}=0$ holds for all 
$n\in\mathcal{N}\setminus(\mathcal{N}_{1}\cup\mathcal{N}_{2})$, with 
$\mathcal{N}_{1}=\mathcal{S}_{+}^{*}$ and 
$\mathcal{N}_{2}=\mathcal{S}_{-}^{*}$. 

We focus on the minimization over the remaining vector 
$\boldsymbol{u}$. Let $y_{(n)}$ denote the 
$n$th maximum of $\{y_{n}: n\in\mathcal{N}\}$. The objective function 
in (\ref{ML_tmp}) minimized over $\mathcal{S}$ is represented as 
\begin{equation}
\sum_{i=1}^{k_{0}}(y_{(i)} - u_{i})_{2}^{2} 
+ \sum_{i=k_{0}+1}^{N - (k - k_{0})}y_{(i)}^{2}
+ \sum_{i=N - (k - k_{0}) + 1}^{N}(y_{(i)} - u_{i})^{2},
\end{equation}
where $\{u_{i}\in\mathcal{U}_{+}: i\in\{1,\ldots,k_{0}\}\}$ and 
$\{u_{i}\in\mathcal{U}_{-}: i\in\{N-(k-k_{0})+1,\ldots,N\}\}$ 
denote the non-zero elements 
associated with $\mathcal{N}_{1}$ and $\mathcal{N}_{2}$, respectively. 
Minimizing this expression over $\boldsymbol{u}\in\mathcal{U}^{k}(k_{0})$, 
we find 
\begin{equation}
\hat{x}_{\mathrm{ML},n} = \argmin_{u\in\mathcal{U}_{+}}(y_{n} - u)^{2}
\end{equation}
for all $n\in\mathcal{N}_{1}$ and 
\begin{equation}
\hat{x}_{\mathrm{ML},n} = \argmin_{u\in\mathcal{U}_{-}}(y_{n} - u)^{2}
\end{equation}
for all $n\in\mathcal{N}_{2}$. 
Furthermore, we minimize the objective function over 
$k_{0}\in\{0,\ldots,k\}$ to obtain 
\begin{align} 
&\|\boldsymbol{y} - \boldsymbol{x}\|_{2}^{2}
\geq  \min_{k_{0}\in\{0,\ldots,k\}}\left\{
 \sum_{i=1}^{k_{0}}\min_{u\in\mathcal{U}_{+}}([\boldsymbol{y}]_{(i)} - u)^{2} 
\right.
\nonumber \\
&\left.
 + \sum_{i=k_{0}+1}^{N - (k - k_{0})}[\boldsymbol{y}]_{(i)}^{2}
 + \sum_{i=N - (k - k_{0})+1}^{N}
 \min_{u\in\mathcal{U}_{-}}([\boldsymbol{y}]_{(i)} - u)^{2}
\right\}.  \label{square_error_ML_tmp}
\end{align}

We are ready to prove Proposition~\ref{proposition_ML}. By relaxing the 
feasible sets $\mathcal{U}_{\pm}$ to $\mathcal{U}$, we find that 
$\xi_{k}^{N}(k_{*},\boldsymbol{y})$ in (\ref{square_error_ML}) with 
$k_{0}=k_{*}$ in (\ref{k_minimization}) is a lower bound on the objective 
function~(\ref{square_error_ML_tmp}). Since $\xi_{k}^{N}(k_{*},\boldsymbol{y})$ 
is achievable, we arrive at Proposition~\ref{proposition_ML}. 

\subsection{Non-Separable Approximate Bayesian Estimator} 
\label{derivation_denoiser}
We present a heuristic derivation for the non-separable approximate Bayesian 
estimator~(\ref{non_separable_denoiser}).  
We start with the optimum posterior mean estimator 
$\hat{\boldsymbol{x}}_{\mathrm{opt}} 
= \mathbb{E}[\boldsymbol{x} | \boldsymbol{y}]$, given by 
\begin{equation}
\hat{x}_{\mathrm{opt},n} 
= \frac{\mathbb{E}_{\boldsymbol{x}}[ x_{n}
p(\boldsymbol{y}|\boldsymbol{x}) ]}
{\mathbb{E}_{\boldsymbol{x}}[ p(\boldsymbol{y}|\boldsymbol{x}) ]}, 
\end{equation}
where the expectation is over $\boldsymbol{x}$ uniformly distributed 
on $\mathcal{X}_{k}^{N}(\mathcal{U})$. 
Using the definition of the AWGN channel~(\ref{AWGN}) yields 
\begin{align}
&\hat{x}_{\mathrm{opt},n} 
= \frac{\mathbb{E}_{\boldsymbol{x}}[ x_{n}
e^{-\|\boldsymbol{y} - \boldsymbol{x}\|_{2}^{2}/(2\sigma_{N/k}^{2})} ]}
{\mathbb{E}_{\boldsymbol{x}}[ 
e^{-\|\boldsymbol{y} - \boldsymbol{x}\|_{2}^{2}/(2\sigma_{N/k}^{2})} ]} \nonumber \\
&= \frac{
 \sum_{m\in\mathcal{M}}u_{m}
 e^{- \frac{(y_{n} - u_{m})^{2}}{2\sigma_{N/k}^{2}}}  
}
{
 \sum_{m\in\mathcal{M}} 
 e^{- \frac{(y_{n} - u_{m})^{2}}{2\sigma_{N/k}^{2}}}  
 + e^{-\frac{y_{n}^{2}}{2\sigma_{N/k}^{2}}}e^{L_{\mathrm{opt}}(\boldsymbol{y}_{\setminus n})}
}, \label{posterior_mean_estimator}
\end{align}
with 
\begin{align}
&L_{\mathrm{opt}}(\boldsymbol{y}_{\setminus n})
= \log\left\{
 \sum_{\boldsymbol{x}_{\setminus n}\in\mathcal{X}_{k}^{N-1}(\mathcal{U})}
 e^{
 - \|\boldsymbol{y}_{\setminus n} - \boldsymbol{x}_{\setminus n}\|_{2}^{2}
 /(2\sigma_{N/k}^{2}) 
 }
\right\} \nonumber \\
&- \log\left( 
 \sum_{\boldsymbol{x}_{\setminus n}\in\mathcal{X}_{k-1}^{N-1}(\mathcal{U})}
 e^{
  - \|\boldsymbol{y}_{\setminus n} - \boldsymbol{x}_{\setminus n}\|_{2}^{2}
  /(2\sigma_{N/k}^{2}) 
 }
\right). \label{L_opt_function} 
\end{align}

To reduce the complexity of $L_{\mathrm{opt}}(\boldsymbol{y}_{\setminus n})$, 
we use the max-log approximation in the low noise limit $\sigma^{2}\to0$ 
to obtain $L_{\mathrm{opt}}(\boldsymbol{y}_{\setminus n}) \approx
L(\boldsymbol{y}_{\setminus n})$, with 
\begin{align}
2\sigma_{N/k}^{2}L(\boldsymbol{y}_{\setminus n}) &= 
\min_{\boldsymbol{x}_{\setminus n}\in\mathcal{X}_{k-1}^{N-1}(\mathcal{U})}
\|\boldsymbol{y}_{\setminus n} - \boldsymbol{x}_{\setminus n}\|_{2}^{2}
\nonumber \\
&-\min_{\boldsymbol{x}_{\setminus n}\in\mathcal{X}_{k}^{N-1}(\mathcal{U})}
\|\boldsymbol{y}_{\setminus n} - \boldsymbol{x}_{\setminus n}\|_{2}^{2}. 
\end{align}
Using Proposition~\ref{proposition_ML} yields 
\begin{align}
2\sigma_{N/k}^{2}L(\boldsymbol{y}_{\setminus n}) 
&= \min_{k_{0}\in\{0,\ldots,k-1\}}\xi_{k-1}^{N-1}(k_{0}; \boldsymbol{y}_{\setminus n})
\nonumber \\
&- \min_{k_{0}\in\{0,\ldots,k\}}\xi_{k}^{N-1}(k_{0}; \boldsymbol{y}_{\setminus n}), 
\end{align}
with $\xi_{k}^{N}(k_{0}; \boldsymbol{y})$ in 
(\ref{square_error_ML}). Thus, (\ref{non_separable_denoiser}) is obtained.

\section{Proof of Lemma~\ref{lemma_Gallager_bound}}
\label{proof_lemma_Gallager_bound} 
\subsection{Overview}
The proof of Lemma~\ref{lemma_Gallager_bound} consists of three steps. 
A first step is the decomposition of the type 
$\mathcal{T}_{w,\boldsymbol{m}}^{N}(\mathcal{S}_{\boldsymbol{x}})$ in (\ref{type}) 
into two types: A first type is associated with the true signal support 
$\mathcal{S}_{\boldsymbol{x}}$. The other type corresponds to the remaining 
indices $\mathcal{N}\setminus\mathcal{S}_{\boldsymbol{x}}$. For the former 
type, the simple union bound is used. 

A second step is Gallager's bound for the latter type. In this step, 
we follow the standard procedure~\cite{Gallager68} to obtain Gallager's 
bound. 

The last step is the optimization of Gallager's reliability function. 
Elementary calculus can be used to obtain the closed form of 
the reliability function. 

\subsection{Decomposition}
For a subset $\mathcal{S}\in\mathfrak{S}_{k}^{N}$ of size $k$, 
we decompose a $k$-sparse vector $\boldsymbol{x}'
\in\mathcal{T}_{w,\boldsymbol{m}}^{N}(\mathcal{S})$ in (\ref{type}) 
into $\boldsymbol{x}_{\mathcal{S}}'
\in\mathcal{T}_{k-w,\boldsymbol{m}_{1}}^{k,1}$ and 
$\boldsymbol{x}_{\mathcal{N}\setminus\mathcal{S}}'\in
\mathcal{T}_{w,\boldsymbol{m}_{2}}^{N-k,2}$ for $\boldsymbol{m}_{1}
\in\mathcal{M}^{k-w}$ and $\boldsymbol{m}_{2}\in\mathcal{M}^{w}$ 
satisfying $\boldsymbol{m}=(\boldsymbol{m}_{1}, \boldsymbol{m}_{2})$, 
with 
\begin{equation} 
\mathcal{T}_{k-w,\boldsymbol{m}_{1}}^{k,1}
= \left\{
 \boldsymbol{x}'\in\mathcal{X}_{k-w}^{k}(\mathcal{U}): 
 \boldsymbol{x}_{\mathcal{S}_{\boldsymbol{x}'}}' 
 = \boldsymbol{u}_{\boldsymbol{m}_{1}}
\right\}, \label{type1}
\end{equation}
\begin{equation} 
\mathcal{T}_{w,\boldsymbol{m}_{2}}^{N-k,2}
= \left\{
 \boldsymbol{x}'\in\mathcal{X}_{w}^{N-k}(\mathcal{U}): 
 \boldsymbol{x}_{\mathcal{S}_{\boldsymbol{x}'}}' 
 = \boldsymbol{u}_{\boldsymbol{m}_{2}}
\right\}.  \label{type2}
\end{equation}
The former type $\mathcal{T}_{k-w,\boldsymbol{m}_{1}}^{k,1}$ 
contains all possible candidates of sparse vectors 
for the support set $\mathcal{S}$ 
while the latter type $\mathcal{T}_{w,\boldsymbol{m}_{2}}^{N-k,2}$ 
includes them for the remaining $(N-k)$ elements. 
  
Fix $w\in\{0,\ldots,k\}$, $w'\in\{0,\ldots,k-w\}$, 
$\boldsymbol{m}\in\mathcal{M}^{k}$, and 
$\boldsymbol{x}\in\mathcal{X}_{k}^{N}(\mathcal{U})$. 
If $\hat{\boldsymbol{x}}_{\mathrm{ML}}
\in\mathcal{T}_{w,\boldsymbol{m}}^{N}(\mathcal{S}_{\boldsymbol{x}})$ and 
$N_{\boldsymbol{x},\hat{\boldsymbol{x}}_{\mathrm{ML}}} = w'$ hold, 
there is some $\boldsymbol{x}'\in
\mathcal{T}_{w,\boldsymbol{m}}^{N}(\mathcal{S}_{\boldsymbol{x}})$ such that 
$p(\boldsymbol{y}|\boldsymbol{x}')\geq p(\boldsymbol{y}|\boldsymbol{x})$ and 
$N_{\boldsymbol{x},\boldsymbol{x}'}=w'$ are satisfied. 
Decompose $\boldsymbol{x}'$ into the two vectors 
$\boldsymbol{x}_{1}'=\boldsymbol{x}'_{\mathcal{S}_{\boldsymbol{x}}}$ and 
$\boldsymbol{x}_{2}'=\boldsymbol{x}'_{\mathcal{N}\setminus\mathcal{S}_{\boldsymbol{x}}}$. 
Since $N_{\boldsymbol{x},\boldsymbol{x}'}$ is a deterministic function of 
$\boldsymbol{x}_{1}=\boldsymbol{x}_{\mathcal{S}_{\boldsymbol{x}}}$ and 
$\boldsymbol{x}_{1}'$, we rewrite 
$N_{\boldsymbol{x},\boldsymbol{x}'}$ as $N_{\boldsymbol{x}_{1}, \boldsymbol{x}_{1}'}$. 

We evaluate $\mathbb{P}(
\hat{\boldsymbol{x}}_{\mathrm{ML}}
\in\mathcal{T}_{w,\boldsymbol{m}}^{N}(\mathcal{S}_{\boldsymbol{x}}), 
N_{\boldsymbol{x},\hat{\boldsymbol{x}}_{\mathrm{ML}}} = w' | \boldsymbol{x})$, in which 
the distribution is induced from that of $\boldsymbol{y}$ 
conditioned on $\boldsymbol{x}$ in (\ref{AWGN}).  
Using the union bound with respect to $\boldsymbol{x}_{1}'$ yields 
\begin{align} 
&\mathbb{P}(
\hat{\boldsymbol{x}}_{\mathrm{ML}}
\in\mathcal{T}_{w,\boldsymbol{m}}^{N}(\mathcal{S}_{\boldsymbol{x}}), 
N_{\boldsymbol{x},\hat{\boldsymbol{x}}_{\mathrm{ML}}} = w' | \boldsymbol{x}) \nonumber \\
&\leq \sum_{\boldsymbol{x}_{1}'\in\mathcal{T}_{k-w,\boldsymbol{m}_{1}}^{k,1}}
1(N_{\boldsymbol{x}_{1}, \boldsymbol{x}_{1}'} = w')
P_{\boldsymbol{x}}(\boldsymbol{x}_{1}'), 
\label{inclusion_probability_bound}
\end{align}
with 
\begin{equation}
P_{\boldsymbol{x}}(\boldsymbol{x}_{1}') 
= \mathbb{P}\left(
 \left. 
 \bigcup_{\boldsymbol{x}_{2}'\in\mathcal{T}_{w,\boldsymbol{m}_{2}}^{N-k,2}}
 \left\{
  \frac{p(\boldsymbol{y} | \boldsymbol{x}')}
  {p(\boldsymbol{y} | \boldsymbol{x})} \geq 1
 \right\}
 \right| \boldsymbol{x} 
\right).
\end{equation} 

\subsection{Gallager's Bound}
We follow Gallager~\cite{Gallager68} to evaluate 
$P_{\boldsymbol{x}}(\boldsymbol{x}_{1}')$ as
\begin{align}
&P_{\boldsymbol{x}}(\boldsymbol{x}_{1}')
= \mathbb{E}\left[
 \left.
  \mathbb{P}\left(
   \left.
    \bigcup_{\boldsymbol{x}_{2}'\in\mathcal{T}_{w,\boldsymbol{m}_{2}}^{N-k,2}}\left\{
     \frac{p(\boldsymbol{y} | \boldsymbol{x}')}
     {p(\boldsymbol{y} | \boldsymbol{x})} \geq 1
    \right\}
   \right| \boldsymbol{x}, \boldsymbol{y}  
  \right)
 \right| \boldsymbol{x} 
\right] \nonumber \\
&\leq \mathbb{E}\left[
 \left.
  \left\{
   \sum_{\boldsymbol{x}_{2}'\in\mathcal{T}_{w,\boldsymbol{m}_{2}}^{N-k,2}}
   \mathbb{P}\left(
    \left.
     \frac{p(\boldsymbol{y} | \boldsymbol{x}')}
     {p(\boldsymbol{y} | \boldsymbol{x})} \geq 1
    \right| \boldsymbol{x}, \boldsymbol{y} 
   \right)
  \right\}^{\rho}
 \right| \boldsymbol{x}
\right] \label{Gallager_bound1}
\end{align}
for all $\rho\in[0, 1]$. The upper bound~(\ref{Gallager_bound1}) is trivial 
when the summation inside the curly brackets $\{\cdots\}$ is larger than $1$. 
Otherwise, it follows from the inequality $a\leq a^{\rho}$ for all 
$a\in[0, 1]$ and $\rho\in[0, 1]$, as well as the union bound. In particular, 
the upper bound~(\ref{Gallager_bound1}) is equivalent 
to the union bound for $\rho=1$. Using the representation 
$\mathbb{P}( p(\boldsymbol{y} | \boldsymbol{x}') 
/p(\boldsymbol{y} | \boldsymbol{x}) 
\geq 1 | \boldsymbol{x}, \boldsymbol{y})
=1(p(\boldsymbol{y} | \boldsymbol{x}') 
\geq p(\boldsymbol{y} | \boldsymbol{x}))$ and the inequality 
$1(p'\geq p)\leq(p'/p)^{s}$ for all $s\geq0$, $p>0$, and $p'\geq0$, we obtain 
\begin{equation} \label{Gallager_bound2} 
P_{\boldsymbol{x}}(\boldsymbol{x}_{1}')
\leq \mathbb{E}\left[
 \left.
  \left\{
   \sum_{\boldsymbol{x}_{2}'\in\mathcal{T}_{w,\boldsymbol{m}_{2}}^{N-k,2}}
   \left[
    \frac{p(\boldsymbol{y} | \boldsymbol{x}')}
    {p(\boldsymbol{y} | \boldsymbol{x})} 
   \right]^{s}
  \right\}^{\rho}
 \right| \boldsymbol{x}
\right]
\end{equation} 
for all $\rho\in[0, 1]$ and $s\geq0$.

To evaluate the upper bound~(\ref{Gallager_bound2}), we focus on each term 
in the summation. Using the definition of the AWGN channel~(\ref{AWGN}), 
as evaluated in \cite[Chapter~7]{Gallager68}, we obtain 
\begin{align}
&\left[
 \frac{p(\boldsymbol{y} | \boldsymbol{x}')}
 {p(\boldsymbol{y} | \boldsymbol{x})} 
\right]^{s}
= \exp\left(
 \frac{s\|\boldsymbol{\omega}\|_{2}^{2}}{2\sigma_{N/k}^{2}}
 - \frac{s\|\boldsymbol{\omega} + \boldsymbol{x} 
 - \boldsymbol{x}'\|_{2}^{2}}{2\sigma_{N/k}^{2}}
\right) \nonumber \\
&= \exp\left\{
 - \frac{s(\|\boldsymbol{x}_{1} - \boldsymbol{x}_{1}'\|_{2}^{2}
 + \|\boldsymbol{u}_{\boldsymbol{m}_{2}}\|_{2}^{2})}{2\sigma_{N/k}^{2}} 
 + \frac{s\boldsymbol{\omega}^{\mathrm{T}}(\boldsymbol{x}' - \boldsymbol{x})}
 {\sigma_{N/k}^{2}}
\right\}, \label{Gallager_bound2_term}
\end{align}
with $\|\boldsymbol{u}_{\boldsymbol{m}_{2}}\|_{2}^{2}=0$ for $w=0$, 
where the last equality is due to $\|\boldsymbol{x} - \boldsymbol{x}'\|_{2}^{2}
= \|\boldsymbol{x}_{1} - \boldsymbol{x}_{1}'\|_{2}^{2} 
+ \|\boldsymbol{u}_{\boldsymbol{m}_{2}}\|_{2}^{2}$ with 
$\boldsymbol{x}_{1}=\boldsymbol{x}_{\mathcal{S}_{\boldsymbol{x}}}$, obtained from 
the definitions of $\boldsymbol{x}_{1}'
\in\mathcal{T}_{k-w,\boldsymbol{m}_{1}}^{k,1}$ 
and $\boldsymbol{x}_{2}'\in\mathcal{T}_{w,\boldsymbol{m}_{2}}^{N-k,2}$ 
in (\ref{type1}) and (\ref{type2}). 
Substituting (\ref{Gallager_bound2_term}) into the upper 
bound~(\ref{Gallager_bound2}) and using the representation 
$\boldsymbol{\omega}^{\mathrm{T}}(\boldsymbol{x}' - \boldsymbol{x})
= \boldsymbol{\omega}_{1}^{\mathrm{T}}(\boldsymbol{x}_{1}' - \boldsymbol{x}_{1})
+ \boldsymbol{\omega}_{2}^{\mathrm{T}}\boldsymbol{x}_{2}'$, with 
$\boldsymbol{\omega}_{1}=\boldsymbol{\omega}_{\mathcal{S}_{\boldsymbol{x}}}$  
and $\boldsymbol{\omega}_{2}=\boldsymbol{\omega}_{\mathcal{N}
\setminus\mathcal{S}_{\boldsymbol{x}}}$, we arrive at 
\begin{align}
&\exp\left\{
 \frac{\rho s(\|\boldsymbol{x}_{1} - \boldsymbol{x}_{1}'\|_{2}^{2}
 + \|\boldsymbol{u}_{\boldsymbol{m}_{2}}\|_{2}^{2})}{2\sigma_{N/k}^{2}} 
\right\}P_{\boldsymbol{x}}(\boldsymbol{x}_{1}') \nonumber \\
&\leq \mathbb{E}\left[
 \left.
  e^{\frac{\rho s\boldsymbol{\omega}_{1}^{\mathrm{T}}
  (\boldsymbol{x}_{1}' - \boldsymbol{x}_{1})}{\sigma_{N/k}^{2}}}
 \right| \boldsymbol{x} 
\right]
\mathbb{E}\left[
 \left.
  \left\{
   \sum_{\boldsymbol{x}_{2}'\in\mathcal{T}_{w,\boldsymbol{m}_{2}}^{N-k,2}}
   e^{\frac{s\boldsymbol{\omega}_{2}^{\mathrm{T}}\boldsymbol{x}_{2}'}{\sigma_{N/k}^{2}}}
  \right\}^{\rho}
 \right| \boldsymbol{x}
\right] \nonumber \\
&\leq e^{
 \frac{\rho^{2} s^{2}\|\boldsymbol{x}_{1}' - \boldsymbol{x}_{1}\|_{2}^{2}}
 {2\sigma_{N/k}^{2}}
}\left\{
 \sum_{\boldsymbol{x}_{2}'\in\mathcal{T}_{w,\boldsymbol{m}_{2}}^{N-k,2}}
 e^{
  s^{2}\|\boldsymbol{x}_{2}'\|_{2}^{2}/(2\sigma_{N/k}^{2})
 }
\right\}^{\rho} \nonumber \\
&= \binom{N-k}{w}^{\rho}
\exp\left(
 \frac{\rho^{2} s^{2}\|\boldsymbol{x}_{1}' - \boldsymbol{x}_{1}\|_{2}^{2}
 + \|\boldsymbol{u}_{\boldsymbol{m}_{2}}\|_{2}^{2}\rho s^{2}}
 {2\sigma_{N/k}^{2}}
\right). \label{Gallager_bound3}
\end{align}
In the derivation of the second inequality, 
we have used Jensen's inequality 
$\mathbb{E}[\{\cdots\}^{\rho} | \boldsymbol{x}]
\leq \{\mathbb{E}[\cdots | \boldsymbol{x}]\}^{\rho}$ for all 
$\rho\in[0, 1]$, as well as the fact that $\boldsymbol{\omega}_{1}^{\mathrm{T}}
(\boldsymbol{x}_{1}' - \boldsymbol{x}_{1})/\sigma_{N/k}$ and 
$\boldsymbol{\omega}_{2}^{\mathrm{T}}\boldsymbol{x}_{2}'/\sigma_{N/k}$ 
conditioned on $\boldsymbol{x}$ follow the zero-mean Gaussian distribution 
with variance $\|\boldsymbol{x}_{1}' - \boldsymbol{x}_{1}\|_{2}^{2}$ and 
$\|\boldsymbol{x}_{2}'\|_{2}^{2}$, respectively. 
The last equality follows from the definition of 
$\mathcal{T}_{w,\boldsymbol{m}_{2}}^{N-k,2}$ in (\ref{type2}) and 
$\|\boldsymbol{x}_{2}'\|_{2}^{2}=\|\boldsymbol{u}_{\boldsymbol{m}_{2}}\|_{2}^{2}$ 
for all $\boldsymbol{x}_{2}'\in\mathcal{T}_{w,\boldsymbol{m}_{2}}^{N-k,2}$. 

We are ready to evaluate $\mathbb{P}(\hat{\boldsymbol{x}}_{\mathrm{ML}}
\in\mathcal{T}_{w,\boldsymbol{m}}^{N}(\mathcal{S}_{\boldsymbol{x}}), 
N_{\boldsymbol{x},\hat{\boldsymbol{x}}_{\mathrm{ML}}} = w' 
| \boldsymbol{x})$ in (\ref{inclusion_probability_bound}). Applying 
the upper bound~(\ref{Gallager_bound3}) to 
(\ref{inclusion_probability_bound}), we have  
\begin{align}
&\mathbb{P}(
\hat{\boldsymbol{x}}_{\mathrm{ML}}
\in\mathcal{T}_{w,\boldsymbol{m}}^{N}(\mathcal{S}_{\boldsymbol{x}}),
N_{\boldsymbol{x},\hat{\boldsymbol{x}}_{\mathrm{ML}}} = w' 
| \boldsymbol{x})\nonumber \\
&\leq \binom{N - k}{w}^{\rho} 
\sum_{\boldsymbol{x}_{1}'\in\mathcal{T}_{k-w,\boldsymbol{m}_{1}}^{k,1}}
1(N_{\boldsymbol{x}_{1}, \boldsymbol{x}_{1}'} = w')
(N/k)^{\frac{\rho f(s, \rho)}{2\sigma^{2}}}, 
\end{align}
where $f(s, \rho)$ is given by 
\begin{equation}
f(s, \rho) = (a\rho + b)s^{2} - (a + b)s, 
\end{equation} 
with $a = \|\boldsymbol{x}_{1}' - \boldsymbol{x}_{1}\|_{2}^{2}$ and 
$b=\|\boldsymbol{u}_{\boldsymbol{m}_{2}}\|_{2}^{2}$. 
Minimizing $f(s, \rho)$ over $s\geq0$ for fixed $\rho\in[0, 1]$ 
yields $f(s, \rho) \geq -(a + b)^{2}/\{4(a\rho + b)\}$ 
at $s=(a + b)/\{2(a\rho + b)\}$. Using this lower bound and the upper bound 
$\binom{N-k}{w}\leq\binom{N}{w}$, we obtain 
\begin{align}
&\mathbb{P}(
\hat{\boldsymbol{x}}_{\mathrm{ML}}
\in\mathcal{T}_{w,\boldsymbol{m}}^{N}(\mathcal{S}_{\boldsymbol{x}}),
N_{\boldsymbol{x},\hat{\boldsymbol{x}}_{\mathrm{ML}}} = w'  
| \boldsymbol{x}) \nonumber \\
&\leq \binom{N}{w}^{\rho}
\sum_{\boldsymbol{x}_{1}'\in\mathcal{T}_{k-w,\boldsymbol{m}_{1}}^{k,1}}
1(N_{\boldsymbol{x}_{1}, \boldsymbol{x}_{1}'} = w')
(N/k)^{-\frac{(a + b)^{2}\rho}{8\sigma^{2}(a\rho+b)}}.
 \label{Gallager_bound4}
\end{align}

For $w=0$ we recall $b=\|\boldsymbol{u}_{\boldsymbol{m}_{2}}\|_{2}^{2}=0$ to find 
\begin{align}
&\mathbb{P}(
\hat{\boldsymbol{x}}_{\mathrm{ML}}
\in\mathcal{T}_{0,\boldsymbol{m}}^{N}(\mathcal{S}_{\boldsymbol{x}}),
N_{\boldsymbol{x},\hat{\boldsymbol{x}}_{\mathrm{ML}}} = w'  
| \boldsymbol{x}) \nonumber \\
&\leq 1(N_{\boldsymbol{x}_{1}, \boldsymbol{u}_{\boldsymbol{m}}} = w')
(N/k)^{-\frac{\|\boldsymbol{u}_{\boldsymbol{m}} - \boldsymbol{x}_{1}\|_{2}^{2}}{8\sigma^{2}}}.
\end{align}
Let $d_{\mathrm{min}}=\min_{m,m'\in\mathcal{M}: m\neq m'}|u_{m} - u_{m'}|>0$ denote 
the minimum 
distance between different discrete points in $\mathcal{U}$. The 
condition $N_{\boldsymbol{x}_{1}, \boldsymbol{u}_{\boldsymbol{m}}} = w'$ holds only when 
the elements of $\boldsymbol{u}_{\boldsymbol{m}}$ are different from those of 
$\boldsymbol{x}_{1}$ in $w'$ positions. Thus, we have the lower bound 
$\|\boldsymbol{u}_{\boldsymbol{m}} - \boldsymbol{x}_{1}\|_{2}^{2} 
\geq d_{\mathrm{min}}^{2}w'$. Evaluating the summation over 
$\boldsymbol{m}\in\mathcal{M}^{k}$ yields 
\begin{align}
&\sum_{\boldsymbol{m}\in\mathcal{M}^{k}}\mathbb{P}(
\hat{\boldsymbol{x}}_{\mathrm{ML}}
\in\mathcal{T}_{0,\boldsymbol{m}}^{N}(\mathcal{S}_{\boldsymbol{x}}),
N_{\boldsymbol{x},\hat{\boldsymbol{x}}_{\mathrm{ML}}} = w'  
| \boldsymbol{x}) \nonumber \\
&\leq \binom{k}{w'}(M - 1)^{w'}
(N/k)^{-\frac{d_{\mathrm{min}}^{2}w'}{8\sigma^{2}}},
\end{align}
which is equivalent to (\ref{Gallager_bound_w0}) for $w=0$. 

For $w>0$, on the other hand, we use the upper bound 
$\binom{n}{m}\leq(en/m)^{m}$ for (\ref{Gallager_bound4}) to obtain 
\begin{align}
&\mathbb{P}(
\hat{\boldsymbol{x}}_{\mathrm{ML}}
\in\mathcal{T}_{w,\boldsymbol{m}}^{N}(\mathcal{S}_{\boldsymbol{x}}),
N_{\boldsymbol{x},\hat{\boldsymbol{x}}_{\mathrm{ML}}} = w'  
| \boldsymbol{x}) \nonumber \\
&\leq(ek/w)^{w}\sum_{\boldsymbol{x}_{1}'\in\mathcal{T}_{k-w,\boldsymbol{m}_{1}}^{k,1}}
1(N_{\boldsymbol{x}_{1}, \boldsymbol{x}_{1}'} = w')
(N/k)^{-E(\sigma^{2}; \rho, a, b)} 
 \label{Gallager_bound5}
\end{align}
for all $w\in\{1,\ldots,k\}$ and $\rho\in[0, 1]$, with 
\begin{equation}
E(\sigma^{2}; \rho, a, b) 
= \frac{(a + b)^{2}\rho}{8\sigma^{2}(a\rho + b)} - \rho w, 
\end{equation}
where the last inequality is due to $w\leq k$ and $\rho\leq 1$. 

\subsection{Reliability Function}
For $w>0$, let $a = \|\boldsymbol{x}_{1}' - \boldsymbol{x}_{1}\|_{2}^{2}$ and 
$b=\|\boldsymbol{u}_{\boldsymbol{m}_{2}}\|_{2}^{2}$. For all 
$\boldsymbol{x}_{1}'\in\mathcal{T}_{k-w,\boldsymbol{m}_{1}}^{k,1}$, 
$\boldsymbol{x}\in\mathcal{X}_{k}^{N}(\mathcal{U})$, and 
$\boldsymbol{m}\in\mathcal{M}^{k}$ under the condition 
$N_{\boldsymbol{x}_{1}, \boldsymbol{x}_{1}'} = w'$, we have 
$a\geq u_{\mathrm{min}}^{2}w + d_{\mathrm{min}}^{2}w'$ and $b\geq u_{\mathrm{min}}^{2}w$. 
Define the reliability function as 
\begin{equation}
E_{w,w'}(\sigma^{2})
=\min_{a\geq u_{\mathrm{min}}^{2}w + d_{\mathrm{min}}^{2}w', b\geq u_{\mathrm{min}}^{2}w}
\max_{\rho\in[0, 1]}E(\sigma^{2}; \rho, a, b). 
\end{equation} 
Applying this definition to the upper bound~(\ref{Gallager_bound5}) yields 
\begin{align}
&\sum_{\boldsymbol{m}\in\mathcal{M}^{k}}\mathbb{P}(
\hat{\boldsymbol{x}}_{\mathrm{ML}}
\in\mathcal{T}_{w,\boldsymbol{m}}^{N}(\mathcal{S}_{\boldsymbol{x}}), 
N_{\boldsymbol{x},\hat{\boldsymbol{x}}_{\mathrm{ML}}} = w'  
| \boldsymbol{x}) \nonumber \\
&\leq(ek/w)^{w}M^{w}
\binom{k - w}{w'}(M - 1)^{w'}(N/k)^{-E_{w,w'}(\sigma^{2})}.
\end{align}
To complete the proof of 
Lemma~\ref{lemma_Gallager_bound}, it is sufficient to prove that 
$E_{w,w'}(\sigma^{2})$ reduces to (\ref{reliability_function}). 

We evaluate $E_{w,w'}(\sigma^{2})$. 
Let $\tilde{a} = (a + b)^{2}/(8\sigma^{2}a)$ and $\tilde{b}=b/a$. The function 
$E(\sigma^{2}; \rho, a, b)$ is represented as 
\begin{equation}
E(\sigma^{2}; \rho, a, b) = \frac{\tilde{a}\rho}{\rho + \tilde{b}} - w\rho. 
\end{equation}
It is straightforward to confirm the concavity of $E$ 
with respect to $\rho\in[0, 1]$. Thus, $E$ is maximized 
with respect to $\rho$ at its stationary point, $\rho=0$, or $\rho=1$. 
Evaluating the derivative of $E$ yields 
\begin{equation}
\frac{\partial E}{\partial\rho} 
= \frac{\tilde{a}}{\rho + \tilde{b}} 
- \frac{\tilde{a}\rho}{(\rho + \tilde{b})^{2}} - w
= - \frac{w(\rho + \tilde{b})^{2} - \tilde{a}\tilde{b}}
{(\rho + \tilde{b})^{2}}. 
\end{equation}

The condition $\partial E/\partial\rho\leq 0$ for all 
$\rho\in[0, 1]$ is equivalent to $\tilde{a}\leq w\tilde{b}$. In this case, 
we find $E\leq E(\sigma^{2}; 0, a, b)=0$. 
Furthermore, $\tilde{a}\leq w\tilde{b}$ 
is equivalent to $8\sigma^{2}w\geq (a + b)^{2}/b$. Minimizing the right-hand 
side (RHS) over $a\geq u_{\mathrm{min}}^{2}w + d_{\mathrm{min}}^{2}w'$ 
and $b\geq u_{\mathrm{min}}^{2}w$, we obtain 
\begin{align}
\frac{(a + b)^{2}}{b} 
&\geq \left(
 \frac{u_{\mathrm{min}}^{2}w + d_{\mathrm{min}}^{2}w'}{\sqrt{b}} + \sqrt{b}
\right)^{2} \nonumber \\
&\geq 4(u_{\mathrm{min}}^{2}w  + d_{\mathrm{min}}^{2}w'). 
\end{align}
Here, the equality for the first inequality holds at 
$a= u_{\mathrm{min}}^{2}w + d_{\mathrm{min}}^{2}w'$. 
The second inequality is due to the arithmetic-mean and geometric-mean 
(AM-GM) inequality, in which the equality holds at 
$b = u_{\mathrm{min}}^{2}w + d_{\mathrm{min}}^{2}w'$. 
Thus, we find $E_{w,w'}(\sigma^{2})=0$ for 
$\sigma^{2}\geq (u_{\mathrm{min}}^{2}w + d_{\mathrm{min}}^{2}w')/(2w)$. 

The condition $\partial E/\partial\rho>0$ for all 
$\rho\in[0, 1]$ is equivalent to $(\tilde{a}\tilde{b}/w)^{1/2} 
> \tilde{b} + 1$. In this case we have 
\begin{equation}
E\leq E(\sigma^{2}; 1, a, b) = \frac{\tilde{a}}{1 + \tilde{b}} - w 
= \frac{a + b}{8\sigma^{2}} - w. 
\end{equation} 
Minimizing the upper bound over $a\geq u_{\mathrm{min}}^{2}w + d_{\mathrm{min}}^{2}w'$ 
and $b\geq u_{\mathrm{min}}^{2}w$, we have 
$E\leq (2u_{\mathrm{min}}^{2}w + d_{\mathrm{min}}^{2}w')/(8\sigma^{2}) - w$ at 
$a = u_{\mathrm{min}}^{2}w + d_{\mathrm{min}}^{2}w'$ and $b = u_{\mathrm{min}}^{2}w$. 
Furthermore, 
$(\tilde{a}\tilde{b}/w)^{1/2} > \tilde{b} + 1$ is equivalent to 
$b> 8\sigma^{2}w$, which reduces to $\sigma^{2}< u_{\mathrm{min}}^{2}/8$ 
for the minimizer $b = u_{\mathrm{min}}^{2}w$.  

Finally, assume $\tilde{a}>w\tilde{b}$ and $(\tilde{a}\tilde{b}/w)^{1/2} 
\leq\tilde{b} + 1$. In this case, $E$ is maximized at 
$\rho=(\tilde{a}\tilde{b}/w)^{1/2} - \tilde{b}\equiv \rho_{\mathrm{s}}$ 
satisfying $\partial E/\partial \rho=0$ for $\rho=\rho_{\mathrm{s}}$ to obtain 
\begin{align}
&E \leq E(\sigma^{2}; \rho_{\mathrm{s}}, a, b) 
= \{\tilde{a}^{1/2} - (w\tilde{b})^{1/2}\}^{2}
\nonumber \\
&= \frac{\{a + (\sqrt{b} - \sqrt{2\sigma^{2}w})^{2} - 2\sigma^{2}w\}^{2}}
{8\sigma^{2}a}
\equiv E_{*}. \label{reliability_function_opt}
\end{align}
Furthermore, $\tilde{a}>w\tilde{b}$ and $(\tilde{a}\tilde{b}/w)^{1/2} 
\leq\tilde{b} + 1$ are equivalent to $8\sigma^{2}w\in[b, (a + b)^{2}/b)$. 

We minimize the upper bound $E_{*}$ 
over $a\geq u_{\mathrm{min}}^{2}w + d_{\mathrm{min}}^{2}w'$ 
and $b\geq u_{\mathrm{min}}^{2}w$. Since we have already evaluated 
$E_{w,w'}(\sigma^{2})$ for 
$\sigma^{2}\geq (u_{\mathrm{min}}^{2}w + d_{\mathrm{min}}^{2}w')/(2w)$ and 
$\sigma^{2}< u_{\mathrm{min}}^{2}/8$, 
we can assume $\sigma^{2}\in[u_{\mathrm{min}}^{2}/8, 
(u_{\mathrm{min}}^{2}w + d_{\mathrm{min}}^{2}w')/(2w))$, 
which implies $a \geq u_{\mathrm{min}}^{2}w + d_{\mathrm{min}}^{2}w' > 2\sigma^{2}w$. 
For $\sigma^{2}\in[u_{\mathrm{min}}^{2}/2, 
(u_{\mathrm{min}}^{2}w + d_{\mathrm{min}}^{2}w')/(2w))$, we use 
$b\geq u_{\mathrm{min}}^{2}w$ and 
$2\sigma^{2}w \geq u_{\mathrm{min}}^{2}w$ to have 
\begin{equation}
E_{*} \geq \frac{(a - 2\sigma^{2}w)^{2}}
{8\sigma^{2}a}, 
\end{equation}
where the equality holds at $b = 2\sigma^{2}w$. 
The lower bound is monotonically increasing with respect to 
$a\geq 2\sigma^{2}w$. Thus, the minimum is attained at 
$a= u_{\mathrm{min}}^{2}w + d_{\mathrm{min}}^{2}w'\geq2\sigma^{2}w$, 
\begin{equation}
E_{*} \geq \frac{(u_{\mathrm{min}}^{2}w + d_{\mathrm{min}}^{2}w' - 2\sigma^{2}w)^{2}}
{8\sigma^{2}(u_{\mathrm{min}}^{2}w + d_{\mathrm{min}}^{2}w')}. 
\end{equation}

For $\sigma^{2}\in[u_{\mathrm{min}}^{2}/8, u_{\mathrm{min}}^{2}/2)$, we find 
$2\sigma^{2}w < u_{\mathrm{min}}^{2}w \leq b$ to lower-bound $E_{*}$ in 
(\ref{reliability_function_opt}) as 
\begin{equation}
E_{*}
\geq \frac{\{a 
- (2\sqrt{2\sigma^{2}u_{\mathrm{min}}^{2}} - u_{\mathrm{min}}^{2})w\}^{2}}
{8\sigma^{2}a}
\end{equation}
where the equality holds at $b=u_{\mathrm{min}}^{2}w$. The lower 
bound is monotonically increasing with respect to 
$a\geq (2\sqrt{2\sigma^{2}u_{\mathrm{min}}^{2}} - u_{\mathrm{min}}^{2})w\geq 0$ 
for $\sigma^{2}\geq u_{\mathrm{min}}^{2}/8$. 
Since the assumption $\sigma^{2}< u_{\mathrm{min}}^{2}/2$ implies 
$(2\sqrt{2\sigma^{2}u_{\mathrm{min}}^{2}} - u_{\mathrm{min}}^{2})w < 
u_{\mathrm{min}}^{2}w$, the minimum is attained at $a= u_{\mathrm{min}}^{2}w
+ d_{\mathrm{min}}^{2}w'$. Thus, we arrive at 
\begin{equation}
E_{*}
\geq \frac{\{2(u_{\mathrm{min}}^{2} - \sqrt{2\sigma^{2}u_{\mathrm{min}}^{2}})w
+ d_{\mathrm{min}}^{2}w'\}^{2}}
{8\sigma^{2}(u_{\mathrm{min}}^{2}w
+ d_{\mathrm{min}}^{2}w')}
\end{equation}
for $\sigma^{2}\in[u_{\mathrm{min}}^{2}/8, u_{\mathrm{min}}^{2}/2)$. 
Thus, $E_{w,w'}(\sigma^{2})$ reduces to (\ref{reliability_function}). 

\section{Evaluation of (\ref{probability_event}) for $M>1$}
\label{evaluation_probability_event} 
Assume $M>1$. 
For $w=0$ we use Lemma~\ref{lemma_Gallager_bound}, $M - 1 < M$, and 
$\binom{k}{w'}\leq(ek/w')^{w'}$ to obtain 
\begin{align}
&\sum_{w' = d}^{k}\sum_{\boldsymbol{m}\in\mathcal{M}^{k}}
\mathbb{P}(\hat{\boldsymbol{x}}_{\mathrm{ML}}
\in\mathcal{T}_{0,\boldsymbol{m}}^{N}(\mathcal{S}_{\boldsymbol{x}}),
N_{\boldsymbol{x},\hat{\boldsymbol{x}}_{\mathrm{ML}}} = w' | \boldsymbol{x}) 
\nonumber \\
&\leq \sum_{w' = d}^{k}(ek/d)^{w'}M^{w'}
(N/k)^{-\frac{d_{\mathrm{min}}^{2}w'}{8\sigma^{2}}} \to 0,  
\end{align} 
where the last convergence follows from 
Proposition~\ref{proposition_geometric} with 
$r_{k,N}=(ek/d)M(N/k)^{-d_{\mathrm{min}}^{2}/(8\sigma^{2})}$. 

For $w>0$ we use Lemma~\ref{lemma_Gallager_bound} and 
the trivial upper bound $M-1<M$ to have 
\begin{align}
&\sum_{w + w' \geq d}\sum_{\boldsymbol{m}\in\mathcal{M}^{k}}
\mathbb{P}(\hat{\boldsymbol{x}}_{\mathrm{ML}}
\in\mathcal{T}_{w,\boldsymbol{m}}^{N}(\mathcal{S}_{\boldsymbol{x}}),
N_{\boldsymbol{x},\hat{\boldsymbol{x}}_{\mathrm{ML}}} = w' | \boldsymbol{x}) 
\nonumber \\ 
&\leq \sum_{w + w' \geq d}
(ek/w)^{w}\binom{k - w}{w'}
M^{w + w'}(N/k)^{-E_{w,w'}(\sigma^{2})}. \label{probability_convergence_w}
\end{align}
In particular, for $w'=0$ we use the linearity of $E_{w,0}(\sigma^{2})$ 
with respect to $w$ to obtain  
\begin{align}
&\sum_{w = d}^{k}\sum_{\boldsymbol{m}\in\mathcal{M}^{k}}
\mathbb{P}(\hat{\boldsymbol{x}}_{\mathrm{ML}}
\in\mathcal{T}_{w,\boldsymbol{m}}^{N}(\mathcal{S}_{\boldsymbol{x}}),
N_{\boldsymbol{x},\hat{\boldsymbol{x}}_{\mathrm{ML}}} = 0 | \boldsymbol{x}) 
\nonumber \\ 
&\leq \sum_{w = d}^{k}
(ek/d)^{w}M^{w}(N/k)^{-wE_{1,0}(\sigma^{2})} \to 0,  
\end{align}
where the last convergence follows from 
Proposition~\ref{proposition_geometric} with 
$r_{k,N}=(ek/d)M(N/k)^{-E_{1,0}(\sigma^{2})}$. 

To evaluate the upper bound~(\ref{probability_convergence_w}) for $w>0$ and 
$w'>0$, we prove two technical results. 
\begin{proposition}
Let $\mathcal{W}_{i}=\{(w, w'): w\in\{1,\ldots,k\}, 
w'\in\{1,\ldots,k-w\}, w + w'= i\}$ for $i\in\{2,\ldots, k\}$. Then, 
\begin{equation} \label{minimization_lower_bound}
\min_{(w, w')\in\mathcal{W}_{i}}w^{w}w'^{w'} \geq (i/2)^{i}.  
\end{equation}
\end{proposition}
\begin{IEEEproof}
Define $f(w, w')=\log(w^{w}w'^{w'}) = w\log w + w'\log w'$. The minimizer 
of $f(w, w')$ over $(w, w')\in\mathcal{W}_{i}$ is also the minimizer of the 
original minimization problem. Relax the feasible region $\mathcal{W}_{i}$ 
to the linear constraints $\mathcal{W}_{i}^{\mathrm{L}} 
= \{(w, w')\in\mathbb{R}^{2}: w\in[1, k], w'\in[1, k-w], w + w'= i\}$. 
Since $f(w, w')$ is convex on $(w, w')\in\mathcal{W}_{i}^{\mathrm{L}}$, 
the minimization of $f$ over $\mathcal{W}_{i}^{\mathrm{L}}$ is 
convex programming. 

We substitute $w'=i-w$ into $f(w,w')$ to obtain $f(w,i-w)
=w\log w + (i - w)\log(i - w)\equiv f(w)$ for $w\in[1, i -1]$. 
Differentiating $f(w)$ yields 
\begin{equation}
f'(w) = \log w - \log(i - w), 
\end{equation} 
which implies $f'(w)=0$ only at $w=i/2$. Thus, $f(w, w')$ is bounded from 
below by $f(i/2)$, which is equivalent to the lower 
bound~(\ref{minimization_lower_bound}).  
\end{IEEEproof}

\begin{proposition}
Let  $a=2(u_{\mathrm{min}}^{2} 
- \sqrt{2\sigma^{2}u_{\mathrm{min}}^{2}})/u_{\mathrm{min}}^{2}$. Then, 
\begin{equation} \label{reliability_function2_linear}
8\sigma^{2}E_{w,w'}^{(2)}(\sigma^{2})
\geq a^{2}u_{\mathrm{min}}^{2}w + ad_{\mathrm{min}}^{2}w'
\end{equation} 
holds for all $w>0$, $w'>0$, and 
$\sigma^{2}\in[u_{\mathrm{min}}^{2}/8, u_{\mathrm{min}}^{2}/2)$. 
\end{proposition}
\begin{IEEEproof}
Let $\tilde{w}=u_{\mathrm{min}}^{2}w$ and $\tilde{w}'=d_{\mathrm{min}}^{2}w'$. 
We rewrite $E_{w,w'}^{(2)}(\sigma^{2})$ in (\ref{reliability_function2}) as 
\begin{equation}
E_{w,w'}^{(2)}(\sigma^{2}) 
= \frac{(a\tilde{w} + \tilde{w}')^{2}}
{8\sigma^{2}(\tilde{w} + \tilde{w}')}. 
\end{equation}
Since $a\in(0, 1]$ holds for all 
$\sigma^{2}\in[u_{\mathrm{min}}^{2}/8, u_{\mathrm{min}}^{2}/2)$, we have 
$\tilde{w}'\geq a\tilde{w}'$, which is equivalent to  
\begin{equation}
\frac{a\tilde{w} + \tilde{w}'}{\tilde{w} + \tilde{w}'} \geq a. 
\end{equation}
Multiplying both sides by $a\tilde{w}+\tilde{w}'$, we obtain 
(\ref{reliability_function2_linear}).  
\end{IEEEproof}

We evaluate the upper bound~(\ref{probability_convergence_w}) for $w>0$ and 
$w'>0$. 
Let $\mathcal{W}=\{(w, w'): w\in\{1,\ldots,k\}, 
w'\in\{1,\ldots,k-w\}, w + w'\geq d\}$. By definition, we have 
$\mathcal{W}=\cup_{i=d}^{k}\mathcal{W}_{i}$. Furthermore, 
define $E_{\mathrm{min}}
=(8\sigma^{2})^{-1}\min\{a^{2}u_{\mathrm{min}}^{2}, ad_{\mathrm{min}}^{2}\}>0$. 
Using the upper bound $\binom{k-w}{w'}\leq(ek/w')^{w'}$ and 
the lower bounds~(\ref{minimization_lower_bound}) and 
(\ref{reliability_function2_linear}) yields 
\begin{align}
&\sum_{(w, w')\in\mathcal{W}}\sum_{\boldsymbol{m}\in\mathcal{M}^{k}}
\mathbb{P}(\hat{\boldsymbol{x}}_{\mathrm{ML}}
\in\mathcal{T}_{w,\boldsymbol{m}}^{N}(\mathcal{S}_{\boldsymbol{x}}),
N_{\boldsymbol{x},\hat{\boldsymbol{x}}_{\mathrm{ML}}} = w' | \boldsymbol{x}) 
\nonumber \\ 
&\leq \sum_{i=d}^{k}(i/2)^{-i}\sum_{(w, w')\in\mathcal{W}_{i}}
(ekM)^{w+w'}(N/k)^{-E_{\mathrm{min}}(w + w')} \nonumber \\
&\leq \sum_{i=d}^{k}(i - 1)r_{k,N}^{i}, 
\label{probability_convergence_w_tmp}
\end{align}
with $r_{k,N}=(2ek/d)M(N/k)^{-E_{\mathrm{min}}}$, 
where the last inequality follows from the upper bound 
$(2ekM/i)^{i}(N/k)^{-E_{\mathrm{min}}i}\leq r_{k,N}^{i}$ for all $i\in\{d,\ldots,k\}$. 

We prove that the upper bound~(\ref{probability_convergence_w_tmp}) 
converges to zero in the sublinear sparsity limit. Consider the following 
geometric series: 
\begin{equation}
S(r) = \sum_{i=d}^{k}r^i = r^{d}\frac{1 - r^{k - d + 1}}{1 - r}. 
\end{equation}
Using $S'(r) = \sum_{i=d}^{k}ir^{i-1}$, we find 
$\sum_{i=d}^{k}ir^{i} = rS'(r)$. We evaluate the derivative of $S(r)$ at 
$r=r_{k,N}$ to obtain 
\begin{equation}
\sum_{i=d}^{k}ir_{k,N}^{i}
= \frac{dr_{k,N}^{d} - (k + 1)r_{k,N}^{k + 1}}{1 - r_{k,N}} 
+ \frac{r_{k,N}^{d + 1} - r_{k,N}^{k + 2}}{(1 - r_{k,N})^{2}}\to 0
\end{equation}
in the sublinear sparsity limit, where the last convergence follows from 
the definitions $d=\lceil k/\sqrt{\log(N/k)} \rceil$ and 
$r_{k,N}=(2ek/d)M(N/k)^{-E_{\mathrm{min}}}$. This implies the convergence of 
the upper bound~(\ref{probability_convergence_w_tmp}) to zero 
in the sublinear sparsity limit. Combining these results and the upper 
bound~(\ref{probability_event}), we arrive at 
$\mathbb{P}(\mathcal{E}_{d} | \boldsymbol{x})\to0$ in the sublinear 
sparsity limit for $M>1$.

\section{Proof of Lemma~\ref{lemma_KL_divergence}}
\label{proof_lemma_KL_divergence}
\subsection{Overview}
Without loss of generality, 
we assume that $\min_{m\in\mathcal{M}}|u_{m}|$ is attained at positive $u_{m}>0$.  
The proof of Lemma~\ref{lemma_KL_divergence} consists of three steps: 
A first step is the formulation of the KL divergence $D(p \| q)$. 
To obtain a tight upper bound on the KL divergence, we utilize 
Jensen's inequality only for Gaussian noise samples in positions where 
the signal vector $\boldsymbol{x}$ has zeros. 
If we applied Jensen's inequality to all Gaussian noise samples, we would 
have a loose upper bound on the KL divergence. 

In a second step, a truncation technique is used for the remaining $k$ 
Gaussian noise samples. The threshold in the truncation separates the 
Gaussian noise samples into typical samples and non-typical samples. 
The set of typical samples 
contains all $k$ Gaussian noise samples that occur jointly with high 
probability, while the set of non-typical samples consists of the remaining 
rare samples. In the second step, the upper bound in the first step is 
evaluated for the set of typical samples. This evaluation itself is not 
challenging because typical noise samples are bounded. 

The last step is the technically most challenging part in the proof of 
Lemma~\ref{lemma_KL_divergence}---evaluation for non-typical noise samples. 
Let $w\in\{1,\ldots,k\}$ denote the number of incorrectly detected 
positions, as considered in the direct part. The main challenge is to 
upper-bound a Gaussian mixture associated with $k$-dimensional and 
$(k-w)$-sparse vectors by a single Gaussian pdf for a dense 
vector. This upper bound is proved via a graphical technique based on 
Hall's marriage theorem~\cite{Hall35}. It is enough tight 
to evaluate the upper bound in the first step for non-typical noise samples.  

\subsection{KL Divergence}
We recall that the true marginal pdf of $\boldsymbol{y}$ is induced from 
the conditional pdf $p(\boldsymbol{y} | \boldsymbol{x})$ defined via 
(\ref{AWGN}) and the true distribution of $\boldsymbol{x}$---uniform 
distribution on $\mathcal{X}_{k}^{N}(\mathcal{U})$. Let 
\begin{equation} \label{lambda}
\lambda = 1 + \frac{\mathbb{E}[\|\boldsymbol{x}\|_{2}^{2}]}{N\sigma_{N/k}^{2}}.  
\end{equation}
From the definition of $\boldsymbol{y}$ in (\ref{AWGN}) and 
the independence between $\boldsymbol{x}$ and $\boldsymbol{y}$, 
the true expectation $\mathbb{E}_{p}[\|\boldsymbol{y}\|_{2}^{2}]$ 
is equal to $\mathbb{E}[\|\boldsymbol{x}\|_{2}^{2}] + N\sigma_{N/k}^{2}
=N\lambda\sigma_{N/k}^{2}$. By definition, $q(\boldsymbol{y})$ is represented as 
\begin{equation} \label{q_y}
q(\boldsymbol{y}) 
= \frac{1}{(2\pi\lambda\sigma_{N/k}^{2})^{N/2}}
\exp\left(
 - \frac{\|\boldsymbol{y}\|_{2}^{2}}{2\lambda\sigma_{N/k}^{2}}
\right).
\end{equation}

To evaluate the KL divergence $D(p \| q)$, 
we first evaluate $\log\{ p(\boldsymbol{y})/q(\boldsymbol{y})\}$. 
We use the definition of $p(\boldsymbol{y} | \boldsymbol{x})$ 
in (\ref{AWGN}) to obtain 
\begin{equation}
p(\boldsymbol{y}) 
= \frac{e^{-\|\boldsymbol{y}\|_{2}^{2}/(2\sigma_{N/k}^{2})}}
{(2\pi\sigma_{N/k}^{2})^{N/2}|\mathcal{X}_{k}^{N}(\mathcal{U})|}
\sum_{\boldsymbol{x}\in\mathcal{X}_{k}^{N}(\mathcal{U})}
e^{(\boldsymbol{x}^{\mathrm{T}}\boldsymbol{y} - \|\boldsymbol{x}\|_{2}^{2}/2)
/\sigma_{N/k}^{2}}. \label{p_y}
\end{equation}
Combining the two results~(\ref{q_y}) and (\ref{p_y}), we arrive at 
\begin{align}
&\log\frac{p(\boldsymbol{y})}
{q(\boldsymbol{y})} 
= \frac{N}{2}\log\lambda - \frac{(\lambda - 1)\|\boldsymbol{y}\|_{2}^{2}}
{2\lambda\sigma_{N/k}^{2}}
\nonumber \\
&+ \log\left\{
 \frac{1}{|\mathcal{X}_{k}^{N}(\mathcal{U})|}
 \sum_{\boldsymbol{x}\in\mathcal{X}_{k}^{N}(\mathcal{U})}
 e^{(\boldsymbol{x}^{\mathrm{T}}\boldsymbol{y} - \|\boldsymbol{x}\|_{2}^{2}/2)
 /\sigma_{N/k}^{2}}
\right\}. \label{log_likelihood}
\end{align}

We upper-bound the KL divergence $D(p \| q)$. 
Evaluating the expectation of (\ref{log_likelihood}) over 
$p(\boldsymbol{y})$ and subsequently using the definition of 
$\lambda$ in (\ref{lambda}) and the upper bound 
$\log(1 + x)\leq x$ for all $x\geq0$, we have 
\begin{equation} \label{KL_divergence_bound}
D(p \| q)
\leq \frac{1}{|\mathcal{X}_{k}^{N}(\mathcal{U})|}
\sum_{\boldsymbol{x}\in\mathcal{X}_{k}^{N}(\mathcal{U})}
g(\boldsymbol{x}),   
\end{equation}
with 
\begin{equation}
g(\boldsymbol{x}) 
= \mathbb{E}\left[
 \log\left\{
  \frac{1}{|\mathcal{X}_{k}^{N}(\mathcal{U})|}
  \sum_{\boldsymbol{x}'\in\mathcal{X}_{k}^{N}(\mathcal{U})}
  e^{
   \frac{\boldsymbol{x}'^{\mathrm{T}}(\boldsymbol{x} + \boldsymbol{\omega}) 
   - \|\boldsymbol{x}'\|_{2}^{2}/2}{\sigma_{N/k}^{2}}}
 \right\}
\right]. 
\end{equation}

To derive a tight upper bound on $g(\boldsymbol{x})$ for all 
$\boldsymbol{x}\in\mathcal{X}_{k}^{N}(\mathcal{U})$, we use the 
representation $\mathcal{T}_{w,\boldsymbol{m}}^{N}(\mathcal{S}_{\boldsymbol{x}})$
of $\mathcal{X}_{k}^{N}(\mathcal{U})$ in (\ref{X_set_representation}) 
to obtain 
\begin{equation} \label{g_function}
g(\boldsymbol{x}) = \mathbb{E}\left[
 \log\left\{
 \frac{1}{|\mathcal{X}_{k}^{N}(\mathcal{U})|}\sum_{w=0}^{k}
  g_{k-w}(\boldsymbol{x}, \boldsymbol{\omega})
 \right\}
\right], 
\end{equation}
with 
\begin{equation}
g_{k-w}(\boldsymbol{x}, \boldsymbol{\omega})
= \sum_{\boldsymbol{m}\in\mathcal{M}^{k}}
e^{- \frac{\|\boldsymbol{u}_{\boldsymbol{m}}\|_{2}^{2}}{2\sigma_{N/k}^{2}}}
\sum_{\boldsymbol{x}'\in\mathcal{T}_{k,\boldsymbol{m}}^{N}
(\mathcal{S}_{\boldsymbol{x}})}
e^{
 \frac{\boldsymbol{x}'^{\mathrm{T}}(\boldsymbol{x} + \boldsymbol{\omega})}
{\sigma_{N/k}^{2}}}. 
\end{equation}
Let $|\boldsymbol{x}|$ denote the element-wise absolute values of 
$\boldsymbol{x}$, i.e.\ $[|\boldsymbol{x}|]_{n} = |[\boldsymbol{x}]_{n}|$. 
Decomposing $\boldsymbol{x}'\in\mathcal{T}_{k,\boldsymbol{m}}^{N}
(\mathcal{S}_{\boldsymbol{x}})$ into 
$\boldsymbol{x}_{1}'\in\mathcal{T}_{k-w,\boldsymbol{m}_{1}}^{k,1}$ and 
$\boldsymbol{x}_{2}'\in\mathcal{T}_{w,\boldsymbol{m}_{2}}^{N-k,2}$ in 
(\ref{type1}) and (\ref{type2}) and subsequently using the upper bound 
$\boldsymbol{x}_{1}'^{\mathrm{T}}\boldsymbol{x}_{1}
\leq|\boldsymbol{x}_{1}'|^{\mathrm{T}}|\boldsymbol{x}_{1}|$ 
with $\boldsymbol{x}_{1} = \boldsymbol{x}_{\mathcal{S}_{\boldsymbol{x}}}$, we have 
\begin{align}
&g_{k-w}(\boldsymbol{x}, \boldsymbol{\omega})
\leq \sum_{\boldsymbol{m}\in\mathcal{M}^{k}}
e^{- \|\boldsymbol{u}_{\boldsymbol{m}}\|_{2}^{2}/(2\sigma_{N/k}^{2})}
\nonumber \\
&\cdot\sum_{\boldsymbol{x}_{1}'\in\mathcal{T}_{k-w,\boldsymbol{m}_{1}}^{k,1}}
e^{
 (|\boldsymbol{x}_{1}'|^{\mathrm{T}}|\boldsymbol{x}_{1}| 
 + \boldsymbol{x}_{1}'^{\mathrm{T}}\boldsymbol{\omega}_{1})/\sigma_{N/k}^{2}
} 
\sum_{\boldsymbol{x}_{2}'\in\mathcal{T}_{w,\boldsymbol{m}_{2}}^{N-k,2}}
e^{
 \boldsymbol{x}_{2}'^{\mathrm{T}}\boldsymbol{\omega}_{2}/\sigma_{N/k}^{2}
}, 
\end{align}
with $\boldsymbol{\omega}_{1} = \boldsymbol{\omega}_{\mathcal{S}_{\boldsymbol{x}}}$ 
and $\boldsymbol{\omega}_{2} 
= \boldsymbol{\omega}_{\mathcal{N}\setminus\mathcal{S}_{\boldsymbol{x}}}
\sim\mathcal{N}(\boldsymbol{0}, 
\sigma_{N/k}^{2}\boldsymbol{I}_{N-k})$. 

We use Jensen's inequality with respect to 
$\boldsymbol{\omega}_{2}$ conditioned on $\boldsymbol{\omega}_{1}$ 
to evaluate $g(\boldsymbol{x})$ in (\ref{g_function}) as  
\begin{equation} \label{g_function_tmp0}
g(\boldsymbol{x}) \leq \mathbb{E}\left[
 \log\left\{
  \sum_{w=0}^{k}\frac{|\mathcal{M}|^{w}}{|\mathcal{X}_{k}^{N}(\mathcal{U})|}
  \binom{N-k}{w}G_{k-w}^{k}(\boldsymbol{x}_{1}, \boldsymbol{\omega}_{1})
 \right\}
\right], 
\end{equation}
with $G_{0}^{k}(\boldsymbol{x}_{1}, \boldsymbol{\omega}_{1})= 1$ and 
\begin{align} 
G_{k-w}^{k} &= \frac{1}{|\mathcal{M}|^{w}}\sum_{\boldsymbol{m}\in\mathcal{M}^{k}}
\sum_{\boldsymbol{x}_{1}'\in\mathcal{T}_{k-w,\boldsymbol{m}_{1}}^{k,1}}
e^{\frac{|\boldsymbol{x}_{1}'|^{\mathrm{T}}|\boldsymbol{x}_{1}|
- 2^{-1}\|\boldsymbol{x}_{1}'\|_{2}^{2} 
+ \boldsymbol{x}_{1}'^{\mathrm{T}}\boldsymbol{\omega}_{1}}
{\sigma_{N/k}^{2}} } \nonumber \\
&= \sum_{\boldsymbol{x}_{1}'\in\mathcal{X}_{k-w}^{k}(\mathcal{U})}
e^{\frac{|\boldsymbol{x}_{1}'|^{\mathrm{T}}|\boldsymbol{x}_{1}|
- 2^{-1}\|\boldsymbol{x}_{1}'\|_{2}^{2} 
+ \boldsymbol{x}_{1}'^{\mathrm{T}}\boldsymbol{\omega}_{1}}
{\sigma_{N/k}^{2}} }
\label{G_w_tmp}
\end{align}
for $w\in\{0,\ldots,k-1\}$, 
where the last equality follows from the definition of 
$\mathcal{T}_{k-w,\boldsymbol{m}_{1}}^{k,1}$ in (\ref{type1}). 

We derive a uniform upper bound on 
$G_{k-w}^{k}(\boldsymbol{x}_{1}, \boldsymbol{\omega}_{1})$ with respect to 
$\boldsymbol{x}_{1}$. Represent (\ref{G_w_tmp}) as 
\begin{equation}
G_{k-w}^{k}(\boldsymbol{x}_{1}, \boldsymbol{\omega}_{1}) 
= \sum_{\boldsymbol{x}_{1}'\in\mathcal{X}_{k-w}^{k}(\{1\})}
\prod_{j\in\mathcal{S}_{\boldsymbol{x}_{1}'}}h([\boldsymbol{x}_{1}]_{j}, 
[\boldsymbol{\omega}_{1}]_{j}), 
\end{equation}
with 
\begin{equation} \label{h_function}
h(u, \omega) = \sum_{m\in\mathcal{M}}\exp\left(
 \frac{|u_{m}||u|}
 {\sigma_{N/k}^{2}}
 - \frac{u_{m}^{2}}{2\sigma_{N/k}^{2}}
 + \frac{u_{m}\omega}{\sigma_{N/k}^{2}}
\right).
\end{equation}
We note that $u$ satisfying $|u|=u_{\mathrm{max}}$ is the maximizer of 
$h(u, \omega)$ over $u\in\mathcal{U}\cup\{0\}$.  
Thus, we have the uniform upper bound 
$G_{k-w}^{k}(\boldsymbol{x}_{1}, \boldsymbol{\omega}_{1}) 
\leq G_{k-w}^{k}(\boldsymbol{\omega}_{1})$, with 
$G_{0}^{k}(\boldsymbol{\omega}_{1})=1$ and 
\begin{equation} \label{G_w}
G_{k-w}^{k}(\boldsymbol{\omega}_{1})
= \sum_{\boldsymbol{x}_{1}'\in\mathcal{X}_{k-w}^{k}(\mathcal{U})}
e^{\frac{u_{\mathrm{max}}\|\boldsymbol{x}_{1}'\|_{1}
- 2^{-1}\|\boldsymbol{x}_{1}'\|_{2}^{2} 
+ \boldsymbol{x}_{1}'^{\mathrm{T}}\boldsymbol{\omega}_{1}}
{\sigma_{N/k}^{2}} }
\end{equation}
for $w\in\{0,\ldots,k-1\}$. 

We are ready to evaluate the upper bound~(\ref{KL_divergence_bound}) on 
the KL divergence $D(p \| q)$. 
Applying the uniform upper bound 
$G_{k-w}^{k}(\boldsymbol{x}_{1}, \boldsymbol{\omega}_{1}) 
\leq G_{k-w}^{k}(\boldsymbol{\omega}_{1})$ to 
$g(\boldsymbol{x})$ in (\ref{g_function_tmp0}) yields 
\begin{equation} 
D(p \| q) \leq \mathbb{E}\left[
 \log\left\{
  \sum_{w=0}^{k}\frac{|\mathcal{M}|^{w}}{|\mathcal{X}_{k}^{N}(\mathcal{U})|}
  \binom{N-k}{w}
  G_{k-w}^{k}(\boldsymbol{\omega}_{1})
 \right\}
\right].
\end{equation}
Using $|\mathcal{X}_{k}^{N}(\mathcal{U})|=|\mathcal{M}|^{k}\binom{N}{k}$ and 
the following upper bound: 
\begin{equation} \label{binomial_bound}
\binom{N}{k}^{-1}\binom{N-k}{w} < 
\frac{k!}{(N-k)^{k}}\frac{(N - k)^{w}}{w!}
< (N/k - 1)^{w - k},
\end{equation} 
we arrive at 
\begin{equation} \label{g_function_tmp}
D(p \| q) < \mathbb{E}\left[
 \log\left\{
  \sum_{w=0}^{k}[|\mathcal{M}|(N/k - 1)]^{w-k}
  G_{k-w}^{k}(\boldsymbol{\omega}_{1})
 \right\}
\right].
\end{equation}

The signal vector $\boldsymbol{x}$ can depend on the upper 
bound~(\ref{g_function_tmp}) only through the Gaussian vector 
$\boldsymbol{\omega}_{1} = \boldsymbol{\omega}_{\mathcal{S}_{\boldsymbol{x}}}$. 
Since $\boldsymbol{\omega}_{1}\sim\mathcal{N}(\boldsymbol{0}, 
\sigma_{N/k}^{2}\boldsymbol{I}_{k})$ is independent of $\boldsymbol{x}$, 
however, (\ref{g_function_tmp}) does not depend on $\boldsymbol{x}$. 
Thus, without loss of generality, we assume $\mathcal{S}_{\boldsymbol{x}}
=\{1,\ldots,k\}$. In this case, 
$\boldsymbol{\omega}_{1}=\boldsymbol{\omega}_{\mathcal{S}_{\boldsymbol{x}}}$ has 
the elements $\{\omega_{j}: j\in\{1,\ldots, k\}\}$.

\subsection{Typical Noise Samples}
To evaluate the expectation in (\ref{g_function_tmp}) over 
$\boldsymbol{\omega}_{1}$, we write the elements of 
$\mathcal{X}_{k-w}^{k}(\mathcal{U})$ for each $w\in\{0,\ldots, k-1\}$
as $\{\boldsymbol{x}_{i}^{w}\in
\mathcal{X}_{k-w}^{k}(\mathcal{U}): i\in\{1,\ldots, 
|\mathcal{X}_{k-w}^{k}(\mathcal{U})|\}$. 
Define the random variable $\Omega_{i}^{w}=(\boldsymbol{x}_{i}^{w})^{\mathrm{T}}
\boldsymbol{\omega}_{1}/\sigma_{N/k}^{2}$. 
We use $\boldsymbol{\omega}_{1}\sim\mathcal{N}(\boldsymbol{0}, 
\sigma_{N/k}^{2}\boldsymbol{I}_{k})$ to find 
$\Omega_{i}^{w}\sim\mathcal{N}(0, \|\boldsymbol{x}_{i}^{w}\|_{2}^{2}
/\sigma_{N/k}^{2})$. 
We focus on the $k$ basis random variables $\boldsymbol{\Omega}_{\mathrm{b}}
=\{\Omega_{i}^{k-1}: i\in\{1,\ldots,k\}\}$ for $k$ different $1$-sparse vectors  
$\boldsymbol{x}_{i}^{k-1}\in\mathcal{X}_{1}^{k}(\mathcal{U})$. We assume that 
$\boldsymbol{x}_{i}^{k-1}$ has a non-zero element 
with amplitude $u_{\mathrm{max}}$ in the $i$th element. 
Since all random variables $\{\Omega_{i}^{w}\}$ are linear combinations of 
the $k$ basis random variables, there are some deterministic linear functions 
$f_{i}^{w}: \mathbb{R}^{k}\to\mathbb{R}$ such that 
$\Omega_{i}^{w} = f_{i}^{w}(\boldsymbol{\Omega}_{\mathrm{b}})$ holds. 

Consider $\mathcal{E}_{i}^{w}=\{\Omega_{i}^{w}: 
|\Omega_{i}^{w}|\leq d_{i}^{w}\}$ for some $d_{i}^{w}\in\mathbb{R}$. In particular, 
we focus on the basis event 
$\mathcal{E}_{\mathrm{b}}=\cap_{i=1}^{k}\mathcal{E}_{i}^{k-1}$.  
Using this basis event $\mathcal{E}_{\mathrm{b}}$, 
we decompose the upper bound on $D(p \| q)$ 
in (\ref{g_function_tmp}) into $D(p \| q)\leq g_{1} + g_{2}$, with 
\begin{equation} 
g_{1} = \mathbb{E}\left[
 1\left(
  \mathcal{E}_{\mathrm{b}}
 \right)
 \log\left\{
  \sum_{w=0}^{k}[|\mathcal{M}|(N/k - 1)]^{w-k}
  G_{k-w}^{k}(\boldsymbol{\omega}_{1})
 \right\}
\right], \label{g_function1} 
\end{equation}
\begin{equation} 
g_{2} = \mathbb{E}\left[
 1\left(
  \mathcal{E}_{\mathrm{b}}^{\mathrm{c}}
 \right)
 \log\left\{
  \sum_{w=0}^{k}[|\mathcal{M}|(N/k - 1)]^{w-k}G_{k-w}^{k}(\boldsymbol{\omega}_{1})
 \right\}
\right].  
\label{g_function2}
\end{equation}
We need to prove that both $g_{1}$ and $g_{2}$ converge to zero in the 
sublinear sparsity limit. 

We define $d_{i}^{w}$ so that $\mathcal{E}_{\mathrm{b}}^{\mathrm{c}}$ 
becomes a rare event. Let 
\begin{equation} 
d_{i}^{w} = \frac{\|\boldsymbol{x}_{i}^{w}\|_{1}}{u_{\mathrm{max}}}
d_{*}, \label{d_i}
\end{equation}
with 
\begin{equation} \label{d_min}
d_{*} =  (1 - \epsilon)\log(N/k)
- \frac{u_{\mathrm{max}}^{2}}{2\sigma_{N/k}^{2}} 
\end{equation}
for any $\epsilon\in(0, 1 - u_{\mathrm{max}}^{2}/(2\sigma^{2}))$. 
In this case, we have $d_{i}^{k - 1} = d_{*}$ for $i\in\{1,\ldots,k\}$.  
It is straightforward to confirm $d_{i}^{w}>0$ in (\ref{d_i}) for all $i$ 
and $w$ or $d_{*}>0$ in (\ref{d_min}). Using 
$\sigma_{N/k}^{2}=\sigma^{2}/\log(N/k)$, we have 
\begin{equation} 
d_{*} = \left(
 1 - \frac{u_{\mathrm{max}}^{2} }{2\sigma^{2}} - \epsilon
\right)\log(N/k)
> 0, \label{d_i_bound}
\end{equation}
where the last positivity is due to 
$\epsilon\in(0, 1 - u_{\mathrm{max}}^{2}/(2\sigma^{2}))$.

To understand the significance of these definitions, we use the definition 
$\Omega_{i}^{k-1}=u_{*}[\boldsymbol{\omega}_{1}]_{i}/\sigma_{N/k}^{2}$ with 
$|u_{*}|=u_{\mathrm{max}}$ for all $i\in\{1,\ldots,k\}$ to represent 
$\Omega_{i}^{w}=(\boldsymbol{x}_{i}^{w})^{\mathrm{T}}
\boldsymbol{\omega}_{1}/\sigma_{N/k}^{2}$ as 
$\Omega_{i}^{w}=u_{*}^{-1}\sum_{j=1}^{k}[\boldsymbol{x}_{i}^{w}]_{j}
\Omega_{j}^{k-1}$. Applying the triangle inequality yields 
the following upper bound: 
\begin{equation}
|\Omega_{i}^{w}|
\leq \frac{1}{u_{\mathrm{max}}}\sum_{j=1}^{k}
|[\boldsymbol{x}_{i}^{w}]_{j}||\Omega_{j}^{k-1}| \leq d_{i}^{w}
\end{equation}
conditioned on $\mathcal{E}_{\mathrm{b}}$, 
where the last inequality follows from the conditioning 
$|\Omega_{i}^{k-1}|\leq d_{i}^{k-1}$ and the definition of $d_{i}^{w}$ 
in (\ref{d_i}). 
Thus, the basis event $\mathcal{E}_{\mathrm{b}}$ implies the 
event $\mathcal{E}_{i}^{w}$ for all $i$ and $w$. 

\begin{lemma} \label{lemma_moment_generating_function}
For $C_{0}$ and $C_{1,1}$ given in (\ref{constants}), we have 
\begin{equation} \label{lemma_moment_generating_function_bound}
\mathbb{P}(|\Omega_{i}^{k-1}| > d_{i}^{k-1}) 
\leq \frac{2e^{C_{0}\epsilon}}{(N/k)^{C_{1,1}}},
\end{equation}
\begin{equation} \label{lemma_moment_generating_function_bound_all}
\mathbb{P}(\mathcal{E}_{\mathrm{b}}) 
> \left\{
 1 - \frac{2e^{C_{0}\epsilon}}{(N/k)^{C_{1,1}}}
\right\}^{k}.  
\end{equation}
\end{lemma}
\begin{IEEEproof}
We first evaluate $\mathbb{P}(|\Omega_{i}^{k-1}| > d_{i}^{k-1})$ with 
$d_{i}^{k-1}=d_{*}$ in (\ref{d_min}).
Since $\Omega_{i}^{k-1}$ is a zero-mean Gaussian random variable 
with variance $u_{\mathrm{max}}^{2}/\sigma_{N/k}^{2}$, we use 
the representation $\mathbb{P}(|\Omega_{i}^{k-1}| > d_{i}^{k-1})
=2Q(\sigma_{N/k}d_{*}/u_{\mathrm{max}})$ and the Chernoff bound  
$Q(x)\leq e^{-x^{2}/2}$ for $x\geq0$ to obtain 
\begin{equation}
\mathbb{P}(|\Omega_{i}^{k-1}| > d_{i}^{k-1})
\leq 2e^{-\sigma_{N/k}^{2}d_{*}^{2}/(2u_{\mathrm{max}}^{2})}. 
\end{equation}
Evaluating the exponent with $d_{*}$ in (\ref{d_i_bound}) as 
\begin{align}
&\frac{\sigma_{N/k}^{2}d_{*}^{2}}{2u_{\mathrm{max}}^{2}} 
> \frac{\sigma^{2}}{2u_{\mathrm{max}}^{2}}\left(
 1 - \frac{u_{\mathrm{max}}^{2} }{2\sigma^{2}}
\right)^{2}\log(N/k) \nonumber \\
&- \frac{\sigma^{2}}{u_{\mathrm{max}}^{2}}\left(
 1 - \frac{u_{\mathrm{max}}^{2} }{2\sigma^{2}}
\right)\epsilon
= C_{1,1}\log(N/k) - C_{0}\epsilon, \label{d_min2}
\end{align}
with $C_{0}$ and $C_{1,1}$ in (\ref{constants}), 
we arrive at (\ref{lemma_moment_generating_function_bound}). 

We next evaluate $\mathbb{P}(\mathcal{E}_{\mathrm{b}})$. 
Since $\{\Omega_{i}^{k-1}\}_{i=1}^{k}$ are i.i.d.\ random variables, 
we represent $\mathbb{P}(\mathcal{E}_{\mathrm{b}})$ as   
\begin{equation}
\mathbb{P}(\mathcal{E}_{\mathrm{b}}) 
= \prod_{i=1}^{k}\mathbb{P}(|\Omega_{i}^{k-1}|\leq d_{i}^{k-1})
= \left\{
 \mathbb{P}(|\Omega_{1}^{k-1}|\leq d_{*})
\right\}^{k}. 
\end{equation}
Using $\mathbb{P}(|\Omega_{1}^{k-1}|\leq d_{*})
=1 - \mathbb{P}(|\Omega_{1}^{k-1}|> d_{*})$ and the 
upper bound~(\ref{lemma_moment_generating_function_bound}), 
we arrive at the lower 
bound~(\ref{lemma_moment_generating_function_bound_all}). 
\end{IEEEproof}

Lemma~\ref{lemma_moment_generating_function} implies that 
$\mathcal{E}_{\mathrm{b}}$ is a typical event under the assumption   
$\gamma<C_{1,1}/(C_{1,1} + 1)$:  
$\mathbb{P}(\mathcal{E}_{\mathrm{b}})\to1$ holds in the sublinear sparsity 
limit. Thus, 
$g_{1}$ in (\ref{g_function1}) represents contributions from the typical 
event $\mathcal{E}_{\mathrm{b}}$ while $g_{2}$ in (\ref{g_function2}) represents 
those from the non-typical rare event $\mathcal{E}_{\mathrm{b}}^{\mathrm{c}}$. 

It is straightforward to evaluate $g_{1}$ in (\ref{g_function1}). 
Since $\mathcal{E}_{\mathrm{b}}$ implies $\mathcal{E}_{i}^{w}$, 
we use $|\Omega_{i}^{w}|\leq d_{i}^{w}$ to evaluate 
$G_{k-w}^{k}(\boldsymbol{\omega}_{1})$ in (\ref{G_w}) as 
\begin{equation}
G_{k-w}^{k}(\boldsymbol{\omega}_{1}) 
\leq \sum_{i=1}^{|\mathcal{X}_{k-w}^{k}(\mathcal{U})|}
e^{\frac{u_{\mathrm{max}}\|\boldsymbol{x}_{i}^{w}\|_{1}
- 2^{-1}\|\boldsymbol{x}_{i}^{w}\|_{2}^{2}}
{\sigma_{N/k}^{2}} + d_{i}^{w}} 
\end{equation}
conditioned on $\mathcal{E}_{\mathrm{b}}$. 
Substituting the definition of $d_{i}^{w}$ in (\ref{d_i}) and 
using the lower bound $\|\boldsymbol{x}_{i}^{w}\|_{1}\geq u_{\mathrm{min}}(k-w)$ 
for the coefficient of $\epsilon\log(N/k)$, we have 
\begin{equation}
G_{k-w}^{k}(\boldsymbol{\omega}_{1}) \leq \binom{k}{w}
\sum_{\boldsymbol{m}_{1}\in\mathcal{M}^{k-w}}
(N/k)^{f(\boldsymbol{u}_{\boldsymbol{m}_{1}})
-\frac{u_{\mathrm{min}}}{u_{\mathrm{max}}}\epsilon(k-w)} \label{G_w_conditioning}
\end{equation}
conditioned on $\mathcal{E}_{\mathrm{b}}$ 
for all $w\in\{0,\ldots, k -1\}$, with 
\begin{equation}
f(\boldsymbol{u}_{\boldsymbol{m}_{1}}) 
= \frac{\|\boldsymbol{u}_{\boldsymbol{m}_{1}}\|_{1}}{u_{\mathrm{max}}}
+ \frac{u_{\mathrm{max}}\|\boldsymbol{u}_{\boldsymbol{m}_{1}}\|_{1}
- \|\boldsymbol{u}_{\boldsymbol{m}_{1}}\|_{2}^{2}}{2\sigma^{2}}. 
\end{equation} 

We evaluate $f(\boldsymbol{u}_{\boldsymbol{m}_{1}})$. 
From $u_{\mathrm{max}}\|\boldsymbol{u}_{\boldsymbol{m}_{1}}\|_{1}
- \|\boldsymbol{u}_{\boldsymbol{m}_{1}}\|_{2}^{2}\geq 0$ 
we use the assumption $2\sigma^{2}>u_{\mathrm{max}}^{2}$ to have 
\begin{equation}
f(\boldsymbol{u}_{\boldsymbol{m}_{1}})
< \frac{2}{u_{\mathrm{max}}}\|\boldsymbol{u}_{\boldsymbol{m}_{1}}\|_{1}
- \frac{\|\boldsymbol{u}_{\boldsymbol{m}_{1}}\|_{2}^{2}}{u_{\mathrm{max}}^{2}}. 
\end{equation}
Completing the square with respect to $|\boldsymbol{u}_{\boldsymbol{m}_{1}}|$ 
yields 
\begin{align}
f(\boldsymbol{u}_{\boldsymbol{m}_{1}})
< - \frac{\| |\boldsymbol{u}_{\boldsymbol{m}_{1}}| 
- u_{\mathrm{max}}\boldsymbol{1}_{k-w}\|_{2}^{2}}{u_{\mathrm{max}}^{2}} 
+ k - w
\leq k - w,
\end{align}
where the upper bound is attained at $|\boldsymbol{u}_{\boldsymbol{m}_{1}}|
=u_{\mathrm{max}}\boldsymbol{1}$. 
Applying this upper bound to (\ref{G_w_conditioning}), we arrive at 
\begin{equation} \label{G_w_bound1}
G_{k-w}^{k}(\boldsymbol{\omega}_{1}) < |\mathcal{M}|^{k-w}\binom{k}{w}
\left(
 N/k
\right)^{(1 - u_{\mathrm{min}}\epsilon/u_{\mathrm{max}})(k - w)}
\end{equation}
conditioned on $\mathcal{E}_{\mathrm{b}}$. 

We are ready to evaluate $g_{1}$ in (\ref{g_function1}). 
Using the upper bound~(\ref{G_w_bound1}) of  
$G_{k-w}^{k}(\boldsymbol{\omega}_{1})$ conditioned on 
$\mathcal{E}_{\mathrm{b}}$, we arrive at 
\begin{align}
g_{1}&< \mathbb{P}(\mathcal{E}_{\mathrm{b}})\log\left\{
 \sum_{w=0}^{k}\binom{k}{w}\left[
 \frac{(N/k)^{ - u_{\mathrm{min}}\epsilon/u_{\mathrm{max}}}}{1 - k/N}
\right]^{k - w}
\right\} \nonumber \\
&\leq k\log\left\{
 1 + \frac{(N/k)^{-u_{\mathrm{min}}\epsilon/u_{\mathrm{max}}}}{1 - k/N}
\right\}, \label{g_function1_upper_bound}
\end{align}
where the last equality follows from the binomial theorem and 
$\mathbb{P}(\mathcal{E}_{\mathrm{b}})\leq 1$. 
The upper bound~(\ref{g_function1_upper_bound}) implies that  
$g_{1}$ converges to zero in the sublinear sparsity limit 
for some $\epsilon$ satisfying   
\begin{equation} \label{epsilon}
\epsilon \in \left(
 \frac{\gamma u_{\mathrm{max}}}{(1 - \gamma)u_{\mathrm{min}}},\;  
 1 - \frac{u_{\mathrm{max}}^{2}}{2\sigma^{2}}
\right). 
\end{equation}
Note that $\epsilon$ in (\ref{epsilon}) exists for 
\begin{equation}
\gamma < \frac{C_{0}}{C_{0} + \sigma^{2}/(u_{\mathrm{max}}u_{\mathrm{min}})}, 
\end{equation}
with $C_{0}$ given in (\ref{constants}).

\subsection{Non-Typical Noise Samples}
For $g_{2}$ in (\ref{g_function2}), the non-typical event 
$\mathcal{E}_{\mathrm{b}}^{\mathrm{c}}$ can be represented as 
$\mathcal{E}_{\mathrm{b}}^{\mathrm{c}} 
=\cup_{i=1}^{k}(\mathcal{E}_{i}^{k-1})^{\mathrm{c}}$. We use the upper bound 
$1(\mathcal{E}_{\mathrm{b}}^{\mathrm{c}})\leq
\sum_{i=1}^{k}1((\mathcal{E}_{i}^{k-1})^{\mathrm{c}})$ to 
obtain $g_{2} = \sum_{i=1}^{k}g_{2,i}$ with 
\begin{align} 
g_{2,i} &= \mathbb{E}\left[
 1\left(
  (\mathcal{E}_{i}^{k-1})^{\mathrm{c}}
 \right)
\right. \nonumber \\
&\left.
 \cdot\log\left\{
  \sum_{w=0}^{k}[|\mathcal{M}|(N/k - 1)]^{w-k}
  G_{k-w}^{k}(\boldsymbol{\omega}_{1})
 \right\}
\right]. \label{g_function2_i}
\end{align}

It is difficult to evaluate the expectation in (\ref{g_function2_i}) 
with respect to $\boldsymbol{\omega}_{1}\sim\mathcal{N}(\boldsymbol{0}, 
\sigma_{N/k}^{2}\boldsymbol{I}_{k})$. Furthermore, using Jensen's 
inequality results in a loose upper bound on $g_{2,i}$. To circumvent 
this difficulty, we prove the following upper bound 
on $G_{k-w}^{k}(\boldsymbol{\omega}_{1})$ 
in (\ref{G_w}) conditioned on $(\mathcal{E}_{i}^{k-1})^{\mathrm{c}}$:
\begin{lemma} \label{lemma_G_w_upper_bound}
Define 
\begin{equation} \label{I_set}
\mathcal{I}=\{j\in\{1,\ldots,k\}: h(\omega_{j})\leq 1\},
\end{equation}
with $h(\omega_{j})=h(u_{\mathrm{max}}, \omega_{j})$ in (\ref{h_function}). 
Then, we have 
\begin{equation} \label{G_w_upper_bound2}
G_{k-w}^{k}(\boldsymbol{\omega}_{1})\leq 
H_{w}(|\mathcal{I}|) 
G_{k-|\mathcal{I}|}^{k-|\mathcal{I}|}(\boldsymbol{\omega}_{1,\mathcal{I}^{\mathrm{c}}})
\end{equation}
for all $w\in\{0, \ldots, k - 1\}$. In (\ref{G_w_upper_bound2}), 
$G_{k-|\mathcal{I}|}^{k-|\mathcal{I}|}(\boldsymbol{\omega}_{1,\mathcal{I}^{\mathrm{c}}})
\geq1$ is given in (\ref{G_w}). Furthermore, $H_{w}(j)$ is defined as  
$H_{0}(j)=1$ and 
\begin{equation} \label{H_w_j}
H_{w}(j) = \sum_{w'=\max\{j - w, 0\}}^{\min\{j, k - w\}}
\binom{j}{w'}\lceil L_{k-w-w',k-j}^{k-j}\rceil 
\end{equation}
for $w\in\{1,\ldots,k-1\}$, with 
\begin{equation} \label{L_ww'}
L_{w,w'}^{k}= \prod_{i=0}^{w'-w-1}\frac{w' - i}{k - w - i}.
\end{equation} 
\end{lemma}
\begin{IEEEproof}
See Appendix~\ref{proof_lemma_G_w_upper_bound}. 
\end{IEEEproof}

Lemma~\ref{lemma_G_w_upper_bound} allows us to replace intractable 
$G_{k-w}^{k}(\boldsymbol{\omega}_{1})$ in (\ref{g_function2_i}) with 
$G_{k-|\mathcal{I}|}^{k-|\mathcal{I}|}(\boldsymbol{\omega}_{1,\mathcal{I}^{\mathrm{c}}})$ 
in (\ref{G_w}). As shown shortly, 
$G_{k-|\mathcal{I}|}^{k-|\mathcal{I}|}(\boldsymbol{\omega}_{1,\mathcal{I}^{\mathrm{c}}})$ 
can be bounded from above by a single exponential function of 
$\boldsymbol{\omega}_{1,\mathcal{I}^{\mathrm{c}}}$. Thus, an upper bound on $g_{2,i}$ 
can be evaluated analytically.  

We evaluate $g_{2,i}$ in (\ref{g_function2_i}). 
For $w=k$ we have $G_{0}^{k}(\boldsymbol{\omega}_{1})
=1\leq G_{k-|\mathcal{I}}^{k-|\mathcal{I}|}
(\boldsymbol{\omega}_{1,\mathcal{I}^{\mathrm{c}}})$ and 
$H_{k}(j)=\lceil L_{0,k-j}^{k-j} \rceil = 1$. 
For $w\in\{0, \ldots, k-1\}$ we use Lemma~\ref{lemma_G_w_upper_bound} to have  
\begin{align}
g_{2,i}&< \mathbb{E}\left[
 1((\mathcal{E}_{i}^{k-1})^{\mathrm{c}})\log\left\{
  |\mathcal{M}|^{-(k - |\mathcal{I}|)}G_{k - |\mathcal{I}|}^{k - |\mathcal{I}|}
  (\boldsymbol{\omega}_{1,\mathcal{I}^{\mathrm{c}}})
 \right\}
\right] 
\nonumber \\
&+ \mathbb{E}\left[
 1((\mathcal{E}_{i}^{k-1})^{\mathrm{c}})H(|\mathcal{I}|)
\right], \label{g_function2_tmp}
\end{align}
with 
\begin{equation} \label{H_j}
H(j) = \log\left\{
 \sum_{w=0}^{k}|\mathcal{M}|^{w - j}(N/k - 1)^{w - k}H_{w}(j)
\right\}.
\end{equation}
The first and second terms on the upper bound~(\ref{g_function2_tmp}) are 
represented as $g_{2,i}^{(1)}$ and $g_{2,i}^{(2)}$, respectively. We use 
the upper bound~(\ref{g_function2_tmp}) to obtain $g_{2}
=\sum_{i=1}^{k}g_{2,i}\leq g_{2}^{(1)} + g_{2}^{(2)}$, with 
$g_{2}^{(l)}=\sum_{i=1}^{k}g_{2,i}^{(l)}$ for $l\in\{1, 2\}$. We need to prove 
that both $g_{2}^{(1)}$ and $g_{2}^{(2)}$ converge to zero in the 
sublinear sparsity limit. 

To evaluate them, 
we investigate statistical properties of $\mathcal{I}$ in (\ref{I_set}). 
Since $\{\omega_{j}\}$ are i.i.d.\ random variables, 
$\{h(\omega_{j})\}$ are also i.i.d. 
Thus, the distribution of $\mathcal{I}$ depends on $\mathcal{I}$ 
only through the cardinality $|\mathcal{I}|$. Furthermore, 
$|\mathcal{I}|\in\{0,\ldots, k\}$ follows the binomial distribution 
$B(k, p)$ with $p=\mathbb{P}(h(\omega_{1})\leq1)$: 
$\mathbb{P}(|\mathcal{I}|=j)=p_{j}^{(k)}$ holds with 
\begin{equation} \label{binomial_probability}
p_{j}^{(k)}=\binom{k}{j}p^{j}(1-p)^{k-j}.
\end{equation} 

In evaluating $g_{2}^{(1)}$ and $g_{2}^{(2)}$, we use two technical results.  
\begin{lemma} \label{lemma_h}
For $\omega\sim\mathcal{N}(0, \sigma_{N/k}^{2})$, we have 
$\mathbb{P}(h(\omega)\leq1) \leq (N/k)^{-C_{1,2}}$, 
with $C_{1,2}$ given in (\ref{constant2}). 
\end{lemma}
\begin{IEEEproof}
For $\omega\sim\mathcal{N}(0, \sigma_{N/k}^{2})$, 
we use the definition of $h(\omega)=h(u_{\mathrm{max}},\omega)$ in 
(\ref{h_function}) to find that $h(\omega)\leq1$ implies 
\begin{equation}
\frac{|u_{m}|u_{\mathrm{max}}}{\sigma_{N/k}^{2}} - \frac{u_{m}^{2}}{2\sigma_{N/k}^{2}} 
+ \frac{u_{m}\omega}{\sigma_{N/k}^{2}} \leq 0 
\end{equation}
for all $m\in\mathcal{M}$. In particular, selecting 
$u_{m}=u_{\mathrm{min}}$---without 
loss of generality $u_{\mathrm{min}}\in\mathcal{U}$ has been assumed---yields 
$\omega \leq -(u_{\mathrm{max}} - u_{\mathrm{min}}/2)$. We use 
$\omega\sim\mathcal{N}(0, \sigma_{N/k}^{2})$ to have 
\begin{align}
&\mathbb{P}(h(\omega)\leq1) \leq \mathbb{P}\left(
 \omega\leq -\left(
  u_{\mathrm{max}} - \frac{u_{\mathrm{min}}}{2}
 \right)
\right) \nonumber \\
&= Q\left(
 \frac{u_{\mathrm{max}} - u_{\mathrm{min}}/2}{\sigma_{N/k}}
\right) 
\leq (N/k)^{-\frac{(u_{\mathrm{max}} - u_{\mathrm{min}}/2)^{2}}{2\sigma^{2}}},
\end{align}
where the last inequality follows from the Chernoff bound for 
$Q(x)$. Thus, Lemma~\ref{lemma_h} holds. 
\end{IEEEproof}

Lemma~\ref{lemma_h} implies that $(1 + p)^{k}$ converges to zero in the 
sublinear sparsity limit for $\gamma<C_{1,2}/(C_{1,2} + 1)$. 

\begin{proposition} \label{proposition_summation}
For all $x\in\mathbb{R}$, we have 
\begin{equation} \label{summation}
S(x) = \sum_{j=0}^{k}\binom{k}{j}jx^{j} = kx(1 + x)^{k}. 
\end{equation}
\end{proposition}
\begin{IEEEproof}
Define $f(x)=\sum_{j=0}^{k}\binom{k}{j}x^{j}$. Differentiating $f(x)$ 
yields $f'(x) = \sum_{j=1}^{k}\binom{k}{j}jx^{j-1}=x^{-1}S(x)$ or equivalently 
$S(x)=xf'(x)$.  
Since the binomial theorem implies $f(x)=(1 + x)^{k}$, we have 
$f'(x) = k(1 + x)^{k-1}$. Thus,  
$S(x)=xf'(x)$ reduces to (\ref{summation}). 
\end{IEEEproof}

We prove the convergence of $g_{2}^{(2)}$ to zero in 
Appendix~\ref{proof_second_term} while $g_{2}^{(1)}$ is evaluated in 
Appendix~\ref{proof_first_term}. 

\subsection{Second Term in (\ref{g_function2_tmp})}
\label{proof_second_term} 
For the second term $g_{2,i}^{(2)}$ on the upper bound~(\ref{g_function2_tmp}), 
we evaluate $g_{2}^{(2)}=\sum_{i=1}^{k}g_{2,i}^{(2)}$. 
We use the representation $g_{2,i}^{(2)}=\sum_{j=0}^{k}g_{2,i,j}^{(2)}$ with 
\begin{equation} \label{g_2_i_j_2}
g_{2,i,j}^{(2)} = \mathbb{P}\left(
 (\mathcal{E}_{i}^{k-1})^{\mathrm{c}}\cap\{|\mathcal{I}|=j\}
\right)H(j).
\end{equation}

We first evaluate $H(j)$ in (\ref{H_j}). 
For $w\in\{1,\ldots,k-1\}$, we use $\max\{j - w, 0\}\geq 0$ and 
$\min\{j, k - w\}\leq j$ to upper-bound $H_{w}(j)$ in (\ref{H_w_j}) as  
\begin{equation}
H_{w}(j) \leq \sum_{w'=0}^{j}
\binom{j}{w'}\lceil L_{k-w-w',k-j}^{k-j}\rceil, 
\end{equation}
with 
\begin{equation}
L_{k - w - w',k - j}^{k - j}= \prod_{i=0}^{w + w' - j - 1}
\frac{k - j - i}{w + w' - j - i}
< k^{w + w' - j}. 
\end{equation}
The binomial theorem implies 
\begin{equation} \label{H_w_j_bound}
H_{w}(j) < k^{w - j}(k + 1)^{j}. 
\end{equation}
In particular, (\ref{H_w_j_bound}) is also correct for $w=0$ and $w=k$, 
because of $H_{0}(j)=H_{k}(j)=1$. We substitute the upper 
bound~(\ref{H_w_j_bound}) into $H(j)$ in (\ref{H_j}) to arrive at  
\begin{align}
H(j) &< \log\left\{
 \frac{(|\mathcal{M}|k)^{-j}(k + 1)^{j}}{(N/k - 1)^{k}}
 \frac{\{|\mathcal{M}|(N - k)\}^{k+1} - 1}{|\mathcal{M}|(N - k) - 1}
\right\} \nonumber \\
&< j\log\left(
 \frac{k + 1}{|\mathcal{M}|k} 
\right)
+ \log\frac{(|\mathcal{M}|k)^{k+1}(N/k - 1)}
{|\mathcal{M}|k(N/k - 1) - 1}.  \label{H_j_bound}
\end{align}

We next focus on $\mathbb{P}( (\mathcal{E}_{i}^{k-1})^{\mathrm{c}}
\cap\{|\mathcal{I}|=j\} )$, in which 
the probability distribution is induced from 
$\boldsymbol{\omega}_{1}\sim\mathcal{N}(\boldsymbol{0}, 
\sigma_{N/k}^{2}\boldsymbol{I}_{k})$ through $\mathcal{I}$ in (\ref{I_set}) 
and $\mathcal{E}_{i}^{k-1}=\{\Omega_{i}^{k-1}: |\Omega_{i}^{k-1}|\leq d_{i}^{k-1}\}$ 
with $\Omega_{i}^{k-1}=(\boldsymbol{x}_{i}^{k-1})^{\mathrm{T}}
\boldsymbol{\omega}_{1}/\sigma_{N/k}^{2}$. Decomposing the event 
$(\mathcal{E}_{i}^{k-1})^{\mathrm{c}}\cap\{|\mathcal{I}|=j\}$ yields 
\begin{align}
&\mathbb{P}\left(
 (\mathcal{E}_{i}^{k-1})^{\mathrm{c}}\cap\{|\mathcal{I}|=j\}
\right) \nonumber \\
&= \mathbb{P}\left(
 (\mathcal{E}_{i}^{k-1})^{\mathrm{c}}\cap\{|\mathcal{I}|=j\}
 \cap\{i\in\mathcal{I}\}
\right) \nonumber \\
&+ \mathbb{P}\left(
 (\mathcal{E}_{i}^{k-1})^{\mathrm{c}}\cap\{|\mathcal{I}|=j\}
 \cap\{i\notin\mathcal{I}\}
\right). \label{target_probability}
\end{align}
We evaluate the first term in (\ref{target_probability}).  
Since $h(\omega_{i'})\leq1$ holds for all $i'\in\mathcal{I}\setminus\{i\}$, 
as well as $(\mathcal{E}_{i}^{k-1})^{\mathrm{c}}\cap\{h(\omega_{i})\leq1\}$, 
we have $\mathbb{P}((\mathcal{E}_{i}^{k-1})^{\mathrm{c}}
\cap\{|\mathcal{I}|=0\}\cap\{i\in\mathcal{I}\})=0$, 
$\mathbb{P}((\mathcal{E}_{i}^{k-1})^{\mathrm{c}}
\cap\{|\mathcal{I}|=1\}\cap\{i\in\mathcal{I}\})=
\mathbb{P}((\mathcal{E}_{i}^{k-1})^{\mathrm{c}}\cap
\{h(\omega_{i})\leq 1\})\leq\mathbb{P}((\mathcal{E}_{i}^{k-1})^{\mathrm{c}})$, and 
\begin{align}
&\mathbb{P}((\mathcal{E}_{i}^{k-1})^{\mathrm{c}}
\cap\{|\mathcal{I}|=j\}\cap\{i\in\mathcal{I}\})
\nonumber \\
&\leq p_{j-1}^{(k-1)}\mathbb{P}((\mathcal{E}_{i}^{k-1})^{\mathrm{c}}) 
\leq \binom{k-1}{j-1}p^{j-1}\mathbb{P}((\mathcal{E}_{i}^{k-1})^{\mathrm{c}}) 
\label{target_probability_bound1}
\end{align}
for all $j\geq2$, with $p_{j}^{(k)}$ defined in (\ref{binomial_probability}) and 
$p=\mathbb{P}(h(\omega_{1})\leq1)$. In particular, this upper bound is also 
correct for $j=1$. 
For the second term in (\ref{target_probability}), similarly, we obtain 
$\mathbb{P}((\mathcal{E}_{i}^{k-1})^{\mathrm{c}}\cap\{|\mathcal{I}|=k\}
\cap\{i\notin\mathcal{I}\})=0$ and 
\begin{align}
&\mathbb{P}\left(
 (\mathcal{E}_{i}^{k-1})^{\mathrm{c}}\cap\{|\mathcal{I}|=j\}
 \cap\{i\notin\mathcal{I}\}
\right) \nonumber \\
&\leq p_{j}^{(k-1)}\mathbb{P}\left(
 (\mathcal{E}_{i}^{k-1})^{\mathrm{c}}
\right) 
\leq \binom{k-1}{j}p^{j}\mathbb{P}\left(
 (\mathcal{E}_{i}^{k-1})^{\mathrm{c}}
\right). \label{target_probability_bound2}
\end{align}
for all $j<k$. 

We are ready to evaluate $g_{2,i}^{(2)}=\sum_{j=0}^{k}g_{2,i,j}^{(2)}$. 
Substituting the upper bounds~(\ref{H_j_bound}), 
(\ref{target_probability_bound1}), and (\ref{target_probability_bound2}) 
into $g_{2,i,j}^{(2)}$ in (\ref{g_2_i_j_2}), 
from the binomial theorem we obtain 
\begin{align}
&\frac{g_{2,i}^{(2)}}{\mathbb{P}( (\mathcal{E}_{i}^{k-1})^{\mathrm{c}} )} 
< \left\{
 \sum_{j=1}^{k}\binom{k-1}{j-1}jp^{j-1} 
 + \sum_{j=0}^{k-1}\binom{k-1}{j}jp^{j}
\right\} \nonumber \\
&\cdot \log\left(
 \frac{k + 1}{|\mathcal{M}|k}
\right)
+ 2(1 + p)^{k-1}\log\frac{(|\mathcal{M}|k)^{k+1}(N/k - 1)}
{|\mathcal{M}|k(N/k - 1) - 1}. \label{g_2_i_2_bound}
\end{align}
Applying Proposition~\ref{proposition_summation} to 
the upper bound~(\ref{g_2_i_2_bound}) and using   
Lemmas~\ref{lemma_moment_generating_function} and \ref{lemma_h}, we find that 
$g_{2}^{(2)}=\sum_{i=1}^{k}g_{2,i}^{(2)}$ reduces to 
\begin{align}
&g_{2}^{(2)} < \frac{2ke^{C_{0}\epsilon}}{(N/k)^{C_{1,1}}}\left\{
 \left[
  2(k - 1)(N/k)^{-C_{1,2}} + 1
 \right]\log\left(
  \frac{k + 1}{|\mathcal{M}|k}
 \right)
\right. \nonumber \\
&\left.
 + 2\log\frac{(|\mathcal{M}|k)^{k+1}(N/k - 1)}
 {|\mathcal{M}|k(N/k - 1) - 1}
\right\}\left\{
 1 + (N/k)^{-C_{1,2}}
\right\}^{k-1}. 
\end{align}
This upper bound implies that $g_{2}^{(2)}$ converges to zero in the 
sublinear sparsity limit for $\epsilon$ satisfying (\ref{epsilon}) and 
\begin{align}
\gamma &< \min\left\{
 \frac{C_{1,1} + 2C_{1,2}}{C_{1,1} + 2C_{1,2} + 3},\; 
 \frac{C_{1,1} + C_{1,2}}{C_{1,1} + C_{1,2} + 3}
\right\} \nonumber \\
&= \frac{C_{1,1} + C_{1,2}}{C_{1,1} + C_{1,2} + 3}. 
\end{align}

\subsection{First Term in (\ref{g_function2_tmp})}
\label{proof_first_term}
For the first term $g_{2,i}^{(1)}$ on the upper bound~(\ref{g_function2_tmp}), 
we evaluate $g_{2}^{(1)}=\sum_{i=1}^{k}g_{2,i}^{(1)}$. 
We use the representation $g_{2,i}^{(1)}=\sum_{j=0}^{k}g_{2,i,j}^{(1)}$ with 
\begin{equation}
g_{2,i,j}^{(1)} = \mathbb{E}\left[
 1((\mathcal{E}_{i}^{k-1})^{\mathrm{c}}\cap\{|\mathcal{I}|=j\})
 \log\left\{
  |\mathcal{M}|^{-(k - j)}G_{k - j}^{k - j}
 \right\}
\right],
\end{equation}
where the probability distribution is induced from 
$\boldsymbol{\omega}_{1}\sim\mathcal{N}(\boldsymbol{0}, 
\sigma_{N/k}^{2}\boldsymbol{I}_{k})$ through $\mathcal{I}$ in (\ref{I_set}) 
and $\mathcal{E}_{i}^{k-1}=\{\Omega_{i}^{k-1}: |\Omega_{i}^{k-1}|\leq d_{i}^{k-1}\}$ 
with $\Omega_{i}^{k-1}=(\boldsymbol{x}_{i}^{k-1})^{\mathrm{T}}
\boldsymbol{\omega}_{1}/\sigma_{N/k}^{2}$. 

We decompose $(\mathcal{E}_{i}^{k-1})^{\mathrm{c}}\cap\{|\mathcal{I}|=j\}$ 
into two disjoint events 
$(\mathcal{E}_{i}^{k-1})^{\mathrm{c}}\cap\{|\mathcal{I}|=j\}
\cap\{i\in\mathcal{I}\}$ and $(\mathcal{E}_{i}^{k-1})^{\mathrm{c}}
\cap\{|\mathcal{I}|=j\}\cap\{i\notin\mathcal{I}\}$. Representing  
$g_{2,i,j}^{(1)}$ with $(\mathcal{E}_{i}^{k-1})^{\mathrm{c}}\cap\{|\mathcal{I}|=j\}$ 
replaced by the former and latter events as 
$g_{2,i,j}^{(1,1)}$ and $g_{2,i,j}^{(1,2)}$, respectively, 
we have $g_{2,i,j}^{(1)} = g_{2,i,j}^{(1,1)} + g_{2,i,j}^{(1,2)}$. The goal is 
evaluation of $g_{2}^{(1)}=g_{2}^{(1,1)} + g_{2}^{(1,2)}$ with 
$g_{2}^{(1,l)}=\sum_{i=1}^{k}g_{2,i}^{(1,l)}$ and 
$g_{2,i}^{(1,l)} = \sum_{j=0}^{k}g_{2,i,j}^{(1,l)}$ for $l\in\{1, 2\}$. 

For $g_{2,i,j}^{(1,1)}$, we use the definition of 
$G_{k-j}^{k-j}(\boldsymbol{\omega}_{1,\mathcal{I}^{\mathrm{c}}})$ in (\ref{G_w}) and 
Jensen's inequality to obtain 
\begin{align}
g_{2,i,j}^{(1,1)} &\leq \mathbb{P}((\mathcal{E}_{i}^{k-1})^{\mathrm{c}}
\cap\{|\mathcal{I}|=j\}\cap\{i\in\mathcal{I}\})
\nonumber \\
&\cdot\log\left\{
 \frac{1}{|\mathcal{M}|^{k - j}}\sum_{\boldsymbol{m}\in\mathcal{M}^{k-j}}
 e^{u_{\mathrm{max}}\|\boldsymbol{u}_{\boldsymbol{m}}\|_{1}/\sigma_{N/k}^{2}}
\right\}.
\end{align}
From $\mathbb{P}((\mathcal{E}_{i}^{k-1})^{\mathrm{c}}
\cap\{|\mathcal{I}|=0\}\cap\{i\in\mathcal{I}\})=0$, we have 
$g_{2,i,0}^{(1,1)}=0$. For all $j\in\{1,\ldots,k\}$, we use the upper 
bound~(\ref{target_probability_bound1}) and  
$\|\boldsymbol{u}_{\boldsymbol{m}}\|_{1}\leq u_{\mathrm{max}}(k-j)$ to obtain 
\begin{equation}
g_{2,i,j}^{(1,1)} 
\leq \frac{u_{\mathrm{max}}^{2}}{\sigma_{N/k}^{2}}(k - j)
\binom{k-1}{j-1}p^{j-1}\mathbb{P}((\mathcal{E}_{i}^{k-1})^{\mathrm{c}}). 
\end{equation}
Utilizing Proposition~\ref{proposition_summation} and the binomial theorem, 
we find that $g_{2,i}^{(1,1)}=\sum_{j=1}^{k}g_{2,i,j}^{(1,1)}$ 
is bounded from above by 
\begin{align}
g_{2,i}^{(1,1)} &\leq \frac{u_{\mathrm{max}}^{2}}{\sigma_{N/k}^{2}}\left\{
 k\sum_{j=1}^{k}\binom{k-1}{j-1}p^{j-1}
 - \sum_{j=1}^{k}\binom{k-1}{j-1}jp^{j-1}
\right\} \nonumber \\
&\cdot
\mathbb{P}((\mathcal{E}_{i}^{k-1})^{\mathrm{c}}) \nonumber \\
&= \frac{u_{\mathrm{max}}^{2}(k - 1)}{\sigma_{N/k}^{2}}
\left(
 1 - p
\right) 
\left(
 1 + p
\right)^{k-1}\mathbb{P}((\mathcal{E}_{i}^{k-1})^{\mathrm{c}}). 
\end{align}
Applying Lemmas~\ref{lemma_moment_generating_function} and \ref{lemma_h}
to this upper bound, we have the following upper bound on  
$g_{2}^{(1,1)}=\sum_{i=1}^{k}g_{2,i}^{(1,1)}$:
\begin{equation}
g_{2}^{(1,1)} < \frac{2u_{\mathrm{max}}^{2}k^{2}}{\sigma_{N/k}^{2}}
\left\{
 1 + (N/k)^{-C_{1,2}}
\right\}^{k-1}\frac{e^{C_{0}\epsilon}}{(N/k)^{C_{1,1}}}. 
\end{equation}
This upper bound implies that $g_{2}^{(1,1)}$ converges to zero in the 
sublinear sparsity limit for $\epsilon$ satisfying 
(\ref{epsilon}) and 
\begin{equation}
\gamma < \frac{C_{1,1} + C_{1,2}}{C_{1,1} + C_{1,2} + 3}.
\end{equation}

For $g_{2,i,j}^{(1,2)}$, similarly, we use the definition of 
$G_{k-j}^{k-j}(\boldsymbol{\omega}_{1,\mathcal{I}^{\mathrm{c}}})$ 
in (\ref{G_w}) and Jensen's inequality with respect to 
$\{\omega_{i'}: i'\in\mathcal{I}^{\mathrm{c}}\setminus\{i\}\}$ 
to obtain 
\begin{align}
g_{2,i,j}^{(1,2)}&\leq \mathbb{E}\left[
 1((\mathcal{E}_{i}^{k-1})^{\mathrm{c}}\cap\{|\mathcal{I}| = j\}
 \cap\{i\notin\mathcal{I}\})
\right. \nonumber \\
&\left.\cdot\log\left\{
  \frac{1}{|\mathcal{M}|^{k - j}}\sum_{\boldsymbol{m}\in\mathcal{M}^{k - j}}
  F_{i,\boldsymbol{m}}(\omega_{i})
 \right\}
\right],
\end{align}
with 
\begin{equation}
F_{i,\boldsymbol{m}}(\omega)
= \exp\left(
 \frac{u_{\mathrm{max}}\|\boldsymbol{u}_{\boldsymbol{m}}\|_{1}
 - 2^{-1}u_{m_{i}}^{2} + u_{m_{i}}\omega}{\sigma_{N/k}^{2}}
\right).
\end{equation}
By definition, $g_{2,i,k}^{(1,2)}=0$ and 
\begin{equation}
g_{2,i,j}^{(1,2)}\leq p_{j}^{(k-1)}\mathbb{E}\left[
 1((\mathcal{E}_{i}^{k-1})^{\mathrm{c}})
\log\left\{
  \sum_{\boldsymbol{m}\in\mathcal{M}^{k - j}}
  \frac{F_{i,\boldsymbol{m}}(\omega_{i})}{|\mathcal{M}|^{k - j}}
 \right\}
\right]
\end{equation}
hold for all $j<k$, with $p_{j}^{(k)}$ defined in (\ref{binomial_probability}). 
Using the upper bound $F_{i,\boldsymbol{m}}(\omega)
< e^{\{u_{\mathrm{max}}^{2}(k-j) + u_{m_{i}}\omega\}/\sigma_{N/k}^{2}}$ and the definition 
$\Omega_{i}^{k-1}=u_{*}\omega_{i}/\sigma_{N/k}^{2}$ with $|u_{*}|=u_{\mathrm{max}}$, 
as well as $d_{i}^{k-1}=d_{*}$, we obtain  
\begin{align}
&\frac{g_{2,i,j}^{(1,2)}}{p_{j}^{(k-1)}} 
- \frac{u_{\mathrm{max}}^{2}(k - j)}{\sigma_{N/k}^{2}}
\mathbb{P}((\mathcal{E}_{i}^{k-1})^{\mathrm{c}})
\nonumber \\
&< \int_{|\Omega_{i}^{k-1}|>d_{*}}\log\left(
 \sum_{\boldsymbol{m}\in\mathcal{M}^{k-j}}
 \frac{e^{\frac{u_{m_{i}}\Omega_{i}^{k-1}}{u_{*}}}}{|\mathcal{M}|^{k - j}}
\right)p(\Omega_{i}^{k-1})d\Omega_{i}^{k-1} \nonumber \\
&\leq  2\int_{d_{*}}^{\infty}\Omega_{i}^{k-1}p(\Omega_{i}^{k-1})
d\Omega_{i}^{k-1} 
= \frac{u_{\mathrm{max}}}{\sigma_{N/k}}\sqrt{\frac{2}{\pi}}
e^{-\frac{\sigma_{N/k}^{2}d_{*}^{2}}{2u_{\mathrm{max}}^{2}}} 
\nonumber \\
&\leq\frac{u_{\mathrm{max}}}{\sigma_{N/k}}\sqrt{\frac{2}{\pi}}
\frac{e^{C_{0}\epsilon}}{(N/k)^{C_{1,1}}}. 
\end{align}
In the derivation of the second inequality, we have used 
$|u_{m_{i}}\Omega_{i}^{k-1}/u_{*}|\leq |\Omega_{i}^{k-1}|$ 
while the equality is obtained from 
$\Omega_{i}^{k-1}\sim\mathcal{N}(0, u_{\mathrm{max}}^{2}/\sigma_{N/k}^{2})$. 
The last inequality follows from the 
lower bound~(\ref{d_min2}) on $d_{*}$. 
Thus, we use Lemma~\ref{lemma_moment_generating_function} to arrive at 
\begin{align} 
g_{2,i,j}^{(1,2)}
&< p_{j}^{(k-1)}\left\{
 \frac{2u_{\mathrm{max}}^{2}(k - j)}{\sigma_{N/k}^{2}} 
 + \frac{u_{\mathrm{max}}}{\sigma_{N/k}}\sqrt{\frac{2}{\pi}}
\right\}
\frac{e^{C_{0}\epsilon}}{(N/k)^{C_{1,1}}}. 
\label{g_function2_1_bound}
\end{align}

We are ready to evaluate $g_{2}^{(1,2)}=\sum_{i=1}^{k}g_{2,i}^{(1,2)}$ with 
$g_{2,i}^{(1,2)}=\sum_{j=0}^{k-1}g_{2,i,j}^{(1,2)}$. Utilizing  
$p_{j}^{(k-1)}\leq \binom{k-1}{j}p^{j}$ in (\ref{binomial_probability}), 
the binomial theorem, and Proposition~\ref{proposition_summation}, we have 
\begin{equation}
\frac{g_{2}^{(1,2)}}{ke^{C_{0}\epsilon}} < \left\{
 \frac{2u_{\mathrm{max}}^{2}[k - (k - 1)p]}{\sigma_{N/k}^{2}}
 + \frac{u_{\mathrm{max}}}{\sigma_{N/k}}\sqrt{\frac{2}{\pi}}
\right\} \frac{(1 + p)^{k-1}}{(N/k)^{C_{1,1}}}.
\end{equation}
We use Lemma~\ref{lemma_h} to arrive at 
\begin{align}
g_{2}^{(1,2)} &< \left(
 \frac{2u_{\mathrm{max}}^{2}k^{2}}{\sigma_{N/k}^{2}}
 + \frac{u_{\mathrm{max}}k}{\sigma_{N/k}}\sqrt{\frac{2}{\pi}}
\right) 
\nonumber \\
&\cdot\left\{
 1 + (N/k)^{-C_{1,2}}
\right\}^{k-1}\frac{e^{C_{0}\epsilon}}{(N/k)^{C_{1,1}}}. 
\end{align}
This upper bound implies that $g_{2}^{(1,2)}$ converges to zero in the 
sublinear sparsity limit for $\epsilon$ satisfying (\ref{epsilon}) and 
\begin{equation}
\gamma < \frac{C_{1,1} + C_{1,2}}{C_{1,1} + C_{1,2} + 3}.
\end{equation}
Thus, Lemma~\ref{lemma_KL_divergence} holds.

\section{Proof of Lemma~\ref{lemma_G_w_upper_bound}}
\label{proof_lemma_G_w_upper_bound}
For $w=0$, we represent $G_{k}^{k}(\boldsymbol{\omega}_{1})$ in (\ref{G_w}) as 
\begin{equation}
G_{k}^{k}(\boldsymbol{\omega}_{1})
= \prod_{j=1}^{k}h(\omega_{j}),
\end{equation}
with $h(\omega)=h(u_{\mathrm{max}}, \omega)$ in (\ref{h_function}). Thus, 
we use $\mathcal{I}=\{j\in\{1,\ldots,k\}: h(\omega_{j})\leq1\}$ to find 
$G_{k}^{k}(\boldsymbol{\omega}_{1})\leq G_{k-|\mathcal{I}|}^{k-|\mathcal{I}|}
(\boldsymbol{\omega}_{1,\mathcal{I}^{\mathrm{c}}})$. 
In particular, from $H_{0}(j) = 1$ we obtain 
the upper bound~(\ref{G_w_upper_bound2}) for $w=0$.  

For $w\in\{1,\ldots, k -1\}$, we consider the following decomposition of 
$\mathcal{X}_{k-w}^{k}(\mathcal{U})$: 
\begin{equation}
\mathcal{X}_{k-w}^{k}(\mathcal{U})
= \bigcup_{w'=\max\{|\mathcal{I}| - w, 0\}}^{\min\{|\mathcal{I}|, k - w\}}
\mathcal{X}_{w'}^{|\mathcal{I}|}(\mathcal{U})\times 
\mathcal{X}_{k-w-w'}^{k - |\mathcal{I}|}(\mathcal{U}),  
\end{equation}
where $\mathcal{X}_{w'}^{|\mathcal{I}|}(\mathcal{U})$ corresponds to 
vectors in positions $\mathcal{I}$ while 
$\mathcal{X}_{k-w-w'}^{k-|\mathcal{I}|}(\mathcal{U})$ 
is associated with $\mathcal{I}^{\mathrm{c}}$. 
We use this decomposition to represent $G_{k - w}^{k}(\boldsymbol{\omega}_{1})$ 
in (\ref{G_w}) as 
\begin{align}
G_{k-w}^{k}(\boldsymbol{\omega}_{1})
&= \sum_{w'=\max\{|\mathcal{I}| - w, 0\}}^{\min\{|\mathcal{I}|, k - w\}}
\tilde{G}_{k-w-w'}^{k - |\mathcal{I}|}(\boldsymbol{\omega}_{1,2}) \nonumber \\
&\cdot\sum_{\boldsymbol{x}_{1,1}'\in\mathcal{X}_{w'}^{|\mathcal{I}|}(\{1\})}
\prod_{j\in\mathcal{S}_{\boldsymbol{x}_{1,1}'}}
h([\boldsymbol{\omega}_{1,1}]_{j}), 
\end{align}
with 
\begin{equation} \label{G_tilde_w}
\tilde{G}_{k-w-w'}^{k - |\mathcal{I}|}(\boldsymbol{\omega}_{1,2})
= \sum_{\boldsymbol{x}_{1,2}'\in\mathcal{X}_{k-w-w'}^{k-|\mathcal{I}|}(\{1\})}
\prod_{j\in\mathcal{S}_{\boldsymbol{x}_{1,2}'}}
h([\boldsymbol{\omega}_{1,2}]_{j}),
\end{equation}
where $\boldsymbol{\omega}_{1}\in\mathbb{R}^{k}$ satisfies 
$\boldsymbol{\omega}_{1,\mathcal{I}}=\boldsymbol{\omega}_{1,1}$ and 
$\boldsymbol{\omega}_{1,\mathcal{I}^{\mathrm{c}}}=\boldsymbol{\omega}_{1,2}$. 
From the definition of $\mathcal{I}$, as well as $h(\omega_{j})
=h(u_{\mathrm{max}}, \omega_{j})$ in 
(\ref{h_function}), we obtain 
\begin{equation} 
G_{k-w}^{k}(\boldsymbol{\omega}_{1})
\leq \sum_{w'=\max\{|\mathcal{I}| - w, 0\}}^{\min\{|\mathcal{I}|, k - w\}}
\binom{|\mathcal{I}|}{w'}
\tilde{G}_{k-w-w'}^{k - |\mathcal{I}|}(\boldsymbol{\omega}_{1,2}). 
\end{equation}
In particular, we use the definition of $G_{k}^{k}$ in (\ref{G_w}) to 
find $\tilde{G}_{k - |\mathcal{I}|}^{k - |\mathcal{I}|}(\boldsymbol{\omega}_{1,2})
=G_{k - |\mathcal{I}|}^{k - |\mathcal{I}|}(\boldsymbol{\omega}_{1,2})\geq1$. 

To upper-bound $\tilde{G}_{k-w-w'}^{k - |\mathcal{I}|}(\boldsymbol{\omega}_{1,2})$, 
we represent (\ref{G_tilde_w}) with $h(\omega)=h(u_{\mathrm{max}}, \omega)$ 
in (\ref{h_function}) as 
\begin{align}
\tilde{G}_{k-w-w'}^{k - |\mathcal{I}|}(\boldsymbol{\omega}_{1,2})
= \sum_{\boldsymbol{x}_{1,2}'\in\mathcal{X}_{k-w-w'}^{k-|\mathcal{I}|}(\mathcal{U})}
\prod_{i=1}^{k - |\mathcal{I}|}f_{i}([\boldsymbol{x}_{1,2}']_{i}),  
\end{align}
with 
\begin{equation} \label{f_function}
f_{i}(x) 
= \exp\left(
 \frac{u_{\mathrm{max}}|x|}
 {\sigma_{N/k}^{2}}
 - \frac{|x|^{2}}{2\sigma_{N/k}^{2}}
 + \frac{x[\boldsymbol{\omega}_{1,2}]_{i}}{\sigma_{N/k}^{2}}
\right). 
\end{equation}
\begin{lemma} \label{lemma_perfect_matching_bound}
For $i\in\{1,\ldots,k\}$, suppose that $f_{i}: \mathbb{R}\to[0, \infty)$ 
satisfies $f_{i}(0)\leq\sum_{m\in\mathcal{M}}f_{i}(u_{m})$. 
Then, we have 
\begin{equation} \label{perfect_matching_bound}
\sum_{\boldsymbol{x}\in\mathcal{X}_{w}^{k}(\mathcal{U})}
\prod_{i=1}^{k}f_{i}(x_{i})
\leq \lceil L_{w,w'}^{k}\rceil\sum_{\boldsymbol{x}'\in\mathcal{X}_{w'}^{k}(\mathcal{U})}
\prod_{i=1}^{k}f_{i}(x_{i}')
\end{equation}
for $w<w'$, with $L_{w,w'}^{k}$ defined in (\ref{L_ww'}). 
\end{lemma}
\begin{IEEEproof}
See Appendix~\ref{proof_perfect_matching_bound}. 
\end{IEEEproof}

From the condition $h(\omega_{i})>1$ for $i\notin\mathcal{I}$, as well as 
the definition of $h(\omega_{i})=h(u_{\mathrm{max}}, \omega_{i})$ in 
(\ref{h_function}), we find $f_{i}(0)=1$ and $\sum_{m\in\mathcal{M}}f_{i}(u_{m})
=h(\omega_{i})>1$ for $f_{i}$ in (\ref{f_function}). Thus, we can use 
Lemma~\ref{lemma_perfect_matching_bound} to obtain 
$\tilde{G}_{k-w-w'}^{k-|\mathcal{I}|}\leq 
\lceil L_{k-w-w',k-|\mathcal{I}|}^{k-|\mathcal{I}|}\rceil
\tilde{G}_{k-|\mathcal{I}|}^{k-|\mathcal{I}|}$. 
Combining these results, we arrive at 
the upper bound~(\ref{G_w_upper_bound2}). 

\section{Proof of Lemma~\ref{lemma_perfect_matching_bound}}
\label{proof_perfect_matching_bound} 
We introduce a graphical relationship between 
$\mathcal{X}_{w}^{k}(\mathcal{U})$ and $\mathcal{X}_{w'}^{k}(\mathcal{U})$ 
for $w<w'$. Let $\mathcal{V}_{w}^{k}=\mathcal{X}_{w}^{k}(\{1\})$ denote 
the set of all $k$-dimensional, $w$-sparse, and binary vectors. 
We consider an undirected bipartite graph 
$G(\mathcal{V}_{w}^{k}\cup\mathcal{V}_{w'}^{k}, \mathcal{E})$, in which there is 
an edge $(v, v')\in\mathcal{E}$ between $v\in\mathcal{V}_{w}^{k}$ and 
$v'\in\mathcal{V}_{w'}^{k}$ only when the support $\mathcal{S}_{v}$ of 
the $w$-sparse binary vector $v\in\mathcal{X}_{w}^{k}(\{1\})$ is a subset of 
the support $\mathcal{S}_{v'}$ of $v'\in\mathcal{X}_{w'}^{k}(\{1\})$. 

The significance of this edge definition can be understood as follows: 
For an edge $(v, v')\in\mathcal{E}=\mathcal{V}_{w}^{k}\times
\mathcal{V}_{w'}^{k}$, 
suppose that $\boldsymbol{x}\in\mathcal{X}_{w}^{k}(\mathcal{U})$ and 
$\boldsymbol{x}'\in\mathcal{X}_{w'}^{k}(\mathcal{U})$ satisfy 
$\mathcal{S}_{\boldsymbol{x}}=\mathcal{S}_{v}$, 
$\mathcal{S}_{\boldsymbol{x}'}=\mathcal{S}_{v'}$, and 
$x_{i}=x_{i}'$ for all $i\in\mathcal{S}_{\boldsymbol{x}}$. By definition, 
each term in (\ref{perfect_matching_bound}) satisfies 
\begin{equation}
\prod_{i=1}^{k}f_{i}(x_{i})
\leq \sum_{i'\in\mathcal{S}_{\boldsymbol{x}'}\setminus
\mathcal{S}_{\boldsymbol{x}}}\sum_{x_{i'}'\in\mathcal{U}}
\prod_{i=1}^{k}f_{i}(x_{i}'), 
\end{equation} 
because of the assumption 
$f_{i}(0)\leq\sum_{m\in\mathcal{M}}f_{i}(u_{m})$ for all $i$. 
Thus, we can upper-bound the left-hand side (LHS) in 
(\ref{perfect_matching_bound}) by replacing 
$\boldsymbol{x}$ with $\boldsymbol{x}'$, if 
there is a one-to-one correspondence between 
$\boldsymbol{x}$ and $\boldsymbol{x}'$. To prove the existence 
of a one-to-one correspondence, we need the notion of a perfect matching 
in graph theory. 

\begin{definition}
For $w<w'\leq \lceil k/2\rceil$, a subset $\mathcal{M}\subset\mathcal{E}$ of 
edges in the undirected bipartite graph 
$G(\mathcal{V}_{w}^{k}\cup\mathcal{V}_{w'}^{k}, \mathcal{E})$ is called a 
perfect matching with respect to $\mathcal{V}_{w}^{k}$ if $\mathcal{M}$ has 
$|\mathcal{V}_{w}^{k}|$ elements and if different edges in $\mathcal{M}$ 
are not adjacent to each other: 
Different edges $(v_{1}, v_{1}'), (v_{2}, v_{2}')\in\mathcal{M}$ satisfy 
$v_{1}\neq v_{2}$ and $v_{2}\neq v_{2}'$.  
\end{definition}

When the graph $G(\mathcal{V}_{w}^{k}\cup\mathcal{V}_{w'}^{k}, \mathcal{E})$ has 
a perfect matching with respect to $\mathcal{V}_{w}^{k}$, there is a 
one-to-one correspondence between 
$\boldsymbol{x}\in\mathcal{X}_{w}^{k}(\mathcal{U})$ 
and $\boldsymbol{x}'\in\mathcal{X}_{w'}^{k}(\mathcal{U})$. Thus, we have 
the upper bound~(\ref{perfect_matching_bound}) with $L_{w,w'}^{k}=1$. 

To prove the existence of a perfect matching, we utilize a general result  
for a regular bipartite graph 
$G(\mathcal{V}_{1}\cup\mathcal{V}_{2}, \mathcal{E})$. 
Let $\mathcal{N}_{1}(v)\subset\mathcal{V}_{2}$ denote the neighborhood of 
$v\in\mathcal{V}_{1}$, i.e.\ the set of all $v'\in\mathcal{V}_{2}$ 
adjacent to $v$. Similarly, the neighborhood of $v'\in\mathcal{V}_{2}$ is 
written as $\mathcal{N}_{2}(v')\subset\mathcal{V}_{1}$. The graph is 
said to be $(d_{1}, d_{2})$-regular when $|\mathcal{N}_{1}(v)|=d_{1}$ and 
$|\mathcal{N}_{2}(v')|=d_{2}$ are satisfied for all 
$v\in\mathcal{V}_{1}$ and $v'\in\mathcal{V}_{2}$. Since the number of outgoing 
edges from $\mathcal{V}_{1}$ is equal to that of incoming edges to 
$\mathcal{V}_{2}$, we have $d_{1}|\mathcal{V}_{1}| = d_{2}|\mathcal{V}_{2}|$. 

\begin{lemma} \label{lemma_perfect_matching}
Suppose that $G(\mathcal{V}_{1}\cup\mathcal{V}_{2}, \mathcal{E})$ is 
$(d_{1}, d_{2})$-regular. 
Then, there is a perfect matching with respect to $\mathcal{V}_{1}$ if and 
only if $|\mathcal{V}_{1}|\leq |\mathcal{V}_{2}|$ holds.  
\end{lemma}
\begin{IEEEproof}
Hall's marriage theorem~\cite{Hall35} implies that there is a perfect matching 
with respect to $\mathcal{V}_{1}$ 
in $G(\mathcal{V}_{1}\cup\mathcal{V}_{2}, \mathcal{E})$ if and only if 
all subsets $\mathcal{V}\subset\mathcal{V}_{1}$ satisfy the marriage 
condition $|\mathcal{V}|\leq|\mathcal{N}_{1}(\mathcal{V})|$ 
with $\mathcal{N}_{1}(\mathcal{V})=\cup_{v\in\mathcal{V}}\mathcal{N}_{1}(v)$. 
Thus, it is sufficient to confirm that the marriage condition is equivalent 
to $|\mathcal{V}_{1}|\leq |\mathcal{V}_{2}|$. The marriage condition 
for $\mathcal{V}=\mathcal{V}_{1}$ implies 
$|\mathcal{V}_{1}|\leq |\mathcal{V}_{2}|$. 

We prove that $|\mathcal{V}_{1}|\leq |\mathcal{V}_{2}|$ implies 
the marriage condition. For any subset 
$\mathcal{V}\subset\mathcal{V}_{1}$, there are $d_{1}|\mathcal{V}|$ edges 
that are outgoing from $\mathcal{V}$ to $\mathcal{N}_{1}(\mathcal{V})$. 
Since $v'\in\mathcal{V}_{2}$ has $d_{2}$ edges, we have 
the inequality $d_{1}|\mathcal{V}|\leq d_{2}|\mathcal{N}_{1}(\mathcal{V})|$. 
From the identity $d_{1}|\mathcal{V}_{1}| = d_{2}|\mathcal{V}_{2}|$ and 
the assumption $|\mathcal{V}_{1}|\leq|\mathcal{V}_{2}|$, we find 
$d_{1}\geq d_{2}$. Combining these two inequalities, 
we arrive at the marriage condition 
$|\mathcal{N}_{1}(\mathcal{V})|\geq (d_{1}/d_{2})|\mathcal{V}|
\geq |\mathcal{V}|$.  
\end{IEEEproof}

\begin{corollary} \label{corollary_covering} 
If $|\mathcal{V}_{w}^{k}|\leq|\mathcal{V}_{w'}^{k}|$ holds, then 
the graph $G(\mathcal{V}_{w}^{k}\cup\mathcal{V}_{w'}^{k}, \mathcal{E})$ has 
a perfect matching with respect to $\mathcal{V}_{w}^{k}$. 
\end{corollary}
\begin{IEEEproof}
From the assumption $|\mathcal{V}_{w}^{k}|\leq|\mathcal{V}_{w'}^{k}|$ and 
Lemma~\ref{lemma_perfect_matching}, it is sufficient to confirm that 
$G(\mathcal{V}_{w}^{k}\cup\mathcal{V}_{w'}^{k}, \mathcal{E})$ is regular. 

For any $v\in\mathcal{V}_{w}^{k}$, the neighborhood $\mathcal{N}_{w}(v)$ of $v$ 
consists of $v'\in\mathcal{V}_{w'}^{k}$ 
with $\mathcal{S}_{v}\subset\mathcal{S}_{v'}$. In other words, 
$v'\in\mathcal{V}_{w'}^{k}$ is obtained by flipping $w'-w$ zeros in the 
$k$-dimensional $w$-sparse vector $v$ to $1$. Thus, 
$|\mathcal{N}_{w}(v)|=\binom{k-w}{w'-w}$ holds. Similarly, we have 
$|\mathcal{N}_{w'}(v')|=\binom{w'}{w'-w}$ for the neighborhood 
$\mathcal{N}_{w'}(v')$ of $v'\in\mathcal{V}_{w'}^{k}$. 
These results imply that 
$G(\mathcal{V}_{w}^{k}\cup\mathcal{V}_{w'}^{k}, \mathcal{E})$ is 
$(d_{1}, d_{2})$-regular with $d_{1}=\binom{k-w}{w'-w}$ and 
$d_{2}=\binom{w'}{w'-w}$. 
\end{IEEEproof}

For $|\mathcal{V}_{w}^{k}|>|\mathcal{V}_{w'}^{k}|$, there is no perfect matching 
with respect to $\mathcal{V}_{w}^{k}$. For this case, 
we consider $\lceil L_{w,w'}^{k} \rceil$ copies $\tilde{\mathcal{V}}_{w'}^{k}
=\{\mathcal{V}_{w',l}^{k}: l\in\{1,\ldots,\lceil L_{w,w'}^{k} \rceil\}\}$ 
of $\mathcal{V}_{w'}^{k}$. We define 
$G(\mathcal{V}_{w}^{k}\cup\tilde{\mathcal{V}}_{w'}^{k}, \tilde{\mathcal{E}})$ 
with $\tilde{\mathcal{E}}=\cup_{l=1}^{\lceil L_{w,w'}^{k} \rceil}\mathcal{E}_{l}$ 
as the bipartite graph of which each subgraph  
$G(\mathcal{V}_{w}^{k}\cup\mathcal{V}_{w',l}^{k}, \mathcal{E}_{l})$ has  
the same edge structure as the original graph 
$G(\mathcal{V}_{w}^{k}\cup\mathcal{V}_{w'}^{k}, \mathcal{E})$. 
When $G(\mathcal{V}_{w}^{k}\cup \tilde{\mathcal{V}}_{w'}^{k}, 
\tilde{\mathcal{E}})$ has 
a perfect matching with respect to $\mathcal{V}_{w}^{k}$, 
the upper bound~(\ref{perfect_matching_bound}) holds. Thus, we prove 
the following lemma to complete the proof of 
Lemma~\ref{lemma_perfect_matching_bound}. 

\begin{lemma} \label{lemma_covering}
Define $L_{w,w'}^{k}$ in (\ref{L_ww'}). The graph 
$G(\mathcal{V}_{w}^{k}\cup \tilde{\mathcal{V}}_{w'}^{k}, \tilde{\mathcal{E}})$  
has a perfect matching with respect to 
$\mathcal{V}_{w}^{k}$. 
\end{lemma}
\begin{IEEEproof}
Since the original graph 
$G(\mathcal{V}_{w}^{k}\cup \mathcal{V}_{w'}^{k}, \mathcal{E})$ is a  
$(d_{1}, d_{2})$-regular graph with $d_{1}=\binom{k-w}{w'-w}$ and 
$d_{2}=\binom{w'}{w'-w}$,  
$G(\mathcal{V}_{w}^{k}\cup \tilde{\mathcal{V}}_{w'}^{k}, \tilde{\mathcal{E}})$  
is $(\lceil L_{w,w'}^{k} \rceil d_{1}, d_{2})$-regular. 
From Lemma~\ref{lemma_perfect_matching} it is sufficient to confirm 
$|\mathcal{V}_{w}^{k}|\leq |\tilde{\mathcal{V}}_{w'}^{k}|$: 
\begin{align}
\frac{|\tilde{\mathcal{V}}_{w'}^{k}|}{|\mathcal{V}_{w}^{k}|}
&= \lceil L_{w,w'}^{k} \rceil\binom{k}{w'}\binom{k}{w}^{-1} \nonumber \\
&= \frac{\lceil L_{w,w'}^{k}\rceil (k - w)\ldots(k-w'+1)}
{w'\cdots(w+1)} \geq 1,
\end{align}
where the inequality follows from the definition of $L_{w,w'}^{k}$ 
in (\ref{L_ww'}). Thus, Lemma~\ref{lemma_covering} holds. 
\end{IEEEproof}

Lemma~\ref{lemma_covering} contains Corollary~\ref{corollary_covering}, 
because of $L_{w,w'}^{k}\leq 1$ for 
$|\mathcal{V}_{w}^{k}|\leq|\mathcal{V}_{w'}^{k}|$. Furthermore, 
the constant $L_{w,w'}^{k}$ is tight because 
$(\lceil L_{w,w'}^{k} \rceil - 1)|\mathcal{V}_{w'}^{k}|
<|\mathcal{V}_{w}^{k}|$ holds. 

\section{Proof of Lemma~\ref{lemma_square_error}}
\label{proof_lemma_square_error}
We know that the KL divergence can be represented with mutual information. 
To connect the mutual information with the square error, 
we focus on an identity between them~\cite{Guo05}. In particular, 
we utilize the identity by introducing a perturbation channel, 
\begin{equation} \label{MIMO}
y_{N+1} = (\alpha/k)^{1/2}\boldsymbol{a}^{\mathrm{T}}\boldsymbol{x} + \omega_{N+1}, 
\quad 
\omega_{N+1}\sim\mathcal{N}(0, 1). 
\end{equation}
In (\ref{MIMO}), $\boldsymbol{a}\in\mathbb{R}^{N}$ has independent 
standard Gaussian elements while the noise $\omega_{N+1}$ is independent of 
$\{\boldsymbol{x}, \boldsymbol{y}, \boldsymbol{a}\}$. 
We recall that $\boldsymbol{x}$ is uniformly distributed on 
$\mathcal{X}_{k}^{N}(\mathcal{U})$, while $\boldsymbol{y}$ is defined 
in the AWGN channel~(\ref{AWGN}). 
The coefficient $\alpha>0$ has been introduced to control SNR 
in the perturbation channel. The limit $\alpha\to0$ is considered 
to eliminate the influence of the perturbation channel from the final result. 
The overall channel is represented with 
$\tilde{\boldsymbol{y}}=[\boldsymbol{y}^{\mathrm{T}}, y_{N+1}]^{\mathrm{T}}$. 

We first connect $I(\boldsymbol{x}; \tilde{\boldsymbol{y}} | \boldsymbol{a})$ 
with the square error. Consider the posterior mean estimator 
$\hat{\boldsymbol{x}}_{\mathrm{opt}}(\boldsymbol{y}) 
= \mathbb{E}[\boldsymbol{x} | \boldsymbol{y}]$ of 
$\boldsymbol{x}$ given $\boldsymbol{y}$. 
Since $\{\hat{\boldsymbol{x}}_{\mathrm{opt}}(\boldsymbol{y}), y_{N+1}\}$ 
is a sufficient statistic in estimating $\boldsymbol{x}$ 
from $\tilde{\boldsymbol{y}}$ conditioned on $\boldsymbol{a}$, we have 
$I(\boldsymbol{x}; \tilde{\boldsymbol{y}} | \boldsymbol{a}) 
=I(\boldsymbol{x}; \hat{\boldsymbol{x}}_{\mathrm{opt}}(\boldsymbol{y}), 
y_{N+1} | \boldsymbol{a})$. 
Using the chain rule for mutual information~\cite{Cover06}, we have 
\begin{equation} \label{mutual_inf_tmp}
I(\boldsymbol{x}; \tilde{\boldsymbol{y}} | \boldsymbol{a})  
= I(\boldsymbol{x}; \hat{\boldsymbol{x}}_{\mathrm{opt}}(\boldsymbol{y}) 
| \boldsymbol{a})
+ I(\boldsymbol{x}; y_{N+1} | \hat{\boldsymbol{x}}_{\mathrm{opt}}(\boldsymbol{y}), 
\boldsymbol{a}).  
\end{equation}
For the first term, we obtain 
\begin{equation} \label{mutual_inf_tmp_first_term}
I(\boldsymbol{x}; \hat{\boldsymbol{x}}_{\mathrm{opt}}(\boldsymbol{y}) 
| \boldsymbol{a})
=I(\boldsymbol{x}; \hat{\boldsymbol{x}}_{\mathrm{opt}}(\boldsymbol{y}))
=I(\boldsymbol{x}; \boldsymbol{y})
\leq C_{\mathrm{AWGN}}, 
\end{equation}
where $C_{\mathrm{AWGN}}$ denotes the capacity~(\ref{capacity}) 
of the AWGN channel~(\ref{AWGN}).  

We evaluate the second term in (\ref{mutual_inf_tmp}). 
The mutual information 
$I(\boldsymbol{x}; y_{N+1} | \hat{\boldsymbol{x}}_{\mathrm{opt}}(\boldsymbol{y}), 
\boldsymbol{a})$ is equivalent to the mutual 
information between $\Delta\boldsymbol{x} = \boldsymbol{x} 
- \hat{\boldsymbol{x}}_{\mathrm{opt}}(\boldsymbol{y})$ and 
$y_{N+1} - (\alpha/k)^{1/2}\boldsymbol{a}^{\mathrm{T}}
\hat{\boldsymbol{x}}_{\mathrm{opt}}(\boldsymbol{y})
= (\alpha/k)^{1/2}\boldsymbol{a}^{\mathrm{T}}\Delta\boldsymbol{x} + \omega_{N+1}$ 
conditioned on $\hat{\boldsymbol{x}}_{\mathrm{opt}}(\boldsymbol{y})$ and 
$\boldsymbol{a}$. 
Replacing $\Delta\boldsymbol{x}$ with capacity-achieving 
signaling~\cite{Telatar99}---a 
zero-mean Gaussian vector with covariance 
$N^{-1}\mathbb{E}[\|\Delta\boldsymbol{x}\|_{2}^{2}]\boldsymbol{I}_{N}$, 
we have 
\begin{align}
I(\boldsymbol{x}; y_{N+1} | \hat{\boldsymbol{x}}_{\mathrm{opt}}(\boldsymbol{y}), 
\boldsymbol{a})
&\leq \frac{1}{2}\mathbb{E}\left[
 \log\left(
  1 + \frac{\alpha\|\boldsymbol{a}\|_{2}^{2}
  \mathbb{E}[\|\Delta\boldsymbol{x}\|_{2}^{2}]}{kN}
 \right)
\right] \nonumber \\
&\leq \frac{1}{2}\log\left(
 1 + \frac{\alpha}{k}\mathbb{E}[\|\Delta\boldsymbol{x}\|_{2}^{2}]
\right), 
\end{align}
where the last inequality follows from Jensen's inequality and 
$\boldsymbol{a}\sim\mathcal{N}(\boldsymbol{0}, \boldsymbol{I}_{N})$. 
Using the fact that the posterior mean estimator 
$\hat{\boldsymbol{x}}_{\mathrm{opt}}(\boldsymbol{y})$ minimizes the mean-square 
error, we arrive at 
\begin{equation} \label{mutual_inf_tmp_second_term}
I(\boldsymbol{x}; y_{N+1} | \hat{\boldsymbol{x}}_{\mathrm{opt}}(\boldsymbol{y}), 
\boldsymbol{a})
\leq \frac{1}{2}\log\left\{
 1 + \frac{\alpha}{k}\mathbb{E}[\|\boldsymbol{x} 
 - \hat{\boldsymbol{x}}(\boldsymbol{y})\|_{2}^{2}] 
\right\}  
\end{equation}
for any estimator $\hat{\boldsymbol{x}}(\boldsymbol{y})$ of 
$\boldsymbol{x}$ based on the observation $\boldsymbol{y}$. 

We next connect the mutual information with the KL divergence. 
For any pdf $q(\boldsymbol{y})$, we use the factorization 
$p(\tilde{\boldsymbol{y}} | \boldsymbol{x}, 
\boldsymbol{a}) = p_{\boldsymbol{y} | \boldsymbol{x}}
(\boldsymbol{y} | \boldsymbol{x})
p(y_{N+1} | \boldsymbol{x}, \boldsymbol{a})$ 
to represent the conditional mutual information 
$I(\boldsymbol{x}; \tilde{\boldsymbol{y}} | \boldsymbol{a})$ as 
\begin{align} 
I(\boldsymbol{x}; \tilde{\boldsymbol{y}} | \boldsymbol{a}) 
&= \mathbb{E}\left[
 \log\frac{p_{\boldsymbol{y} | \boldsymbol{x}}(\boldsymbol{y} | \boldsymbol{x})}
 {q(\boldsymbol{y})}
\right]
+ \mathbb{E}\left[
 \log\frac{p(\boldsymbol{y})p(y_{N+1} | \boldsymbol{x}, \boldsymbol{a})}
 {p(\tilde{\boldsymbol{y}} | \boldsymbol{a})}
\right]
\nonumber \\
&- D(p \| q), \label{mutual_inf}
\end{align}
with $p(\boldsymbol{y})=\mathbb{E}_{\boldsymbol{x}}[
p_{\boldsymbol{y}|\boldsymbol{x}}(\boldsymbol{y} | \boldsymbol{x})]$ and 
$p(\tilde{\boldsymbol{y}} | \boldsymbol{a}) 
= \mathbb{E}_{\boldsymbol{x}}[p(\tilde{\boldsymbol{y}} | \boldsymbol{x}, 
\boldsymbol{a})]$.  
Combining (\ref{mutual_inf_tmp}), (\ref{mutual_inf_tmp_first_term}), 
(\ref{mutual_inf_tmp_second_term}), and (\ref{mutual_inf}), we arrive at 
\begin{align} 
&\frac{1}{2}\log\left\{
 1 + \frac{\alpha}{k}\mathbb{E}[\|\boldsymbol{x} 
 - \hat{\boldsymbol{x}}(\boldsymbol{y})\|_{2}^{2} 
\right\} \nonumber \\
&\geq \mathbb{E}\left[
 \log\frac{p(\boldsymbol{y})p(y_{N+1} | \boldsymbol{x}, \boldsymbol{a})}
 {p(\tilde{\boldsymbol{y}} | \boldsymbol{a})}
\right]
- J(q(\boldsymbol{y})), 
\label{mutual_inf_bound}
\end{align}
with $J(q(\boldsymbol{y}))$ in (\ref{J_function}). 

We prove the following lower bound on the first term in 
(\ref{mutual_inf_bound}). 
\begin{lemma} \label{lemma_lower_bound}
Suppose that $p_{\boldsymbol{y} | \boldsymbol{x}}$ is permutation-invariant, i.e.\ 
$p_{\boldsymbol{y}|\boldsymbol{x}}
(\boldsymbol{P}\boldsymbol{y} | \boldsymbol{P}\boldsymbol{x})
=p_{\boldsymbol{y} | \boldsymbol{x}}(\boldsymbol{y} | \boldsymbol{x})$ for any 
permutation matrix $\boldsymbol{P}$. Then, 
for all $\alpha\in(0, (2u_{\mathrm{max}}^{2})^{-1})$ we have 
\begin{align}
&\mathbb{E}\left[
 \log\frac{p(\boldsymbol{y})p(y_{N+1} | \boldsymbol{x}, \boldsymbol{a})}
 {p(\tilde{\boldsymbol{y}} | \boldsymbol{a})}
\right] \geq \frac{1}{2}\log\left(
 1 - u_{\mathrm{max}}^{4}\alpha^{2}
\right)
 \nonumber \\
&+ \frac{\alpha}{2k}\mathbb{E}[\|\boldsymbol{x}\|_{2}^{2}]
- \log\left\{
 1 + f(\alpha)\left[
  (1 - k/N)^{-k} - 1 
 \right]
\right\}, 
\end{align}
with $f(\alpha)\geq0$ given  in (\ref{function_f}). 
\end{lemma}
\begin{IEEEproof}
See Appendix~\ref{proof_lemma_lower_bound}. 
\end{IEEEproof}

We are ready to prove Lemma~\ref{lemma_square_error}. 
Since the AWGN channel~(\ref{AWGN}) implies the permutation invariance of 
$p_{\boldsymbol{y}|\boldsymbol{x}}$, we can apply Lemma~\ref{lemma_lower_bound} 
to the lower bound~(\ref{mutual_inf_bound}), 
\begin{align} 
&\frac{\alpha}{k}\mathbb{E}[\|\boldsymbol{x} 
- \hat{\boldsymbol{x}}(\boldsymbol{y})\|_{2}^{2} 
 \nonumber \\
&\geq \frac{(1 - u_{\mathrm{max}}^{4}\alpha^{2})e^{
\frac{\alpha}{k}\mathbb{E}[\|\boldsymbol{x}\|_{2}^{2}]} }
{\{1 + f(\alpha)[ (1 - k/N)^{-k} - 1 ]\}^{2}}
e^{-2J(q(\boldsymbol{y}))} - 1 
\end{align}
for all $\alpha\in(0, (2u_{\mathrm{max}}^{2})^{-1})$. 
Using $J(q(\boldsymbol{y})) = D(p(\boldsymbol{y}) \| q(\boldsymbol{y}))$ 
in (\ref{J_function}) for the Gaussian pdf $q(\boldsymbol{y})$,
we arrive at Lemma~\ref{lemma_square_error}. 

\section{Proof of Lemma~\ref{lemma_lower_bound}} 
\label{proof_lemma_lower_bound}
Let $\lambda_{N+1} = \mathbb{E}[y_{N+1}^{2} | \boldsymbol{a}]$ 
for $y_{N+1}$ in (\ref{MIMO}), with 
\begin{equation}
\lambda_{N+1} = 1 + \frac{\alpha}{k}
\boldsymbol{a}^{\mathrm{T}}\mathbb{E}\left[
 \boldsymbol{x}\boldsymbol{x}^{\mathrm{T}}
\right]\boldsymbol{a}, 
\end{equation}
where $\boldsymbol{x}$ is uniformly distributed on 
$\mathcal{X}_{k}^{N}(\mathcal{U})$. 
In particular, we use $\boldsymbol{a}\sim\mathcal{N}(\boldsymbol{0}, 
\boldsymbol{I}_{N})$ to have 
\begin{equation} \label{lambda_N_mean}
\mathbb{E}[\lambda_{N+1}] 
= 1 + \frac{\alpha}{k}\mathbb{E}[\|\boldsymbol{x}\|_{2}^{2}]. 
\end{equation}
We recall that $\boldsymbol{y}$ given 
$\boldsymbol{x}$ is represented with the conditional pdf $p_{\boldsymbol{y} 
| \boldsymbol{x}}$. 
Using $p(\tilde{\boldsymbol{y}} | \boldsymbol{a}) 
= \mathbb{E}_{\boldsymbol{x}}[p_{\boldsymbol{y} | \boldsymbol{x}}
(\boldsymbol{y} | \boldsymbol{x})
p(y_{N+1} | \boldsymbol{x}, \boldsymbol{a})]$ and the definition of 
$y_{N+1}$ in (\ref{MIMO}) yields 
\begin{align}
&\mathbb{E}\left[
 \log\frac{p(\boldsymbol{y})p(y_{N+1} | \boldsymbol{x}, \boldsymbol{a})}
 {p(\tilde{\boldsymbol{y}} | \boldsymbol{a})}
\right] \nonumber \\ 
&= \frac{\mathbb{E}[\lambda_{N+1}] - 1}{2} 
+ \mathbb{E}[\log p(\boldsymbol{y})]
- \mathbb{E}\left[
 \log\chi(\tilde{\boldsymbol{y}}, \boldsymbol{a})
\right] \nonumber \\
&= \frac{\alpha}{2k}\mathbb{E}[\|\boldsymbol{x}\|_{2}^{2}]  
+ \mathbb{E}[\log p(\boldsymbol{y})]
- \mathbb{E}\left[
 \log\chi(\tilde{\boldsymbol{y}}, \boldsymbol{a})
\right], \label{mutual_inf_last_term}
\end{align}
with 
\begin{equation}
\chi(\tilde{\boldsymbol{y}}, \boldsymbol{a})
= \sum_{\boldsymbol{x}'\in\mathcal{X}_{k}^{N}(\mathcal{U})}
\frac{p_{\boldsymbol{y} | \boldsymbol{x}}(\boldsymbol{y} | \boldsymbol{x}') 
e^{\sqrt{\alpha/k}\boldsymbol{a}^{\mathrm{T}}
\boldsymbol{x}'y_{N+1} 
- \frac{(\alpha/k)(\boldsymbol{a}^{\mathrm{T}}\boldsymbol{x}')^{2}}{2}}}
{|\mathcal{X}_{k}^{N}(\mathcal{U})|},  
\end{equation}
where the last equality in (\ref{mutual_inf_last_term}) follows from 
the expression of $\mathbb{E}[\lambda_{N+1}]$ in (\ref{lambda_N_mean}). 

We upper-bound the last term in (\ref{mutual_inf_last_term}). 
Substituting the definition of $y_{N+1}$ in (\ref{MIMO}) and 
using Jensen's inequality with respect 
to $\omega_{N+1}\sim\mathcal{N}(0, 1)$ and $\boldsymbol{a}$, we obtain 
$\mathbb{E}[ \log\chi(\tilde{\boldsymbol{y}}, \boldsymbol{a}) ]
\leq\mathbb{E}[\log\bar{\chi}(\boldsymbol{y}, \boldsymbol{x})]$ 
with 
\begin{equation}
\bar{\chi}(\boldsymbol{y}, \boldsymbol{x}) 
= \sum_{\boldsymbol{x}'\in\mathcal{X}_{k}^{N}(\mathcal{U})}
\frac{p_{\boldsymbol{y} | \boldsymbol{x}}(\boldsymbol{y} 
| \boldsymbol{x}') 
\mathbb{E}[e^{(\alpha/k)\boldsymbol{a}^{\mathrm{T}}\boldsymbol{x}'
\boldsymbol{a}^{\mathrm{T}}\boldsymbol{x}} | \boldsymbol{x}]}
{|\mathcal{X}_{k}^{N}(\mathcal{U})|}. 
\end{equation}
We repeat the derivation of (\ref{g_function_tmp0}):  
Using the representation 
of $\mathcal{X}_{k}^{N}(\mathcal{U})$ in (\ref{X_set_representation}) and 
decomposing 
$\boldsymbol{x}'\in \mathcal{T}_{w,\boldsymbol{m}}^{N}(\mathcal{S}_{\boldsymbol{x}})$ 
into $\boldsymbol{x}_{1}'\in \mathcal{T}_{k-w,\boldsymbol{m}_{1}}^{k,1}$ and 
$\boldsymbol{x}_{2}'\in \mathcal{T}_{w,\boldsymbol{m}_{2}}^{N-k,2}$ in 
(\ref{type1}) and (\ref{type2}), we have   
\begin{equation} \label{chi_function}
\bar{\chi}(\boldsymbol{y}, \boldsymbol{x}) 
= \frac{1}{|\mathcal{X}_{k}^{N}(\mathcal{U})|}
\sum_{w=0}^{k}\sum_{\boldsymbol{m}\in\mathcal{M}^{k}}
\bar{\chi}_{w,\boldsymbol{m}}(\boldsymbol{y}, \boldsymbol{x}),  
\end{equation}
where $\bar{\chi}_{w,\boldsymbol{m}}(\boldsymbol{y}, \boldsymbol{x})$ is given by 
\begin{equation}
\bar{\chi}_{w,\boldsymbol{m}}
= \sum_{\boldsymbol{x}_{1}'\in\mathcal{T}_{k-w,\boldsymbol{m}_{1}}^{k,1}}
\sum_{\boldsymbol{x}_{2}'\in\mathcal{T}_{w,\boldsymbol{m}_{2}}^{N-k,2}}
p_{\boldsymbol{y} | \boldsymbol{x}}(\boldsymbol{y} | 
\boldsymbol{x}')e_{w,\boldsymbol{m}}(\boldsymbol{x}_{1}', \boldsymbol{x}_{2}', 
\boldsymbol{x}) 
\end{equation}
for $w\in\{0,\ldots,k\}$, with 
$\boldsymbol{a}_{1}=\boldsymbol{a}_{\mathcal{S}_{\boldsymbol{x}}}$,
$\boldsymbol{a}_{2}=\boldsymbol{a}_{\mathcal{N}\setminus\mathcal{S}_{\boldsymbol{x}}}$,  
$\boldsymbol{x}_{1}=\boldsymbol{x}_{\mathcal{S}_{\boldsymbol{x}}}$, 
$\boldsymbol{x}_{1}'=\boldsymbol{x}_{\mathcal{S}_{\boldsymbol{x}}}'$, 
$\boldsymbol{x}_{2}' = \boldsymbol{x}_{\mathcal{N}\setminus\mathcal{S}_{\boldsymbol{x}}}'$, 
and 
\begin{equation} \label{e_function}
e_{w,\boldsymbol{m}}
= \mathbb{E}\left[
 \left.
  \exp\left\{
   \frac{\alpha}{k}(\boldsymbol{a}_{1}^{\mathrm{T}}\boldsymbol{x}_{1}'
   + \boldsymbol{a}_{2}^{\mathrm{T}}\boldsymbol{x}_{2}')
   \boldsymbol{a}_{1}^{\mathrm{T}}\boldsymbol{x}_{1}
  \right\} 
 \right| \boldsymbol{x}
\right]. 
\end{equation}
In particular, for $w=0$ we have 
$\boldsymbol{x}_{1}'=\boldsymbol{u}_{\boldsymbol{m}}$ and 
$\boldsymbol{x}_{2}'=\boldsymbol{0}$ while $\boldsymbol{x}_{1}'=\boldsymbol{0}$ 
holds for $w=k$.  

To evaluate the expectation over 
$\boldsymbol{a}\sim\mathcal{N}(\boldsymbol{0}, \boldsymbol{I}_{N})$, 
we prove the following proposition: 
\begin{proposition} \label{proposition_Gauss}
For all deterministic $v\in\mathbb{R}$ and 
$\boldsymbol{u}_{1}, \boldsymbol{u}_{2}\in\mathbb{R}^{k}$, 
\begin{align}
&\mathbb{E}\left[
 e^{\boldsymbol{a}_{1}^{\mathrm{T}}\boldsymbol{u}_{1}
 (v\boldsymbol{a}_{1}^{\mathrm{T}}\boldsymbol{u}_{1} 
 + \boldsymbol{a}_{1}^{\mathrm{T}}\boldsymbol{u}_{2})}
\right] \nonumber \\
&= \left\{
 (1 - \boldsymbol{u}_{2}^{\mathrm{T}}\boldsymbol{u}_{1})^{2} 
 - (2v + \|\boldsymbol{u}_{2}\|_{2}^{2})\|\boldsymbol{u}_{1}\|_{2}^{2}
\right\}^{-1/2}
\end{align}
holds if $(1 - \boldsymbol{u}_{2}^{\mathrm{T}}\boldsymbol{u}_{1})^{2} 
> (2v + \|\boldsymbol{u}_{2}\|_{2}^{2})
\|\boldsymbol{u}_{1}\|_{2}^{2}$ is satisfied. 
\end{proposition} 
\begin{IEEEproof}
Let $Z_{i} = \boldsymbol{a}_{1}^{\mathrm{T}}\boldsymbol{u}_{i}$ for $i\in\{1, 2\}$. 
The random variables $(Z_{1}, Z_{2})$ follow the zero-mean Gaussian distribution 
with covariance $\mathbb{E}[Z_{i}Z_{j}]=\boldsymbol{u}_{i}^{\mathrm{T}}
\boldsymbol{u}_{j}$ for $i, j\in\{1, 2\}$. Thus, we can represent $Z_{2}$ as 
\begin{equation}
Z_{2} = \frac{\boldsymbol{u}_{2}^{\mathrm{T}}\boldsymbol{u}_{1}}
{\|\boldsymbol{u}_{1}\|_{2}^{2}}Z_{1} + W, 
\end{equation}
with $W\sim\mathcal{N}(0, \|\boldsymbol{u}_{2}\|_{2}^{2} 
- (\boldsymbol{u}_{2}^{\mathrm{T}}\boldsymbol{u}_{1})^{2}
/\|\boldsymbol{u}_{1}\|_{2}^{2})$ independent of $Z_{1}$. 
Substituting this representation and evaluating the expectation over $W$, 
we have  
\begin{align}
&\mathbb{E}\left[
 e^{\boldsymbol{a}_{1}^{\mathrm{T}}\boldsymbol{u}_{1}
 (v\boldsymbol{a}_{1}^{\mathrm{T}}\boldsymbol{u}_{1} 
 + \boldsymbol{a}_{1}^{\mathrm{T}}\boldsymbol{u}_{2})}
\right] \nonumber \\
&= \mathbb{E}\left[
 \exp\left(
  \left\{
   2v + \frac{2\boldsymbol{u}_{2}^{\mathrm{T}}\boldsymbol{u}_{1}}
   {\|\boldsymbol{u}_{1}\|_{2}^{2}}
   + \|\boldsymbol{u}_{2}\|_{2}^{2} 
   - \frac{(\boldsymbol{u}_{2}^{\mathrm{T}}\boldsymbol{u}_{1})^{2}}
   {\|\boldsymbol{u}_{1}\|_{2}^{2}}
  \right\}\frac{Z_{1}^{2}}{2}
 \right)
\right] \nonumber \\
&= \left\{
 (1 - \boldsymbol{u}_{2}^{\mathrm{T}}\boldsymbol{u}_{1})^{2} 
 - (2v + \|\boldsymbol{u}_{2}\|_{2}^{2})\|\boldsymbol{u}_{1}\|_{2}^{2}
\right\}^{-1/2}, 
\end{align}
where the last equality follows from $Z_{1}\sim\mathcal{N}(0, 
\|\boldsymbol{u}_{1}\|_{2}^{2})$ and the assumption in 
Proposition~\ref{proposition_Gauss}. 
\end{IEEEproof}

For $w=0$, we have $\boldsymbol{x}_{1}'=\boldsymbol{u}_{\boldsymbol{m}}$ and 
$\boldsymbol{x}_{2}'=\boldsymbol{0}$ in $e_{0,\boldsymbol{m}}$ given by 
(\ref{e_function}). 
we use Proposition~\ref{proposition_Gauss} with 
$\boldsymbol{u}_{1}=\sqrt{\alpha/k}\boldsymbol{u}_{\boldsymbol{m}}$, 
$\boldsymbol{u}_{2}=\sqrt{\alpha/k}\boldsymbol{x}_{1}$, and $v=0$ to have 
\begin{align}
e_{0,\boldsymbol{m}} 
&= \left\{
 [1 - (\alpha/k)\boldsymbol{u}_{\boldsymbol{m}}^{\mathrm{T}}\boldsymbol{x}_{1}]^{2}
 - (\alpha/k)^{2}\|\boldsymbol{u}_{\boldsymbol{m}}\|_{2}^{2}
 \|\boldsymbol{x}_{1}\|_{2}^{2}
\right\}^{-1/2} \nonumber \\
&\leq \left(
 1 - 2u_{\mathrm{max}}^{2}\alpha 
\right)^{-1/2}
\end{align}
for all $\alpha\in(0, (2u_{\mathrm{max}}^{2})^{-1})$, where the last inequality 
follows from the upper bounds $\boldsymbol{u}_{\boldsymbol{m}}^{\mathrm{T}}
\boldsymbol{x}_{1}\leq u_{\mathrm{max}}^{2}k$,  
$\|\boldsymbol{u}_{\boldsymbol{m}}\|_{2}^{2}\leq u_{\mathrm{max}}^{2}k$, and 
$\|\boldsymbol{x}_{1}\|_{2}^{2}\leq u_{\mathrm{max}}^{2}k$ for all 
$\boldsymbol{x}\in\mathcal{X}_{k}^{N}(\mathcal{U})$. 

For $w=k$, we have $\boldsymbol{x}_{1}'=\boldsymbol{0}$ in $e_{k,\boldsymbol{m}}$.  
Evaluating the expectation over $\boldsymbol{a}_{2}$ and subsequently 
over $\boldsymbol{a}_{1}$ yields 
\begin{align}
&e_{k,\boldsymbol{m}}
= \mathbb{E}\left[
 \left. 
  e^{\frac{\alpha^{2}}{2k^{2}}\|\boldsymbol{u}_{\boldsymbol{m}}\|_{2}^{2}
  (\boldsymbol{a}_{1}^{\mathrm{T}}\boldsymbol{x}_{1})^{2}}
 \right| \boldsymbol{x}
\right] \nonumber \\
&= \left(
 1 - \|\boldsymbol{u}_{\boldsymbol{m}}\|_{2}^{2}\|\boldsymbol{x}_{1}\|_{2}^{2}
 \alpha^{2}/k^{2}
\right)^{-1/2} \leq \left(
 1 - u_{\mathrm{max}}^{4}\alpha^{2}
\right)^{-1/2}
\end{align}
for all $\boldsymbol{x}_{2}'\in\mathcal{T}_{k,\boldsymbol{m}}^{N-k,2}$ and 
$\alpha\in(0, u_{\mathrm{max}}^{-2})$. 

For $w\in\{1,\ldots,k-1\}$ we evaluate the expectation 
in (\ref{e_function}) over $\boldsymbol{a}_{2}$ as 
\begin{equation} 
e_{w,\boldsymbol{m}}
= \mathbb{E}\left[
 \left.
  \exp\left(
   \frac{\alpha\boldsymbol{a}_{1}^{\mathrm{T}}\boldsymbol{x}_{1}'
   \boldsymbol{a}_{1}^{\mathrm{T}}\boldsymbol{x}_{1}}{k} 
   + \frac{\alpha^{2}\|\boldsymbol{u}_{\boldsymbol{m}_{2}}\|_{2}^{2}
   (\boldsymbol{a}_{1}^{\mathrm{T}}\boldsymbol{x}_{1})^{2}}{2k^{2}}
  \right)
 \right| \boldsymbol{x}
\right]. 
\end{equation}
Using Proposition~\ref{proposition_Gauss} with 
$\boldsymbol{u}_{1}=(\alpha/k)\boldsymbol{x}_{1}$, 
$\boldsymbol{u}_{2}=\boldsymbol{x}_{1}'$, 
and $v=\|\boldsymbol{u}_{\boldsymbol{m}_{2}}\|_{2}^{2}/2$ yields 
\begin{align}
e_{w,\boldsymbol{m}}
&= \left\{
 [1 - (\alpha/k)\boldsymbol{x}_{1}'^{\mathrm{T}}\boldsymbol{x}_{1}]^{2}
 - (\alpha^{2}/k^{2})\|\boldsymbol{u}_{\boldsymbol{m}}\|_{2}^{2}
 \|\boldsymbol{x}_{1}\|_{2}^{2}
\right\}^{-1/2} \nonumber \\
&\leq \left(
 1 - 2u_{\mathrm{max}}^{2}\alpha
\right)^{-1/2}
\end{align}
for all $\boldsymbol{x}_{1}'\in\mathcal{T}_{k-w,\boldsymbol{m}_{1}}^{k,1}$ and 
$\alpha\in(0, (2u_{\mathrm{max}}^{2})^{-1})$, where the last inequality 
follows from the upper bounds 
$\boldsymbol{x}_{1}'^{\mathrm{T}}\boldsymbol{x}_{1}\leq u_{\mathrm{max}}^{2}k$,  
$\|\boldsymbol{u}_{\boldsymbol{m}}\|_{2}^{2}\leq u_{\mathrm{max}}^{2}k$, and 
$\|\boldsymbol{x}_{1}\|_{2}^{2}\leq u_{\mathrm{max}}^{2}k$ for all 
$\boldsymbol{x}\in\mathcal{X}_{k}^{N}(\mathcal{U})$ and 
$\boldsymbol{x}_{1}'\in\mathcal{T}_{k-w,\boldsymbol{m}_{1}}^{k,1}$. 
Applying these results to (\ref{chi_function}), we have 
\begin{align}
\bar{\chi}(\boldsymbol{y}, \boldsymbol{x}) 
\leq& (1 - 2u_{\mathrm{max}}^{2}\alpha)^{-1/2}
\left\{
 p(\boldsymbol{y}) - \Delta_{\mathcal{S}_{\boldsymbol{x}}} p(\boldsymbol{y})
\right\} \nonumber \\
{}&+ (1 - u_{\mathrm{max}}^{4}\alpha^{2})^{-1/2}
\Delta_{\mathcal{S}_{\boldsymbol{x}}} p(\boldsymbol{y}), 
\label{chi_function_bound}
\end{align}
with 
\begin{equation} \label{Delta_p}
\Delta_{\mathcal{S}_{\boldsymbol{x}}} p(\boldsymbol{y})
= \frac{1}{|\mathcal{X}_{k}^{N}(\mathcal{U})|}
\sum_{\boldsymbol{m}\in\mathcal{M}^{k}}
\sum_{\boldsymbol{x}_{2}'\in\mathcal{T}_{k,\boldsymbol{m}}^{N-k,2}}
p_{\boldsymbol{y} | \boldsymbol{x}}(\boldsymbol{y} | \boldsymbol{x}'), 
\end{equation}
with $\boldsymbol{x}_{\mathcal{S}_{\boldsymbol{x}}}'=\boldsymbol{0}$ and 
$\boldsymbol{x}_{2}' = \boldsymbol{x}_{\mathcal{N}\setminus\mathcal{S}_{\boldsymbol{x}}}'$. 
The quantity $\Delta_{\mathcal{S}_{\boldsymbol{x}}} p(\boldsymbol{y})$ 
depends on $\mathcal{S}_{\boldsymbol{x}}$ only through the 
partition of $\boldsymbol{x}'$.
 
We evaluate the quantity~(\ref{mutual_inf_last_term}). 
Applying (\ref{chi_function_bound}) and the inequality 
$\mathbb{E}[\log\chi(\tilde{\boldsymbol{y}}, \boldsymbol{a})]
\leq\mathbb{E}[\log\bar{\chi}(\boldsymbol{y}, \boldsymbol{x})]$   
to (\ref{mutual_inf_last_term}), we obtain 
\begin{align}
&\mathbb{E}\left[
 \log\frac{p(\boldsymbol{y})p(y_{N+1} | \boldsymbol{x}, \boldsymbol{a})}
 {p(\tilde{\boldsymbol{y}} | \boldsymbol{a})}
\right] \geq \frac{1}{2}\log\left(
 1 - u_{\mathrm{max}}^{4}\alpha^{2}
\right)
 \nonumber \\
&- \mathbb{E}\left[
 \log\left\{
  1 + f(\alpha)\left[
   1 - \frac{\Delta_{\mathcal{S}_{\boldsymbol{x}}}p(\boldsymbol{y})}
   {p(\boldsymbol{y})}
  \right]
 \right\}
\right]
+ \frac{\alpha}{2k}\mathbb{E}[\|\boldsymbol{x}\|_{2}^{2}]
  \label{mutual_inf_last_term_bound}
\end{align}
for all $\alpha\in(0, (2u_{\mathrm{max}}^{2})^{-1})$, with 
$f(\alpha)\geq0$ given in (\ref{function_f}).  
Using Jensen's inequality for the second term 
in (\ref{mutual_inf_last_term_bound}) yields 
\begin{align}
&\mathbb{E}\left[
 \log\left\{
  1 + f(\alpha)\left[
   1 - \frac{\Delta_{\mathcal{S}_{\boldsymbol{x}}}p(\boldsymbol{y})}
   {p(\boldsymbol{y})}
  \right]
 \right\}
\right]  \nonumber \\
&\leq \log\left(
 1 + f(\alpha)\left\{
  1 - \mathbb{E}\left[
   \frac{\Delta_{\mathcal{S}_{\boldsymbol{x}}} p(\boldsymbol{y})}{p(\boldsymbol{y})}
  \right]
 \right\}
\right). \label{mutual_inf_last_term_bound_second_term} 
\end{align}

To evaluate the expectation in 
(\ref{mutual_inf_last_term_bound_second_term}), we utilize the 
permutation-invariance of $p(\boldsymbol{y} | 
\boldsymbol{S}_{\boldsymbol{x}})$---written as 
$p_{\boldsymbol{y} | \mathcal{S}_{\boldsymbol{x}}}(\boldsymbol{y} | 
\boldsymbol{S}_{\boldsymbol{x}})$ for clarification, given by 
\begin{equation} \label{p_yS}
p_{\boldsymbol{y} | \mathcal{S}_{\boldsymbol{x}}}(\boldsymbol{y} | \mathcal{S})
= \frac{1}{|\mathcal{M}|^{k}}
\sum_{\boldsymbol{x}'\in\mathcal{X}_{k}^{N}(\mathcal{S}, \mathcal{U})}
p_{\boldsymbol{y} | \boldsymbol{x}}(\boldsymbol{y} | \boldsymbol{x}'),
\end{equation}
with $\mathcal{X}_{k}^{N}(\mathcal{S}, \mathcal{U})
=\{\boldsymbol{x}'\in\mathcal{X}_{k}^{N}(\mathcal{U}): 
\mathcal{S}_{\boldsymbol{x}'}=\mathcal{S}\}$ 
for $\mathcal{S}\in\mathfrak{S}_{k}^{N}$. 
By definition, we have $|\mathcal{X}_{k}^{N}(\mathcal{S}, 
\mathcal{U})| = |\mathcal{M}|^{k}$. 

The permutation invariance of $p_{\boldsymbol{y} | \mathcal{S}_{\boldsymbol{x}}}$ is 
defined via any $N\times N$ permutation matrix $\boldsymbol{P}$, which 
is a bijection from  
$\mathcal{X}_{k}^{N}(\{1\})$ onto $\mathcal{X}_{k}^{N}(\{1\})$. We define  
the one-to-one mapping $\pi_{\boldsymbol{P}}:\mathfrak{S}_{k}^{N}
\to\mathfrak{S}_{k}^{N}$ induced from $\boldsymbol{P}$ as 
\begin{equation}
\boldsymbol{s}'=\boldsymbol{P}\boldsymbol{s} 
\iff \mathcal{S}_{\boldsymbol{s}'} 
= \pi_{\boldsymbol{P}}(\mathcal{S}_{\boldsymbol{s}})
\end{equation}
for all $\boldsymbol{s}, \boldsymbol{s}'\in\mathcal{X}_{k}^{N}(\{1\})$. 
Then, for all $\boldsymbol{y}\in\mathbb{R}^{N}$ and 
$\mathcal{S}\in\mathfrak{S}_{k}^{N}$ 
we have the following permutation invariance: 
\begin{align}
&p_{\boldsymbol{y}|\mathcal{S}_{\boldsymbol{x}}}
(\boldsymbol{P}\boldsymbol{y} | \pi_{\boldsymbol{P}}(\mathcal{S}))
= \frac{1}{|\mathcal{M}|^{k}}
\sum_{\boldsymbol{x}''\in\mathcal{X}_{k}^{N}(\pi_{\boldsymbol{P}}(\mathcal{S}), \mathcal{U})}
p_{\boldsymbol{y} | \boldsymbol{x}}(\boldsymbol{P}\boldsymbol{y} | \boldsymbol{x}'') 
\nonumber \\
&= \frac{1}{|\mathcal{M}|^{k}}
\sum_{\boldsymbol{x}'\in\mathcal{X}_{k}^{N}(\mathcal{S}, \mathcal{U})}
p_{\boldsymbol{y} | \boldsymbol{x}}(\boldsymbol{P}\boldsymbol{y} | 
\boldsymbol{P}\boldsymbol{x}')
=  p_{\boldsymbol{y}|\mathcal{S}_{\boldsymbol{x}}}
(\boldsymbol{y} | \mathcal{S}). 
\label{permutation_invariance_ys}
\end{align} 
In the derivation of the second equality, we have used the change of 
variables $\boldsymbol{x}'=\boldsymbol{P}^{\mathrm{T}}\boldsymbol{x}''$ and 
$\boldsymbol{x}'\in\mathcal{X}_{k}^{N}(\mathcal{S}, \mathcal{U})$, of which 
the latter is obtained from the definition of $\pi_{\boldsymbol{P}}$.  
The last equality follows from the permutation-invariance assumption of 
$p_{\boldsymbol{y}|\boldsymbol{x}}$ in Lemma~\ref{lemma_lower_bound}. 

Similarly, we confirm the permutation invariance of 
$p(\boldsymbol{y})$---written as $p_{\boldsymbol{y}}(\boldsymbol{y})$. 
Since $\mathcal{S}_{\boldsymbol{x}}$ is uniformly distributed on 
$\mathfrak{S}_{k}^{N}$, we have the representation 
$p_{\boldsymbol{y}}(\boldsymbol{y})=|\mathfrak{S}_{k}^{N}|^{-1}
\sum_{\mathcal{S}\in\mathfrak{S}_{k}^{N}}p_{\boldsymbol{y}|\mathcal{S}_{\boldsymbol{x}}}
(\boldsymbol{y} | \mathcal{S})$. Repeating the 
proof~(\ref{permutation_invariance_ys}) with the permutation invariance 
of $p_{\boldsymbol{y} | \mathcal{S}_{\boldsymbol{x}}}$, we find 
the permutation invariance of $p_{\boldsymbol{y}}(\boldsymbol{y})$. 

We are ready to evaluate the expectation in 
(\ref{mutual_inf_last_term_bound_second_term}). 
Let $\mathcal{S}'$ denote the support of $\boldsymbol{x}_{2}'$ in 
(\ref{Delta_p}). By definition, $\mathcal{S}'$ is in  
$\{\mathcal{S}'\in\mathfrak{S}_{k}^{N}: \mathcal{S}\cap\mathcal{S}'
=\emptyset\}$ for given $\mathcal{S}_{\boldsymbol{x}}=\mathcal{S}$. 
Since $\mathcal{S}_{\boldsymbol{x}}$ is uniformly distributed on 
$\mathfrak{S}_{k}^{N}$, we use the definition of 
$\Delta_{\mathcal{S}_{\boldsymbol{x}}} p(\boldsymbol{y})$ in (\ref{Delta_p}) to have 
\begin{align} 
&\mathbb{E}\left[
 \frac{\Delta_{\mathcal{S}_{\boldsymbol{x}}} p(\boldsymbol{y})}
 {p_{\boldsymbol{y}}(\boldsymbol{y})}
\right]
= \frac{1}{|\mathfrak{S}_{k}^{N}|}\sum_{\mathcal{S}\in\mathfrak{S}_{k}^{N}}
\int\frac{\Delta_{\mathcal{S}_{\boldsymbol{x}}} p(\boldsymbol{y})}
{p_{\boldsymbol{y}}(\boldsymbol{y})}
p_{\boldsymbol{y} | \mathcal{S}_{\boldsymbol{x}}}(\boldsymbol{y} | \mathcal{S})
d\boldsymbol{y}
\nonumber \\
&= \sum_{\mathcal{S}\in\mathfrak{S}_{k}^{N}}
\sum_{\mathcal{S}'\in\mathfrak{S}_{k}^{N}: \mathcal{S}\cap\mathcal{S}'=\emptyset}
\frac{f(\mathcal{S}, \mathcal{S}')}
{|\mathfrak{S}_{k}^{N}||\mathcal{X}_{k}^{N}(\mathcal{U})|}, 
\label{mutual_inf_last_term_bound_second_term_tmp}
\end{align}
where $f(\mathcal{S}, \mathcal{S}')$ is given by 
\begin{equation} \label{f_SS}
f(\mathcal{S}, \mathcal{S}') 
= \int
\frac{p_{\boldsymbol{y} | \mathcal{S}_{\boldsymbol{x}}}(\boldsymbol{y} | \mathcal{S})}
{p_{\boldsymbol{y}}(\boldsymbol{y})}
\sum_{\boldsymbol{m}\in\mathcal{M}^{k}}
p_{\boldsymbol{y} | \boldsymbol{x}}(\boldsymbol{y} | \boldsymbol{x}')d\boldsymbol{y}, 
\end{equation}
with $\boldsymbol{x}_{\mathcal{S}'}'=\boldsymbol{u}_{\boldsymbol{m}}$ and 
$\boldsymbol{x}_{\mathcal{N}\setminus\mathcal{S}'}'=\boldsymbol{0}$. 

We confirm that the function~(\ref{f_SS}) does not depend on $\mathcal{S}'$. 
Let $\bar{\mathcal{S}}'\in\mathfrak{S}_{k}^{N}$ with 
$\mathcal{S}\cap\bar{\mathcal{S}}'
=\emptyset$ denote a reference support. We write the set of the 
corresponding reference vectors as 
$\bar{\mathcal{X}}'(\bar{\mathcal{S}}')
=\{\bar{\boldsymbol{x}}'\in\mathcal{X}_{k}^{N}(\mathcal{U}): 
\bar{\boldsymbol{x}}_{\bar{\mathcal{S}}'}'\in\mathcal{M}^{k}\}$. 
For any $\mathcal{S}'\in\mathfrak{S}_{k}^{N}$ with 
$\mathcal{S}\cap\mathcal{S}'=\emptyset$, there is some permutation matrix 
$\boldsymbol{P}$ such that $\boldsymbol{P}$ is a bijection from 
$\bar{\mathcal{X}}'(\mathcal{S}')$ onto 
$\bar{\mathcal{X}}'(\bar{\mathcal{S}}')$. The corresponding mapping 
$\pi_{\boldsymbol{P}}$ satisfies $\pi_{\boldsymbol{P}}(\mathcal{S}')
=\bar{\mathcal{S}}'$. When $\mathcal{S}$ has no 
intersection with $\mathcal{S}'$, we find 
the identity 
$\pi_{\boldsymbol{P}}(\mathcal{S})=\mathcal{S}$. 
Using the permutation invariance of the pdfs 
$p_{\boldsymbol{y} | \mathcal{S}_{\boldsymbol{x}}}$, 
$p_{\boldsymbol{y}}$, and $p_{\boldsymbol{y} | \boldsymbol{x}}$, as well as 
$\pi_{\boldsymbol{P}}(\mathcal{S})=\mathcal{S}$, we evaluate (\ref{f_SS}) as  
\begin{align}
&f(\mathcal{S}, \mathcal{S}') 
= \int\frac{p_{\boldsymbol{y} | \mathcal{S}_{\boldsymbol{x}}}
(\boldsymbol{P}\boldsymbol{y} | \mathcal{S})}
{p_{\boldsymbol{y}}(\boldsymbol{P}\boldsymbol{y})}\sum_{\boldsymbol{m}\in\mathcal{M}^{k}}
p_{\boldsymbol{y} | \boldsymbol{x}}(\boldsymbol{P}\boldsymbol{y} 
| \boldsymbol{P}\boldsymbol{x}')d\boldsymbol{y} 
\nonumber \\
&= \int\frac{p_{\boldsymbol{y} | \mathcal{S}_{\boldsymbol{x}}}
(\boldsymbol{y} | \mathcal{S})}
{p_{\boldsymbol{y}}(\boldsymbol{y})}\sum_{\boldsymbol{m}\in\mathcal{M}^{k}}
p_{\boldsymbol{y} | \boldsymbol{x}}(\boldsymbol{y} 
| \bar{\boldsymbol{x}}')d\boldsymbol{y} 
= f(S, \bar{S}'),  \label{f_SS_identity}
\end{align}
with $\bar{\boldsymbol{x}}_{\bar{\mathcal{S}'}}'=\boldsymbol{u}_{\boldsymbol{m}}$ 
and $\bar{\boldsymbol{x}}_{\mathcal{N}\setminus\bar{\mathcal{S}'}}'=\boldsymbol{0}$, 
where the second equality follows from the change of variables in the 
integral and the definition of the permutation matrix $\boldsymbol{P}$. 

We evaluate the 
quantity~(\ref{mutual_inf_last_term_bound_second_term_tmp}). Taking  
the average of $f(\mathcal{S}, \bar{\mathcal{S}}')$ in (\ref{f_SS}) 
over $\mathcal{S}\in\mathfrak{S}_{k}^{N}$, we use the definition of 
$p_{\boldsymbol{y}}(\boldsymbol{y})$ to have 
\begin{equation}
\frac{1}{|\mathfrak{S}_{k}^{N}|}
\sum_{\mathcal{S}\in\mathfrak{S}_{k}^{N}}f(\mathcal{S}, \bar{\mathcal{S}}') 
= \sum_{\boldsymbol{m}\in\mathcal{M}^{k}}\int
p_{\boldsymbol{y} | \boldsymbol{x}}(\boldsymbol{y} | \boldsymbol{x}')d\boldsymbol{y}
= |\mathcal{M}|^{k}. 
\end{equation}
From this identity and (\ref{f_SS_identity}), we evaluate 
(\ref{mutual_inf_last_term_bound_second_term_tmp}) as 
\begin{align}
\mathbb{E}\left[
 \frac{\Delta_{\mathcal{S}_{\boldsymbol{x}}} p(\boldsymbol{y})}
 {p_{\boldsymbol{y}}(\boldsymbol{y})}
\right]
&= \binom{N-k}{k}\sum_{\mathcal{S}\in\mathfrak{S}_{k}^{N}}
\frac{f(\mathcal{S}, \bar{\mathcal{S}}')}
{|\mathfrak{S}_{k}^{N}||\mathcal{X}_{k}^{N}(\mathcal{U})|} 
\nonumber \\
&=  \binom{N-k}{k}\binom{N}{k}^{-1},  
\end{align}
because of $|\mathcal{X}_{k}^{N}(\mathcal{U})|=|\mathcal{M}|^{k}\binom{N}{k}$. 
Using $\sum_{w=0}^{k}\binom{k}{w}\binom{N-k}{w}=\binom{N}{k}$, 
we arrive at 
\begin{align}
&1 - \mathbb{E}\left[
 \frac{\Delta_{\mathcal{S}_{\boldsymbol{x}}} p(\boldsymbol{y})}
 {p_{\boldsymbol{y}}(\boldsymbol{y})}
\right]
= \binom{N}{k}^{-1}\sum_{w=0}^{k-1}\binom{k}{w}\binom{N-k}{w}
 \nonumber \\
&\leq \{1 + (N/k - 1)^{-1}\}^{k} - 1
= (1 - k/N)^{-k} - 1, \label{mutual_inf_last_term_bound_second_term_end}
\end{align}
where the inequality follows from the upper bound~(\ref{binomial_bound}) 
and the binomial theorem. 

We are ready to prove Lemma~\ref{lemma_lower_bound}.  
Applying (\ref{mutual_inf_last_term_bound_second_term}) and 
(\ref{mutual_inf_last_term_bound_second_term_end}) to 
(\ref{mutual_inf_last_term_bound}), we obtain 
\begin{align}
&\mathbb{E}\left[
 \log\frac{p(\boldsymbol{y})p(y_{N+1} | \boldsymbol{x}, \boldsymbol{a})}
 {p(\tilde{\boldsymbol{y}} | \boldsymbol{a})}
\right] \geq \frac{1}{2}\log\left(
 1 - u_{\mathrm{max}}^{4}\alpha^{2}
\right)
 \nonumber \\
&+ \frac{\alpha}{2k}\mathbb{E}[\|\boldsymbol{x}\|_{2}^{2}]
- \log\left\{
 1 + f(\alpha)\left[
  (1 - k/N)^{-k} - 1 
 \right]
\right\} \label{mutual_inf_last_term_final}
\end{align}
for all $\alpha\in(0, (2u_{\mathrm{max}}^{2})^{-1})$.

\section*{Acknowledgment}
The author thanks the anonymous reviewers for their suggestions that have 
improved the quality of the manuscript greatly. 

\begin{table}[t]
\centering
\caption{List of notation}
\label{table1}
\begin{tabular}{|c|c|}
\hline
Notation & Definition \\
\hline
$\gamma\in[0, 1)$ & $\log k/\log N\to\gamma$ \\
\hline 
$\mathcal{N}$ & $\{1,\ldots,N\}$ \\
\hline 
$\mathcal{M}$ & $\{1,\ldots,M\}$ \\
\hline
$\mathfrak{S}_{k}^{N}$ & $\{\mathcal{S}\subset\mathcal{N}: |\mathcal{S}| = k\}$ 
\\ 
\hline
$\mathcal{S}_{\boldsymbol{x}}$ & 
$\{n\in\mathcal{N}: x_{n}\neq0\}$ \\
\hline
$\mathcal{U}$ &  $\{u_{m}\in\mathbb{R}\setminus\{0\}: 
m\in\mathcal{M}\}$ \\
\hline
$\mathcal{X}_{k}^{N}(\mathcal{U})$ & 
$\{\boldsymbol{x}\in\mathbb{R}^{N}: 
|\mathcal{S}_{\boldsymbol{x}}|=k, x_{n}\in\mathcal{U}~\hbox{for all 
$n\notin\mathcal{S}_{\boldsymbol{x}}$}\}$ \\
\hline
$\mathcal{X}_{k}^{N}(\mathcal{S}, \mathcal{U})$ &
$\{\boldsymbol{x}'\in\mathcal{X}_{k}^{N}(\mathcal{U}): 
\mathcal{S}_{\boldsymbol{x}'}=\mathcal{S}\}$ 
for $\mathcal{S}\in\mathfrak{S}_{k}^{N}$ \\
\hline

$u_{\mathrm{min}}$ & $\min_{m\in\mathcal{M}}|u_{m}|$ \\
\hline
$u_{\mathrm{max}}$ & $\max_{m\in\mathcal{M}}|u_{m}|$ \\
\hline
$d_{\mathrm{max}}$ & $\max_{m,m'\in\mathcal{M}: m\neq m'}|u_{m} - u_{m'}|$ \\
\hline 
$\sigma_{N/k}^{2}$ &  $\sigma^{2}/\log(N/k)$ \\
\hline 
$N_{\boldsymbol{x}, \hat{\boldsymbol{x}}_{\mathrm{ML}}}$ & 
$|\mathcal{S}_{\boldsymbol{x}}\cap\mathcal{S}_{\hat{\boldsymbol{x}}_{\mathrm{ML}}}|
- |\{n\in\mathcal{S}_{\boldsymbol{x}}
\cap\mathcal{S}_{\hat{\boldsymbol{x}}_{\mathrm{ML}}}: x_{n}=\hat{x}_{\mathrm{ML}, n}\}|$ \\
\hline 
$\boldsymbol{u}_{\boldsymbol{m}}\in\mathcal{U}^{k}$ & 
$[u_{m_{1}},\ldots, u_{m_{k}}]^{\mathrm{T}}$ \\
\hline
$\mathcal{T}_{w,\boldsymbol{m}}^{N}(\mathcal{S})$ & $\{
 \boldsymbol{x}'\in\mathcal{X}_{|\mathcal{S}|}^{N}(\mathcal{U}): 
 |\mathcal{S}_{\boldsymbol{x}'}\setminus\mathcal{S}| = w,\; 
 \boldsymbol{x}_{\mathcal{S}_{\boldsymbol{x}'}}'
 = \boldsymbol{u}_{\boldsymbol{m}}  
\}$ \\ 
\hline
$\mathcal{T}_{k-w,\boldsymbol{m}_{1}}^{k,1}$ & 
$\{
 \boldsymbol{x}'\in\mathcal{X}_{k-w}^{k}(\mathcal{U}): 
 \boldsymbol{x}_{\mathcal{S}_{\boldsymbol{x}'}}' 
 = \boldsymbol{u}_{\boldsymbol{m}_{1}}
\}$ for $\boldsymbol{m}_{1}\in\mathcal{M}^{k-w}$ \\
\hline 
$\mathcal{T}_{w,\boldsymbol{m}_{2}}^{N-k,2}$ & $\{
 \boldsymbol{x}'\in\mathcal{X}_{w}^{N-k}(\mathcal{U}): 
 \boldsymbol{x}_{\mathcal{S}_{\boldsymbol{x}'}}' 
 = \boldsymbol{u}_{\boldsymbol{m}_{2}}
\}$ for $\boldsymbol{m}_{2}\in\mathcal{M}^{w}$  \\
\hline
$\boldsymbol{x}_{1}$ & $\boldsymbol{x}_{\mathcal{S}_{\boldsymbol{x}}}$ \\
\hline
$\boldsymbol{\omega}_{1}$ & 
$\boldsymbol{\omega}_{\mathcal{S}_{\boldsymbol{x}}}$ \\
\hline
$\boldsymbol{\omega}_{2}$ & $\boldsymbol{\omega}_{\mathcal{N}
\setminus\mathcal{S}_{\boldsymbol{x}}}$ \\
\hline
$\Omega_{i}^{w}$ & $(\boldsymbol{x}_{i}^{w})^{\mathrm{T}}
\boldsymbol{\omega}_{1}/\sigma_{N/k}^{2}$ for $\boldsymbol{x}_{i}^{w}\in
\mathcal{X}_{k-w}^{k}(\mathcal{U})$ \\
\hline
$d_{i}^{w}$ & $u_{\mathrm{max}}^{-1}\|\boldsymbol{x}_{i}^{w}\|_{1}
d_{*}$ with $d_{*}$ in (\ref{d_min}) \\
\hline
$\mathcal{E}_{i}^{w}$ & $\{\Omega_{i}^{w}: 
|\Omega_{i}^{w}|\leq d_{i}^{w}\}$ \\
\hline
$\mathcal{E}_{\mathrm{b}}$ & $\cap_{i=1}^{k}\mathcal{E}_{i}^{k-1}$ \\
\hline
\end{tabular}
\end{table}

\bibliographystyle{IEEEtran}
\bibliography{IEEEabrv,kt-it2025_2}

\begin{thebibliography}{10}
\providecommand{\url}[1]{#1}
\csname url@samestyle\endcsname
\providecommand{\newblock}{\relax}
\providecommand{\bibinfo}[2]{#2}
\providecommand{\BIBentrySTDinterwordspacing}{\spaceskip=0pt\relax}
\providecommand{\BIBentryALTinterwordstretchfactor}{4}
\providecommand{\BIBentryALTinterwordspacing}{\spaceskip=\fontdimen2\font plus
\BIBentryALTinterwordstretchfactor\fontdimen3\font minus
  \fontdimen4\font\relax}
\providecommand{\BIBforeignlanguage}[2]{{%
\expandafter\ifx\csname l@#1\endcsname\relax
\typeout{** WARNING: IEEEtran.bst: No hyphenation pattern has been}%
\typeout{** loaded for the language `#1'. Using the pattern for}%
\typeout{** the default language instead.}%
\else
\language=\csname l@#1\endcsname
\fi
#2}}
\providecommand{\BIBdecl}{\relax}
\BIBdecl

\bibitem{Takeuchi251}
K.~Takeuchi, ``Generalized approximate message-passing for compressed sensing
  with sublinear sparsity,'' \emph{{IEEE} Trans. Inf. Theory}, vol.~71, no.~6,
  pp. 4602--4636, Jun. 2025.

\bibitem{Takeuchi261}
------, ``Orthogonal approximate message-passing for sublinear sparsity,'' in
  \emph{Proc. IEEE Int. Conf. Acoust. Speech Signal Process.}, Barcelona,
  Spain, May 2026, pp. 22\,552--22\,556.

\bibitem{Donoho92}
D.~L. Donoho, I.~M. Johnstone, J.~C. Hoch, and A.~S. Stern, ``Maximum entropy
  and the nearly black object,'' \emph{J. R. Stat. Soc. B}, vol.~54, no.~1, pp.
  41--81, 1992.

\bibitem{Johnstone04}
I.~M. Johnstone and B.~W. Silverman, ``Needles and straw in haystacks:
  Empirical {Bayes} estimates of possibly sparse sequences,'' \emph{Ann.
  Statist.}, vol.~32, no.~4, pp. 1594--1649, Aug. 2004.

\bibitem{Pas14}
S.~L. {van der Pas}, B.~J.~K. Kleijn, and A.~W. {van der Vaart}, ``The
  horseshoe estimator: Posterior concentration around nearly black vectors,''
  \emph{Electron. J. Statist.}, vol.~8, no.~2, pp. 2585--2618, Dec. 2014.

\bibitem{Rockova18}
V.~Ro\v{c}kov\'{a}, ``Bayesian estimation of sparse signals with a continuous
  spike-and-slab prior,'' \emph{Ann. Statist.}, vol.~46, no.~1, pp. 401--437,
  Feb. 2018.

\bibitem{Wainwright09}
M.~J. Wainwright, ``Information-theoretic limits on sparsity recovery in the
  high-dimensional and noisy setting,'' \emph{{IEEE} Trans. Inf. Theory},
  vol.~55, no.~12, pp. 5728--5741, Dec. 2009.

\bibitem{Fletcher09}
A.~K. Fletcher, S.~Rangan, and V.~K. Goyal, ``Necessary and sufficient
  conditions for sparsity pattern recovery,'' \emph{{IEEE} Trans. Inf. Theory},
  vol.~55, no.~12, pp. 5758--5772, Dec. 2009.

\bibitem{Aeron10}
S.~Aeron, V.~Saligrama, and M.~Zhao, ``Information theoretic bounds for
  compressed sensing,'' \emph{{IEEE} Trans. Inf. Theory}, vol.~56, no.~10, pp.
  5111--5130, Oct. 2010.

\bibitem{Scarlett17}
J.~Scarlett and V.~Cevher, ``Limits on support recovery with probabilistic
  models: An information-theoretic framework,'' \emph{{IEEE} Trans. Inf.
  Theory}, vol.~63, no.~1, pp. 593--620, Jan. 2017.

\bibitem{Aksoylar17}
C.~Aksoylar, G.~K. Atia, and V.~Saligrama, ``Sparse signal processing with
  linear and nonlinear observations: A unified {Shannon}-theoretic approach,''
  \emph{{IEEE} Trans. Inf. Theory}, vol.~63, no.~2, pp. 749--776, Feb. 2017.

\bibitem{Gamarnik17}
D.~Gamarnik and I.~Zadik, ``High dimensional regression with binary
  coefficients. estimating squared error and a phase transtition,'' in
  \emph{Proc. 30th Annu. Conf. Learn. Theory}, Amsterdam, Netherlands, Jul.
  2017.

\bibitem{Reeves20}
G.~Reeves, J.~Xu, and I.~Zadik, ``The all-or-nothing phenomenon in sparse
  linear regression,'' \emph{Math. Stat. Learn.}, vol.~3, no. 3/4, pp.
  259--313, 2020.

\bibitem{Donoho09}
D.~L. Donoho, A.~Maleki, and A.~Montanari, ``Message-passing algorithms for
  compressed sensing,'' \emph{Proc. Nat. Acad. Sci.}, vol. 106, no.~45, pp.
  18\,914--18\,919, Nov. 2009.

\bibitem{Rangan11}
S.~Rangan, ``Generalized approximate message passing for estimation with random
  linear mixing,'' in \emph{Proc. 2011 IEEE Int. Symp. Inf. Theory}, Saint
  Petersburg, Russia, Aug. 2011, pp. 2168--2172.

\bibitem{Reeves191}
G.~Reeves and H.~D. Pfister, ``The replica-symmetric prediction for random
  linear estimation with {Gaussian} matrices is exact,'' \emph{{IEEE} Trans.
  Inf. Theory}, vol.~65, no.~4, pp. 2252--2283, Apr. 2019.

\bibitem{Barbier201}
J.~Barbier, N.~Macris, M.~Dia, and F.~Krzakala, ``Mutual information and
  optimality of approximate message-passing in random linear estimation,''
  \emph{{IEEE} Trans. Inf. Theory}, vol.~66, no.~7, pp. 4270--4303, Jul. 2020.

\bibitem{Barbier19}
J.~Barbier, F.~Krzakala, N.~Macris, L.~Miolane, and L.~Zdeborov\'a, ``Optimal
  errors and phase transitions in high-dimensional generalized linear models,''
  \emph{Proc. Nat. Acad. Sci.}, vol. 116, no.~12, pp. 5451--5460, Mar. 2019.

\bibitem{Bayati11}
M.~Bayati and A.~Montanari, ``The dynamics of message passing on dense graphs,
  with applications to compressed sensing,'' \emph{{IEEE} Trans. Inf. Theory},
  vol.~57, no.~2, pp. 764--785, Feb. 2011.

\bibitem{Bayati15}
M.~Bayati, M.~Lelarge, and A.~Montanari, ``Universality in polytope phase
  transitions and message passing algorithms,'' \emph{Ann. Appl. Probab.},
  vol.~25, no.~2, pp. 753--822, Apr. 2015.

\bibitem{Javanmard13}
A.~Javanmard and A.~Montanari, ``State evolution for general approximate
  message passing algorithms, with applications to spatial coupling,''
  \emph{Inf. Inference: A Journal of the IMA}, vol.~2, no.~2, pp. 115--144,
  Dec. 2013.

\bibitem{Takeuchi242}
K.~Takeuchi, ``Decentralized generalized approximate message-passing for
  tree-structured networks,'' \emph{IEEE Trans. Inf. Theory}, vol.~70, no.~10,
  pp. 7385--7409, Oct. 2024.

\bibitem{Ma17}
J.~Ma and L.~Ping, ``Orthogonal {AMP},'' \emph{IEEE Access}, vol.~5, pp.
  2020--2033, Jan. 2017.

\bibitem{Rangan192}
S.~Rangan, P.~Schniter, and A.~K. Fletcher, ``Vector approximate message
  passing,'' \emph{{IEEE} Trans. Inf. Theory}, vol.~65, no.~10, pp. 6664--6684,
  Oct. 2019.

\bibitem{Barbier18}
J.~Barbier, N.~Macris, A.~Maillard, and F.~Krzakala, ``The mutual information
  in random linear estimation beyond i.i.d. matrices,'' in \emph{Proc. 2018
  IEEE Int. Symp. Inf. Theory}, Vail, CO, USA, Jun. 2018, pp. 1390--1394.

\bibitem{Li24}
Y.~Li, Z.~Fan, S.~Sen, and Y.~Wu, ``Random linear estimation with
  rotationally-invariant designs: Asymptotics at high temperature,''
  \emph{{IEEE} Trans. Inf. Theory}, no.~3, pp. 2118--2153, Mar. 2024.

\bibitem{Takeuchi20}
K.~Takeuchi, ``Rigorous dynamics of expectation-propagation-based signal
  recovery from unitarily invariant measurements,'' \emph{{IEEE} Trans. Inf.
  Theory}, vol.~66, no.~1, pp. 368--386, Jan. 2020.

\bibitem{Wang24}
T.~Wang, X.~Zhong, and Z.~Fan, ``Universality of approximate message passing
  algorithms and tensor networks,'' \emph{Ann. Appl. Probab.}, vol.~34, no.~4,
  pp. 3943--3994, Aug. 2024.

\bibitem{Dudeja24}
R.~Dudeja, S.~Sen, and Y.~M. Lu, ``Spectral universality in regularized linear
  regression with nearly deterministic sensing matrices,'' \emph{{IEEE} Trans.
  Inf. Theory}, vol.~70, no.~11, pp. 7923--7951, Nov. 2024.

\bibitem{Gallager68}
R.~G. Gallager, \emph{Information Theory and Reliable Communication}.\hskip 1em
  plus 0.5em minus 0.4em\relax Hoboken, NJ, USA: Wiley, 1968.

\bibitem{Mesleh08}
R.~Y. Mesleh, H.~Haas, S.~Sinanovi\'c, C.~W. Ahn, and S.~Yun, ``Spatial
  modulation,'' \emph{{IEEE} Trans. Veh. Technol.}, vol.~57, no.~4, pp.
  2228--2241, Jul. 2008.

\bibitem{Jeganathan08}
J.~Jeganathan, A.~Ghrayeb, and L.~Szczecinski, ``Generalized space shift keying
  modulation for {MIMO} channels,'' in \emph{Proc. IEEE 19th Int. Symp.
  Personal, Indoor, Mobile Radio Commun.}, Cannes, France, Sep. 2008.

\bibitem{Basar13}
E.~Ba\c{s}ar, U.~Ayg\"{o}l\"{u}, E.~Panayırcı, and H.~V. Poor, ``Orthogonal
  frequency division multiplexing with index modulation,'' \emph{{IEEE} Trans.
  Signal Process.}, vol.~61, no.~22, pp. 5536--5549, Nov. 2013.

\bibitem{Renzo14}
M.~{Di Renzo}, H.~Haas, A.~Ghrayeb, S.~Sugiura, and L.~Hanzo, ``Spatial
  modulation for generalized {MIMO}: Challenges, oppotunities, and
  implementation,'' \emph{Proc. IEEE}, vol. 102, no.~1, pp. 56--103, Jan. 2014.

\bibitem{Polyanskiy10}
Y.~Polyanskiy, H.~V. Poor, and S.~Verd\'u, ``Channel coding rate in the finite
  blocklength regime,'' \emph{{IEEE} Trans. Inf. Theory}, vol.~56, no.~5, pp.
  2307--2359, May 2010.

\bibitem{Polyanskiy13}
Y.~Polyanskiy, ``Saddle point in the minimax converse for channel coding,''
  \emph{{IEEE} Trans. Inf. Theory}, vol.~59, no.~5, pp. 2576--2595, May 2013.

\bibitem{Guo05}
D.~Guo, S.~{Shamai (Shitz)}, and S.~Verd\'u, ``Mutual information and minimum
  mean-square error in {Gaussian} channels,'' \emph{{IEEE} Trans. Inf. Theory},
  vol.~51, no.~4, pp. 1261--1282, Apr. 2005.

\bibitem{Takeuchi262}
K.~Takeuchi, ``Direct and converse theorems in estimating signals with
  sublinear sparsity,'' submitted to {\em 2026 Int. Symp. Inf. Theory and its
  Appl.}

\bibitem{Cover06}
T.~M. Cover and J.~A. Thomas, \emph{Elements of Information Theory},
  2nd~ed.\hskip 1em plus 0.5em minus 0.4em\relax New Jersey: Wiley, 2006.

\bibitem{David03}
H.~A. David and H.~N. Nagaraja, \emph{Order Statistics}, 3rd~ed.\hskip 1em plus
  0.5em minus 0.4em\relax Hoboken, NJ, USA: Wiley, 2003.

\bibitem{Hall35}
P.~Hall, ``On representatives of subsets,'' \emph{J. Lond. Math. Soc.}, vol.
  s1-10, no.~1, pp. 26--30, Jan. 1935.

\bibitem{Telatar99}
E.~Telatar, ``Capacity of multi-antenna {Gaussian} channels,'' \emph{Euro.
  Trans. Telecommun.}, vol.~10, no.~6, pp. 585--595, Nov.--Dec. 1999.

\end{thebibliography}

\end{document}